\documentclass[10pt]{siamltex}
\usepackage[]{amsmath,amssymb,epsfig}
\usepackage[top=18mm, bottom=18mm, left=22mm, right=22mm]{geometry}
\usepackage{subfigure,graphicx,comment,array,url}
\usepackage{algorithm,algpseudocode}  
\usepackage{mathtools,setspace,multicol,multirow}
\usepackage{hyperref}

\makeatletter
\def\BState{\State\hskip-\ALG@thistlm}
\makeatother

\hypersetup{
    colorlinks=true,
    citecolor=red,
    linkcolor=blue,
    urlcolor=blue
}

\newcommand{\mb}[1]{\mbox{\boldmath$#1$}}
\newcounter{ale}

\newenvironment{liste}{\begin{itemize}}{\end{itemize}}
\newcommand{\aliste}{\begin{liste} \setcounter{ale}{1}}
\newcommand{\zliste}{\end{liste}}

\begin{document}

\title{ADM-CLE approach for detecting slow variables in continuous time Markov chains 
and dynamic data\thanks{Submitted to the journal's Computational Methods 
in Science and Engineering section \today.
The research leading to these results has
received funding from the European Research Council under the {\it
European Community}'s Seventh Framework Programme ({\it
FP7/2007-2013})/ ERC {\it grant agreement} No. 239870. This publication is based on work supported 
in part by Award No KUK-C1-013-04, made by King Abdullah University of Science
and Technology (KAUST).}}
\author{Mihai~Cucuringu\footnotemark[2] \footnotemark[3]  \hskip 6mm
and \hskip 3mm Radek~Erban\footnotemark[4]}

\maketitle

\renewcommand{\thefootnote}{\fnsymbol{footnote}}
\footnotetext[2]{Department of Mathematics, UCLA, 520 Portola Plaza, Mathematical Sciences Building
6363, Los Angeles, CA 90095-1555. Email: mihai@math.ucla.edu}
\footnotetext[3]{Program in Applied and Computational Mathematics (PACM), Princeton University, Fine Hall, Washington Road, Princeton, NJ, 08544-1000 USA. This work was initiated when the author was a Ph.D. student supported by PACM.
Mihai Cucuringu also acknowledges support from AFOSR MURI grant FA9550-10-1-0569,  and is grateful to Amit Singer 
for his support, via Award Number R01GM090200 from the NIGMS, and Award Number FA9550-09-1-0551 from AFOSR.
}
\footnotetext[4]{Mathematical Institute, University of Oxford, Radcliffe Observatory Quarter,
Woodstock Road, Oxford, OX2 6GG, United Kingdom. E-mail: erban@maths.ox.ac.uk; 
Radek Erban would like to thank the Royal Society for a University  Research Fellowship; 
Brasenose College, University of Oxford, for  a Nicholas Kurti Junior Fellowship; and the 
Leverhulme Trust for  a Philip Leverhulme Prize. This prize money was used to support  research 
visits of Mihai Cucuringu in Oxford.}
\renewcommand{\thefootnote}{\arabic{footnote}}

\begin{abstract}
A method for detecting intrinsic slow variables in high-dimensional 
stochastic chemical reaction networks is developed and analyzed. It 
combines anisotropic diffusion maps (ADM) with approximations 
based on the chemical Langevin equation (CLE). The resulting approach, 
called ADM-CLE, has the potential of being more efficient than the 
ADM method for a large class of chemical reaction systems, because 
it replaces the computationally most expensive step of ADM (running 
local short bursts of simulations) by using an approximation based 
on the CLE. The ADM-CLE approach can be used to estimate the stationary 
distribution of the detected slow variable, without any a-priori 
knowledge of it. If the conditional distribution of the fast variables 
can be obtained analytically, then the resulting ADM-CLE approach does 
not make any use of Monte Carlo simulations to estimate the distributions 
of both slow and fast variables.
\end{abstract}

\begin{keywords}
diffusion maps, stochastic chemical reaction networks, slow variables,
stationary distributions
\end{keywords}

\section{Introduction}

The time evolution of a complex chemical reaction network often occurs 
at different time scales, and the observer is interested in tracking 
the evolution of the slowly evolving quantities (i.e., of the so called
\textit{slow variables}) as opposed to recording each and every single
reaction that takes place in the system. Whenever a separation of scales 
exists, one has to simulate a large number of reactions in the system 
in order to capture the evolution of the slowly evolving variables. 
With this observation in mind, it becomes crucial to be able to detect 
and parametrize the underlying slow manifold corresponding to the 
slow variables intrinsic to the system. In the present paper, 
we introduce an unsupervised method of discovering the underlying 
hidden slow variables in chemical reaction networks, and of their 
stationary distributions, using the anisotropic diffusion map (ADM)
framework~\cite{amitslowvars}.

The ADM is a special class of diffusion maps which have gained 
tremendous popularity in machine learning and statistical analysis, 
as a robust nonlinear dimensionality reduction technique, in recent 
years~\cite{RS00,BN03,DonohoGrimes2003,Coifman_Lafon}. Diffusion 
maps have been successfully used as a manifold learning tool, 
where it is assumed that the high dimensional data lies on 
a lower dimensional manifold, and one tries to capture the 
underlying geometric structure of the data, a setup where the 
traditional linear dimensionality reduction techniques (such 
as the Principal Component Analysis) have been shown to fail.
In the diffusion maps setup, one constructs or is given 
a sparse weighted connected graph (usually in the form of 
a weighted k-Nearest-Neighbor graph, with each node 
connected only to its $k$ nearest or most similar neighbors), 
and uses it to build the associated \textit{combinatorial Laplacian}
$\tilde{L} = D-W$, where $W$ denotes the matrix of weights and $D$ 
denotes a diagonal matrix with $D_{ii}$ equal to the sum of all 
weights of the node $i$. Next, one considers the generalized 
eigenvalue problem $ \tilde{L} x = \lambda D x$, whose solutions 
are related to the solutions of the eigenvalue problem $L x = \lambda x$,
where  $L = D^{-1} W$ is a row-stochastic matrix often dubbed as the
\textit{random walk normalized Laplacian}. Whenever the pair 
$(\lambda,x)$ is an eigenvalue-eigenvector solution to $L x = \lambda x$,
then so is $(1-\lambda,x)$ for $\tilde{L} x = \lambda D x$.
The (non-symmetric) matrix $L$ can also be interpreted as 
a transition probability matrix of a Markov chain with state 
space given by the nodes of the graph, and entries $L_{ij}$ denoting 
the one-step transition probability from node $i$ to $j$.

In the diffusion map framework, one exploits a property of the top 
nontrivial eigenvector of the graph Laplacian of being piecewise 
constant on subsets of nodes in the domain that correspond to 
the same state associated to the underlying slow variable.
We make this statement precise in Section \ref{sec:EigBinning}, 
and further use the resulting classification in Section
\ref{sec:MarkovSteady} to propose an unsupervised method for 
computing the stationary distribution of the hidden slow variable, 
without using any prior information on its  structure. Since the top
eigenvectors of the above Laplacian define the coarsest modes 
of variation in the data, and have a natural interpretation 
in terms of diffusion and random walks, they have been used in 
a very wide range of applications, including but not limited to 
partitioning~\cite{spielman07_spectral,spielman96_spectral}, 
clustering and community 
detection~\cite{NJW01_spectral,luxburg05_survey,newman2006finding}, 
image segmentation~\cite{ShiMalik00_NCut}, 
ranking~\cite{laplacianRanking,AspremontRelaxPermutation}, and 
data visualization and learning from data~\cite{CLLMNWZ05a,RS00}.

The main application area studied in this paper are stochastic models 
of chemical reaction networks. They are written in terms of stochastic
simulation algorithms (SSAs)~\cite{Gillespie1,Gillespie2} which have 
been used to model a number of biological systems, including the phage
$\lambda$ lysis-lysogeny decision circuit~\cite{cssa3}, circadian 
rhythms~\cite{cssa39}, and the cell cycle~\cite{cssa26}. The Gillespie
SSA~\cite{Gillespie1} is an exact stochastic method that simulates 
every chemical reaction, sampling from the solution of the 
corresponding chemical master equation (CME). To characterize the
behavior of a chemical system, one needs to simulate a large number 
of reactions and realizations, which leads to very computationally 
intensive algorithms. For suitable classes of chemically reacting 
systems, one can sometimes use exact algorithms which are equivalent 
to the Gillespie SSA, but are less computationally intensive, such as 
the Gibson-Bruck SSA~\cite{cssa19} and the Optimized Direct
Method~\cite{cssa7}. However, these methods also stochastically 
simulate the occurrence of every chemical reaction, which can be 
a computationally challenging task for systems with a very large 
number of species. One way to tackle this problem is to use parallel
stochastic simulations~\cite{radekGPU}. In this work, we discuss an
alternative approach which does not make use of parallel stochastic 
simulations, but at the same time, the proposed approach can also 
benefit from large processing power and parallel computing, as 
many steps of our proposed algorithms are highly parallelizable.

An alternative approach to treating the molecular populations as 
discrete random variables, is to describe them in terms of their 
continuously changing concentration, which can be done via the 
Chemical Langevin equation (CLE), a stochastic differential equation
that links the discrete stochastic simulation algorithm with the
deterministic reaction rate equations~\cite{gillepsie}. Although 
such an approach can be less computationally expensive, it comes 
with the disadvantage that, for certain chemical systems, it can 
lead to negative populations~\cite{cssa40}. In addition, note 
that none of the above approaches takes explicit advantage of the 
separation of scales if one exists, a which we will make use of 
in this paper as detailed in Sections \ref{sec:EigBinning} 
and \ref{sec:MarkovSteady}.

It is often the case that a modeller is not interested in every 
single reaction which takes place in the system, but only in the 
slowly evolving quantities. Certain systems possess multiple time 
scales, meaning that one has to simulate a large number of reactions 
to reveal the slow dynamics. Several  
algorithms for chemical networks with fast and slow variables have
already been developed in the literature. The authors of~\cite{cssa24}
proposed to simulate the fast reactions using Langevin dynamics, 
and the slow reactions using the Gillespie algorithm. This approach 
requires both the time scale separation and a sufficiently large 
system volume; however the latter constraint can be avoided using 
probability densities of the fast species conditioned on the slow 
species, and estimating the effective propensity functions of the 
slow species~\cite{cssa5,cssa6,cssa10,cssa34,cssa38}.
An alternative approach to simulating the evolution of the slow 
variables while avoiding doing so for the fast variables, is to 
estimate the probability distribution of the slow 
variables~\cite{cssa14}. The key point in this approach is to use 
short bursts of appropriately initialized stochastic simulations 
to estimate the drift and diffusion coefficients for an approximating
Fokker-Planck equation written in terms of the slow 
variables~\cite{cssa2006ep}. The success of this approach has already 
been demonstrated in a range of applications including materials 
science \cite{cssa23}, cell motility \cite{cssa15}, and social behavior 
of insects \cite{sun2014}.

Ref.~\cite{cssa} introduces the conditional stochastic simulation algorithm 
(CSSA) that allows one to sample efficiently from the distribution of the 
fast variables conditioned on the slow ones~\cite{cssa}, and to estimate 
the coefficients of the effective stochastic differential equation (SDE) 
on the fly via a proposed constrained multiscale algorithm (CMA) algorithm. 
The CMA can be further modified by estimating the drift and diffusion 
coefficients in the form given by the CLE for the slow subsystem, which 
requires the estimation of effective propensity functions of slow 
reactions~\cite{cssa2014sim}. The main question we plan to address in our present 
work builds on and combines two already existing ideas investigated 
in~\cite{cssa} and~\cite{amitslowvars}, and brings several 
computational and algorithmic improvements. 
The above-mentioned CSSA algorithm explicitly makes use of the knowledge of 
the slow variables (often unavailable in many real applications), 
a drawback we plan to address as explained later in 
Section~\ref{sec:EigBinning}, where, driven by the top 
eigenvector of an appropriately constructed Laplacian, we discover 
the underlying slow variable. In doing so, we make use of the ADM 
framework~\cite{amitslowvars} which modifies the traditional diffusion 
map approach to take into account the time-dependence of the data, 
i.e., the time stamp of each of the data points under consideration. 
By integrating local similarities at different scales, the ADM 
gives a global description of the data set.

The rest of this paper is organized as follows. In 
Section~\ref{sec:ADMLangevin} we provide a mathematical framework 
for multiscale modeling of stochastic chemical reactions networks
and detail the two chemical systems via which we use to 
illustrate our approach. In Section \ref{secADMCLE} we introduce 
the ADM-CLE framework and we highlight its differences from the 
approaches which were previously introduced in the literature. 
In Section \ref{sec:EigBinning} we propose a robust mapping 
from the observable space to the ``dynamically meaningful" 
inaccessible space, that allows us to recover the hidden slow 
variables. In Section \ref{sec:MarkovSteady} we introduce 
a Markov-based approach for approximating the steady distribution 
of the slow variable, and compare our results with another 
recently proposed approach. We conclude with a summary and 
discussion of future work in Section \ref{SummaryDiscussion}.

\section{Problem formulation}   
\label{sec:ADMLangevin}

A multi-scale modeling framework for stochastic chemical reaction 
networks can be formulated as follows. We consider a well-mixed 
system of $\ell$ chemical species, denoted  by ${X_1,X_2,\ldots,X_\ell}$ 
that interact through $m$ reaction channels $R_1,R_2,\ldots,R_m$ 
in a reactor of volume $V$. We denote the state of the system by
$X(t)=[X_1(t), X_2, \ldots, X_\ell(t)]$, where $X_i(t),$ 
$i=1,2,\ldots,\ell$ represents the number of molecules of type $X_i$ in 
the system at time $t$. With a slight abuse of notation, we interchangeably 
use $X_i$ to denote the type $i$ of the molecule. In certain 
scenarios, one may assume that the reactions can be classified as 
either fast or slow, depending on the time scale of 
occurrence~\cite{cssa5}. As expected, the fast reactions occur 
many times on a timescale for which the slow reactions occur 
with very small probability. As defined in~\cite{cssa5}, 
the fast species denoted by \textbf{F} are those species whose 
population gets changed by a fast reaction. Slow species (denoted 
by \textbf{S}) are not changed by fast reactions. Considering that 
slow species are not only species from the set $\{X_1,X_2,\ldots,X_\ell\}$, 
but also their functions which are not changed by fast reactions, 
the components of the fast and slow species can be used as a basis 
for the state space of the system, whose dimension equals the 
number of linearly independent species. 

For each reaction channel $R_j$, $j = 1,2,\ldots,m$, there exists 
a corresponding propensity function $\alpha_j \equiv \alpha_j(\mb{x})$, 
such that $\alpha_j \, \text{d}t$  denotes the probability that, given 
$\mb{X}(t)=\mb{x}$, reaction $R_j$ occurs within the 
infinitesimal time interval $[t,t + \text{d} t)$. We denote by 
$\nu$ the stochiometrix matrix of size $m \times \ell$, 
with entry $\nu_{ji}$ denoting the change in the number 
of molecules of type $X_i$ caused by one occurrence of reaction 
channel $R_j$. The continuous time discrete in space Markov 
chain can be further approximated by the CLE for 
a multivariate continuous Markov process~\cite{gillepsie}.
Using time step $\Delta t$, the Euler-Maruyama discretization of 
the CLE is given by 
\begin{equation}
X_i(t+\Delta t) 
= 
X_i(t) 
+ \Delta t \sum_{j=1}^{m} \nu_{ji} \, \alpha_j(\mb{X}(t)) 
+ \sum_{j=1}^{m} \nu_{ji}
\sqrt{\alpha_j(\mb{X}(t))} \, N_{j}(t) \, \sqrt{\Delta t}, 
\qquad \quad
\mbox{ for all } i=1,2,\ldots,\ell,
\label{gillApprox}
\end{equation}
where $X_i$, with another slight abuse of notation, denotes 
a real-valued approximation of the number of molecules of 
the $i$-th chemical species, $i=1,2,\dots,\ell$. Here, 
$N_j(t)$, $j=1,2,\dots,m$, denote the set of $m$ independent 
normally distributed random variables with zero mean and unit
variance. 

\subsection{Illustrative example CS-I}  
\label{sec:subsectEx1Intro}

\begin{figure}[t]
\centerline{
\hskip 3mm
\raise 4.4cm \hbox{\hbox{(a)}}
\hskip -8mm
% \hskip -4mm
\includegraphics[width=0.43\columnwidth]{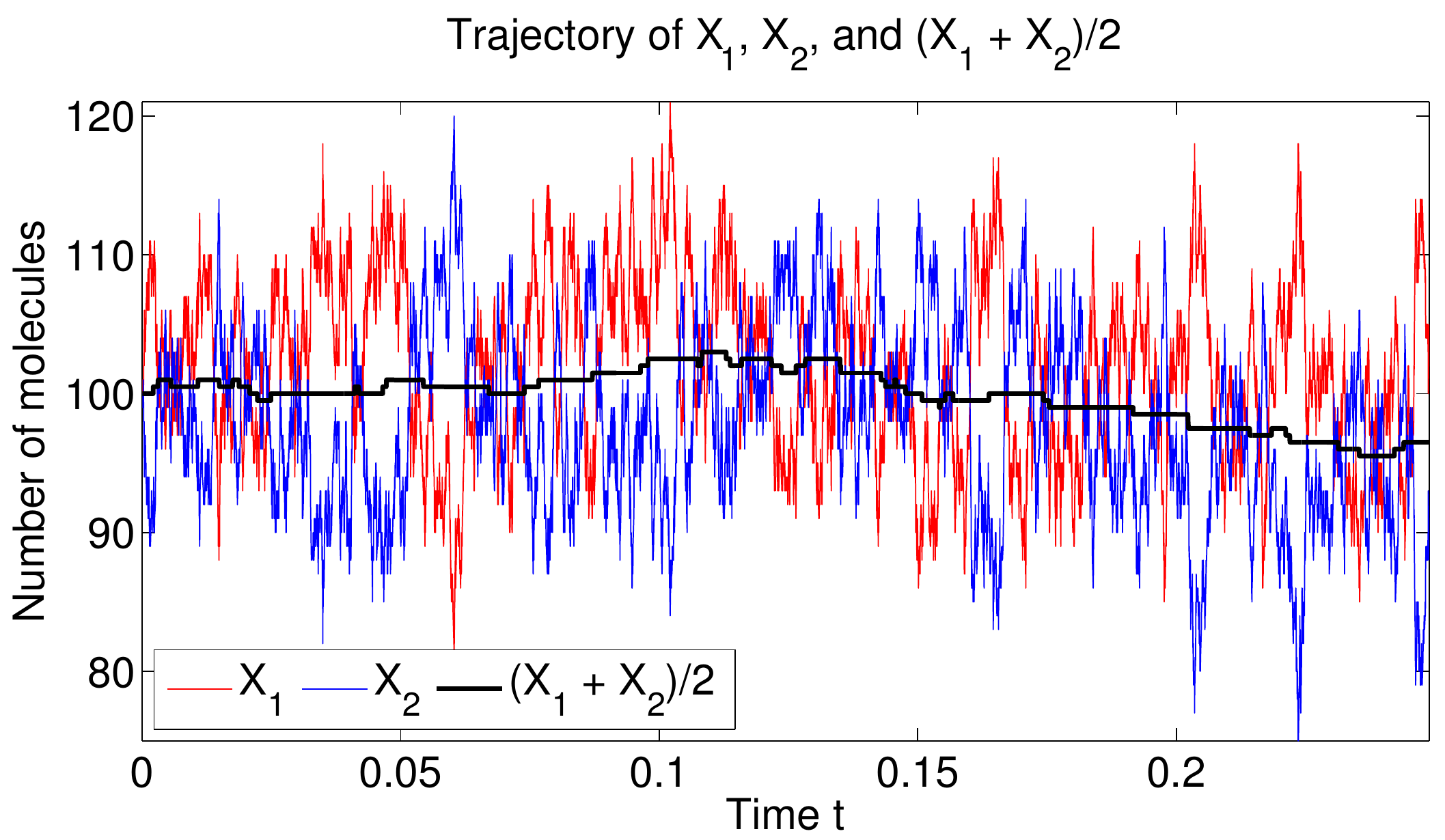}
\hskip 5mm
\raise 4.4cm \hbox{\hbox{(b)}}
\hskip -4mm
\includegraphics[width=0.43\columnwidth]{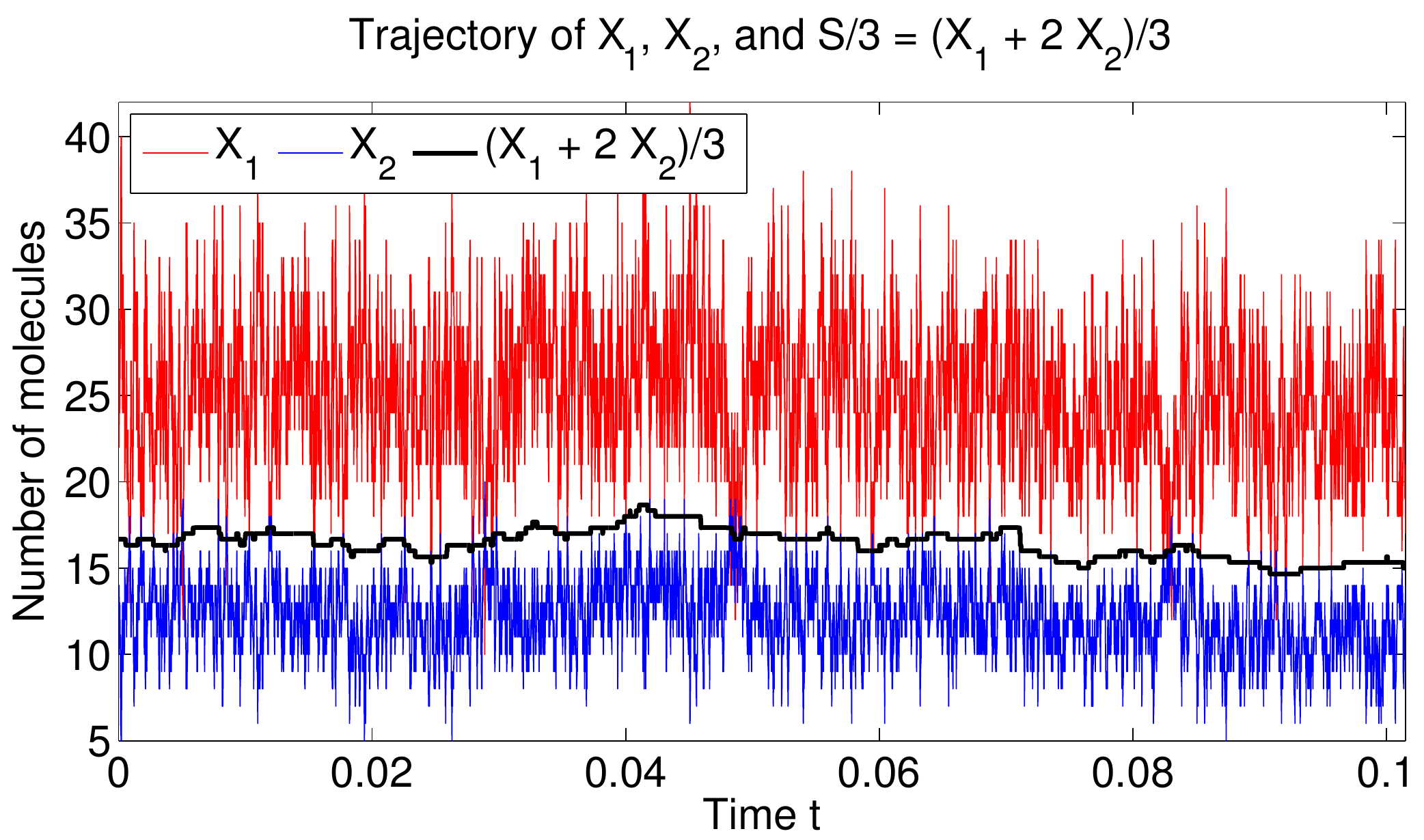}
}
\caption{
{\rm (a)}
{\it Trajectories of the {\rm CS-I} considered in $(\ref{ex1system})$
showing the behavior of the slow variable 
$S = (X_1 + X_2)/2$ in contrast to the behavior of the fast variables 
$X_1$ and $X_2$, where the system propensity functions and parameters 
are given by $(\ref{propExample1})$ and $(\ref{parsExample1})$.}
{\rm (b)}
{\it Trajectories of the {\rm CS-II} considered in $(\ref{ex2system})$, 
showing the slow behavior of the variable $S = X_1 +2 X_2$ in contrast 
to the fast behavior of variables $X_1$ and  $X_2$, where the system 
parameters 
are given by $(\ref{parsExample2})$.}
}
\label{figure2p1}
\end{figure}

As the first illustrative example, we consider the following simple
2-dimensional chemical system, with the two chemical species denoted 
by $X_1$ and $X_2$ (i.e., $\ell=2$) which are subject to four reaction 
channels $R_j,$ $j=1,2,3,4$ (i.e., $m=4$), given by
\begin{equation}
\emptyset  \xrightarrow{k_1} X_1 
\; 
\mbox{\raise -0.9 mm \hbox{$
\displaystyle
\mathop{\stackrel{\displaystyle\longrightarrow}\longleftarrow}^{k_2}_{k_3}
$}}
\; X_2  \xrightarrow{k_4}  \emptyset. 
\label{ex1system}
\end{equation}
Throughout the rest of this paper, we shall refer to the 
chemical system~(\ref{ex1system}) as CS-I (i.e. ``chemical system I").
We label by $R_j$ the reaction corresponding to the 
reaction rate subscript $k_j$, $j=1,2,3,4$, and note that each 
reaction $R_j$ has associated a propensity function $\alpha_j(t)$ 
given by \cite{Gillespie1}
\begin{equation}
\alpha_1(t) = k_1 V, 
\hspace{5mm}  
\alpha_2(t)=k_2 \, X_1(t), 
\hspace{5mm}  
\alpha_3(t)=k_3 \, X_2(t),  
\hspace{5mm}  
\alpha_4(t) = k_4 \, X_2(t), 
\label{propExample1}
\end{equation}
where $V$ denotes the volume of the reactor. We consider the system 
with the following dimensionless parameters  
\begin{equation}
k_1 V=100, \qquad k_2 = k_3 = 200 \qquad
\mbox{and} \qquad k_4=1.
\label{parsExample1}
\end{equation}
We plot in Figure \ref{figure2p1}(a) the time evolution of the two 
different species in system (\ref{ex1system}), together with the 
slow variable $S=(X_1+X_2)/2$, starting from initial conditions 
$X_1(0) = X_2(0) = 100$. As the figure shows, the system variables 
$X_1$ and $X_2$ are changing very frequently (thus we label them 
as fast variables), while the newly defined variable $S$ 
changes very infrequently and can be considered to be 
a slow variable. 

Following \cite{monomolecular}, for the chemical system 
in (\ref{ex1system}) comprised only of monomolecular 
reactions, it is possible to compute analytically the 
stationary distribution of the slow variable $S$, 
since the joint probability distribution of the two 
variables  $X_1$ and $X_2$ is a multivariate Poisson 
distribution 
\begin{equation}
\mathbb{P}(X_1 = n_1, X_2 = n_2) 
= \frac{ \bar{\lambda}_1^{n_1}}{n_1!} 
\frac{\bar{\lambda}_2^{n_2}}{n_2!} 
\, \exp \!
\left( - \bar{\lambda}_1 
- \bar{\lambda}_2   
\right)
\end{equation}
with parameters given by 
\begin{equation}
\bar{\lambda}_1 =\frac{k_1 V (k_3+k_4)}{k_2 k_4}=100.5  
\hspace{5mm} 
\mbox{and}  
\hspace{5mm} 
\bar{\lambda}_2 = \frac{k_1V}{k_4}=100.
\label{analyticPoisson}
\end{equation}

\subsection{Illustrative example CS-II}  
\label{sec:subsectEx2Intro}

The second example is taken from Ref.~\cite{cssa}. We shall refer to 
it as CS-II from now on. We consider the following system
\begin{equation}
X_2  
\; 
\mbox{\raise -0.9 mm \hbox{$
\displaystyle
\mathop{\stackrel{\displaystyle\longrightarrow}\longleftarrow}^{k_1}_{k_2}
$}}
\; 
X_1+X_2, 
\hspace{11 mm} 
\emptyset  
\; 
\mbox{\raise -0.9 mm \hbox{$
\displaystyle
\mathop{\stackrel{\displaystyle\longrightarrow}\longleftarrow}^{k_3}_{k_4}
$}}
\;
X_1,  
\hspace{11 mm} 
X_1+X_1
\; 
\mbox{\raise -0.9 mm \hbox{$
\displaystyle
\mathop{\stackrel{\displaystyle\longrightarrow}\longleftarrow}^{k_5}_{k_6}
$}}
\;
X_2,
\label{ex2system}
\end{equation}
involving two molecular species $X_1$ and $X_2$, whose 
reactions $R_1, R_2, \ldots, R_6$ have the
propensity functions given by
\begin{eqnarray*}
\alpha_1(t)=k_1 X_2(t),  &   
\hspace{11 mm}  \alpha_2(t)=k_2 X_1(t)X_2(t)/V,
&  \hspace{11 mm}  \alpha_3(t)=k_3 V, \\
\alpha_4(t)=k_4 X_1(t),  &  \hspace{11 mm}   
\alpha_5(t)=k_5 \frac{X_1(t) (X_1(t)-1)}{V}, &  
\hspace{11 mm}  \alpha_6(t)=k_6 X_2(t), 
\label{propExample2}
\end{eqnarray*}   
where $V$ denotes the system volume. 
Figure \ref{figure2p1}(b) shows a simulated trajectory 
of this chemical system using the Gillespie algorithm 
for the following dimensionless parameters~\cite{cssa} 
\begin{equation}
k_1=32,  
\hspace{4mm} k_2=0.04 V,   
\hspace{4mm} k_3 V = 1475,    
\hspace{4mm} k_4=19.75,  
\hspace{4mm} k_5=10V,  
\hspace{4mm} k_6=4000, 
\label{parsExample2}
\end{equation}
where we use $V=8$. Note that in this second example, 
reactions $R_5$ and $R_6$ are occurring on a much faster 
timescale than the other four reactions $R_1,$ $R_2,$ $R_3$ 
and $R_4$. A natural choice for the slow variable is  
$S=X_1+2 X_2$, which is invariant with respect to all 
fast reactions~\cite{cssa}, as we illustrate in Figure \ref{figure2p1}(b). 

\subsection{Main problem}
Our end goal in the present paper is to propose an algorithm 
that efficiently and accurately estimates the stationary probability density 
of the hidden slow variable $S$, without any prior knowledge of it. The 
approach we propose builds on the anisotropic diffusion map framework (ADM) 
to implicitly discover the mapping from the observable state space to the 
dynamically meaningful coordinates of the fast and slow variables, as 
previously introduced in \cite{amitslowvars}, and on the CLE
approximation (\ref{gillApprox}).

\section{ADM-CLE approach}
\label{secADMCLE}
Let us consider example CS-II, and assume that 
$s=s(x_1,x_2)=x_1+ 2 x_2$ and $f=f(x_1,x_2)=x_1$ are the slowly 
and rapidly changing variables, respectively. They together 
define a mapping $g : (x_1,x_2) \mapsto (s,f)$ 
from the observable state variables $x_1$ and $x_2$ in the 
accessible space $\mathcal{O}$  
to the ``dynamically meaningful" (but in more complicated 
examples inaccessible)
slow variable $s$ and the fast accessible 
variable $f$, both in space $\mathcal{H}$. In other words, 
$g$ maps $(x_1,x_2) \mapsto (x_1+ 2 x_2, x_1)$, 
and conversely its inverse 
$h := g^{-1} : (s,f) \mapsto (f,\frac{s - f}{2})$.

The approach introduced in~\cite{amitslowvars} exploits the 
local point clouds generated by many local bursts of simulations 
at each point $(x_1,x_2)$ in the observable 
space $\mathcal{O}$. Such observable local point clouds are the 
image under $h$ of similar local point clouds in the inaccessible 
space $\mathcal{H}$ (at corresponding coordinates $(s,f)$ such that
$h(s,f)=(x_1,x_2)$), which, due to the separation of scales between 
the fast and slow variables $f$ and $s$, have the appearance of 
thin elongated ellipses. It is precisely this separation of scales 
that we leverage into building a sparse anisotropic graph Laplacian 
$L$ in the observable space, and use it as an approximation of 
the isotropic graph Laplacian in the inaccessible space $\mathcal{H}$. 
As we shall see, the top nontrivial eigenvector of $L$ will robustly 
indicate all pairs of original states 
$(x_1, x_2)$ that correspond to the same slow variable $S=s$ 
(where $s=x_1+2 x_2$ for CS-II). In other words, we discover 
on the fly the structure of the slow variable $S$, and further 
integrate this information into a Markov-based method for estimating 
its stationary distribution $\mathbb{P}(S=s)$, while also computing 
along the way an analytical expression for the conditional 
distribution of the fast variable given the slow variable
$\mathbb{P}(F=f|S=s)$.

Singer et al.~\cite{amitslowvars} run many local bursts of 
simulations for a short time step $\delta t$ starting 
at $(x_1,x_2)$. Such trajectories end up at random locations forming 
a cloud of points in the observable plane $\mathcal{O}$, with 
a bivariate normal distribution with $2 \times 2$ covariance 
matrix $\Sigma$. The shape of the resulting point cloud is an 
ellipse, whose axes reflect the dynamics of the data points. 
In other words, when there is a separation of scales, the ellipses 
are thin and elongated, 
with the ratio between the axis of the ellipse given by the 
ratio 
\begin{equation}
\tau=\frac{\widehat{\lambda}_1}{\widehat{\lambda}_2}
\label{taudef}
\end{equation} 
of the two eigenvalues of $\Sigma$. The first eigenvector corresponding to $\widehat{\lambda}_1$ 
points in the direction of the fast dynamics on the line 
$x_1 + 2x_2 = s$, while the second one points in the direction 
of the slow dynamics. In particular, $\tau$ is a small parameter,
i.e. $0 < \tau \ll 1$. In general, we wish to piece together 
locally defined components into a globally consistent framework, 
a nontrivial task when the underlying unobservable slow variables 
(or the propensity functions of the system) are complicated nonlinear 
functions of the observable variables in $\mathcal{O}$.

The construction of the ADM framework in \cite{amitslowvars} relates 
the anisotropic graph Laplacian in the observable space $\mathcal{O}$ 
with the isotropic graph Laplacian in the inaccessible space 
$\mathcal{H}$. In that setup, each of the $N$ data points 
${\mathbf x}^{(i)},$ $i =1,2,\ldots,N,$ lives in an $\ell$-dimensional 
data space. For both CS-I and CS-II, the data is two-dimensional, thus 
$\ell=2$. For the former system, we consider each lattice point in the 
domain $[50,150] \times [50,150] $, hence there are $N = 101^2 = 10,201$ 
states, 
while for the latter one we consider the domain $[1,110] \times [1,110]$, 
i.e. $N=110^2=12,100$. Throughout the paper, we will often refer to the 
$N$ data points ${\mathbf x}^{(i)}=(x_1,x_2)^{(i)},$ 
$i=1,2,\ldots,N,$ as $\mathcal{O}$-states of 
the chemical system. The ADM~\cite{amitslowvars} then generates 
ensembles of short simulation bursts at each of the $N$ points in the 
data set, computes the averaged position after statistically 
averaging over the many simulated trajectories, and obtain 
an estimate of the local $2 \times 2$ covariance matrix 
$\Sigma_{(i)}$. For each data point ${\mathbf x}^{(i)}$, 
the inverse of $\Sigma_{(i)}$ is computed and
symmetric $\Sigma$-dependent squared distance between pairs 
of data points in the
two-dimensional observable space $\mathbb{R}^2$
(given by (\ref{defDist}) below) is defined. The ADM framework 
then uses this dynamic distance measure to approximate the 
Laplacian on the underlying hidden slow manifold. 
We provide further details on the anisotropic diffusion maps 
framework in Section \ref{sec:ADMKernel}. We now highlight
the first difference between the approach taken in this paper
and in~\cite{amitslowvars}.

\subsection{Replacing short simulation bursts by the CLE approximation}

The local bursts of simulations initiated at each data point 
in order to estimate the local covariances may be computationally 
expensive to estimate. In this paper, we bypass these short 
bursts of simulations by using an approximation given by the CLE 
(\ref{gillApprox}), which allows for a theoretical derivation 
of the local $2 \times 2$ covariance matrices. 
Using (\ref{gillApprox}), we obtain
\begin{eqnarray}
 \mbox{Cov} (X_i(t+\Delta t),X_k(t+\Delta t)) &=& 
 \mathbb{E}[ X_i(t+\Delta t) X_k(t+\Delta t)]
 -\mathbb{E}[ X_i(t+\Delta t) ]\mathbb{E}[ X_k(t+\Delta t) ]
 \nonumber \\
 &=&
 \Delta t \sum_{j=1}^{m} \nu_{ji} \, \nu_{jk} \, \alpha_j(\mb{X}).
\label{anformula}
\end{eqnarray}
Computing the eigen-decomposition of a local covariance 
matrix is analogous to performing the Principal Component Analysis 
on the local cloud of points, generated by the short simulations bursts. 
The advantage of (\ref{anformula}) over the computational approach used
in \cite{amitslowvars} is that $\Sigma_{(i)}$ can be computed at
each data point without running (computationally
intensive) short bursts of simulations. The error of the CLE
approximation depends on the values of coordinates of the data
point ${\mathbf x}^{(i)}$, i.e. on the system 
volume $V$~\cite{grima,duncan}. In the case of CS-I or CS-II, 
the most probable states contain about one hundred molecules of 
each chemical species and the CLE approximation (\ref{anformula})
is well justified.

\subsection{Anisotropic diffusion kernels}  \label{sec:ADMKernel}

The next task is the integration of all local principal components 
into a global framework, with the purpose of identifying the hidden 
slow variable. We estimate the distance 
(and hence the similarity measure) between the slow variables in the 
underlying inaccessible manifold using the anisotropic graph
Laplacian~\cite{amitslowvars}. We derive a symmetrized second 
order approximation of the (unknown) distances in the inaccessible 
space $\mathcal{H}$, based on the Jacobian of the unknown mapping 
from the inaccessible to the observable space. 
The $\Sigma$-dependent distance between two $\mathcal{O}$-states
is given by 
\begin{equation}
d_{\Sigma}^2 \left( (x_1,x_2)^{(i)} , (x_1,x_2)^{(j)} \right) 
= \frac{1}{2} \left( (x_1,x_2)^{(i)}  - (x_1,x_2)^{(j)} \right)
\left(  
\Sigma_{(x_1,x_2)^{(i)}}^{-1}   
+  \Sigma_{(x_1,x_2)^{(j)}}^{-1}   \right)    
\left( (x_1,x_2)^{(i)}  - (x_1,x_2)^{(j)} \right)^T, \;\;
\label{defDist}
\end{equation}
and represents a second order approximation of the 
Euclidean distance in the 
inaccessible ($s, \tau f$)-space 
\begin{equation}
d_{\Sigma}^2[ (x_1,x_2)^{(i)}  ,  (x_1,x_2)^{(j)} ]  
\approx  ( s^{(i)} - s^{(j)} )^2 + 
\tau^2 ( f^{(i)} - f^{(j)} )^2  
\approx 
( s^{(i)} - s^{(j)} )^2, 
\label{approxChain}
\end{equation}
where the last approximation is due to the fact that 
$\tau$ is a small parameter, see (\ref{taudef}).
Note that it is also possible to extend (\ref{approxChain}) 
to higher dimensions, as long as there exists 
a separation of scales between the set of slow variables and 
the set of fast variables~\cite{amitslowvars}. Using approximation
(\ref{defDist})--(\ref{approxChain}) of the distance 
between states of the slow variable,
we next construct (an approximation of) the Laplacian on the 
underlying hidden slow manifold, using the Gaussian kernel 
as a similarity measure between the slow variable states. 
We build an $N \times N$ similarity matrix $W$ with entries
\begin{equation}
W_{ij} = 
\exp{ \left\{  \frac{  - d_{\Sigma}^2[ (x_1,x_2)^{(i)},
(x_1,x_2)^{(j)} ] } {\varepsilon ^2}  \right\} }
\approx    
\exp{ \left\{  \frac{  - ( s^{(i)} - s^{(j)} )^2} {\varepsilon ^2}  
\right\} }
,
 \qquad i,j=1,2,\dots,N,
\label{defWij}
\end{equation}
where the single smoothing parameter $\varepsilon$ (the kernel scale) 
has a two-fold interpretation. On one hand, $\varepsilon$ denotes 
the squared radius of the neighborhood used to infer local geometric
information, in particular $W_{ij}$ is $O(1)$ when $s^{(i)}$ and 
$s^{(j)}$ are in a ball of radius $\sqrt{\varepsilon}$, thus close 
on the underlying slow manifold, but it is exponentially small 
for states that are more than $\sqrt{\varepsilon}$ apart. 
On the other hand, $\varepsilon$ represents the discrete time step 
at which the underlying random walker jumps from one point to 
another. We refer the reader to~\cite{lovasz1993random} for 
a detailed survey of random walks on graphs, and their applications. 
We normalize $W$ using the diagonal matrix $D$
to define the row-stochastic matrix $L$ by
\begin{equation}
D_{ii} = \sum_{j=1}^{N} W_{ij},
\qquad 
\qquad 
L = D^{-1}W.
\label{ALap}
\end{equation} 
Since $L$ is a row-stochastic matrix, it has eigenvalue $\lambda_0=1$ 
with trivial eigenvector ${\boldsymbol \Phi}_0 = (1, 1, \ldots, 1)^{T}$.
The remaining eigenvalues can be ordered as
$$
1= \lambda_0 \geq \lambda_1 \geq \lambda_2 \geq \ldots  \geq \lambda_{N-1}.
$$ 
We denote by ${\boldsymbol \Phi}_i$ the corresponding eigenvectors,
i.e. $L {\boldsymbol \Phi}_i = \lambda_i {\boldsymbol \Phi}_i.$ 
The top $d$ nontrivial eigenvectors of the random-walk anisotropic 
Laplacian $L$ describe the geometry of the underlying $d$-dimensional
manifold~\cite{varfree}, i.e. the $i$-th data
point ${\mathbf x}^{(i)}$ is represented by the following diffusion
map
\begin{equation} 
(\Phi_1(i),\Phi_2(i),\ldots,\Phi_d(i)), \qquad \qquad i=1,2,\ldots,N,
\label{DifMapEmb}
\end{equation}
where $\Phi_j(i)$ denotes the $i$-th component of the eigenvector
${\boldsymbol \Phi}_j.$
However, note that some of the considered eigenvectors can be 
higher harmonics of the same principal direction along the 
manifold, thus in practice one computes the correlation between 
the computed eigenvectors before selecting the above $d$ eigenvectors 
chosen to parametrize the underlying manifold. For the two chemical 
systems considered in this paper, we show in the remainder of this 
section how the top (i.e., $d=1$) non-trivial eigenvector of $L$ 
can be used to successfully recover the underlying slow variable.

\begin{figure}[t]
\centerline{
\hskip 3mm
\raise 3.7cm \hbox{\hbox{(a)}}
\hskip -3mm
\includegraphics[width=0.23\columnwidth]{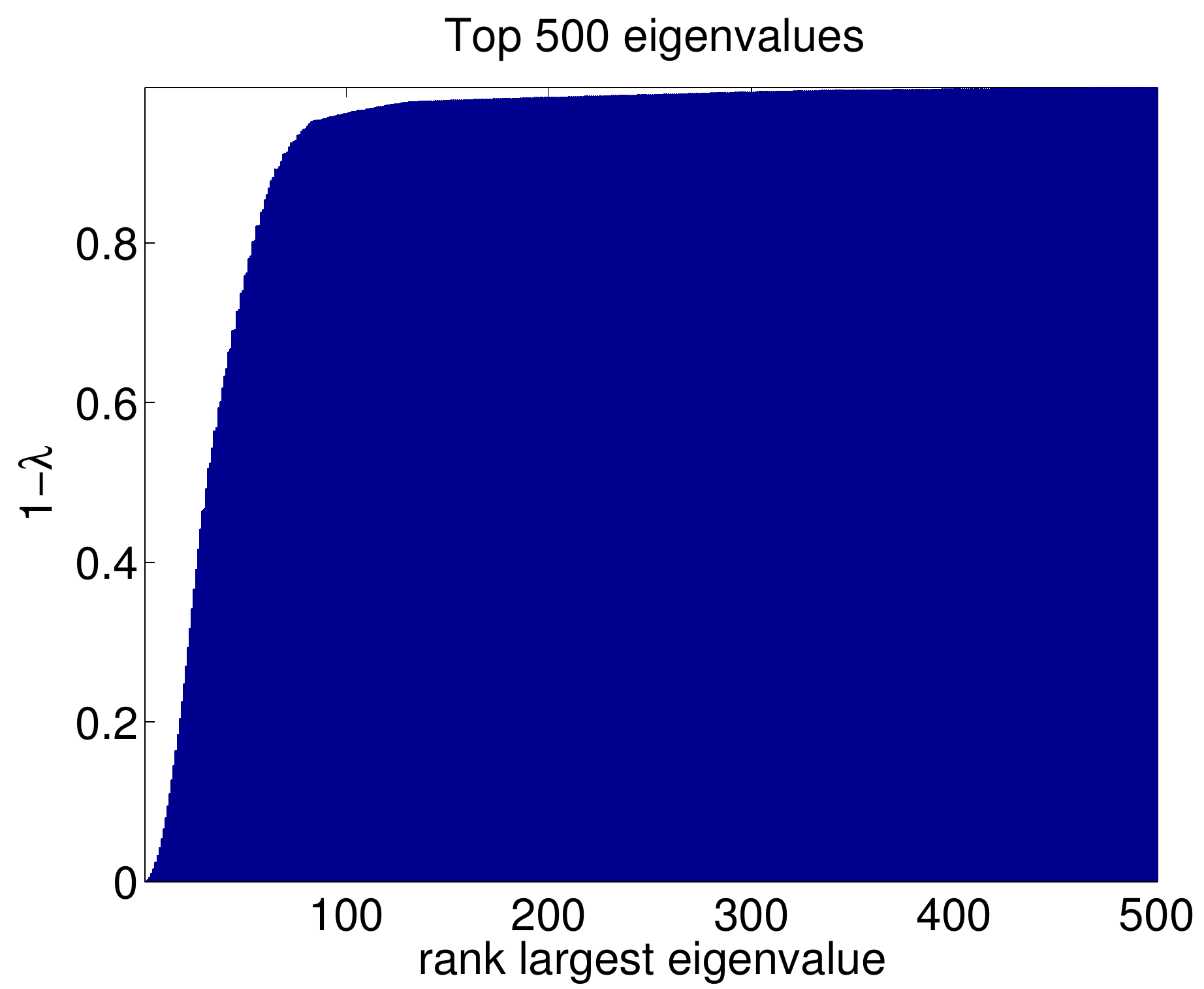}
\raise 3.7cm \hbox{\hbox{(b)}}
\hskip -3mm
\includegraphics[width=0.23\columnwidth]{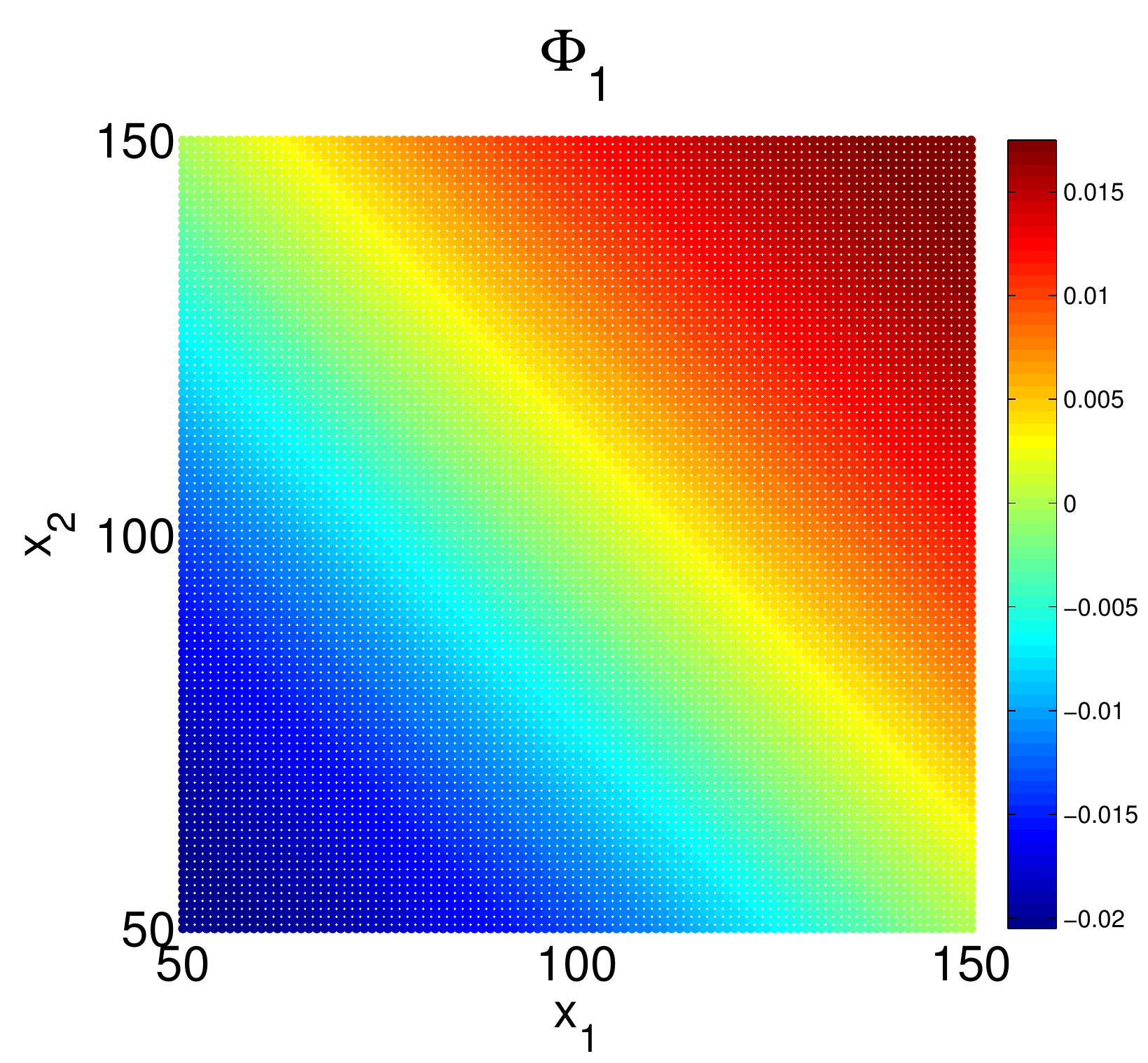}
\raise 3.7cm \hbox{\hbox{(c)}}
\hskip -3mm
\includegraphics[width=0.23\columnwidth]{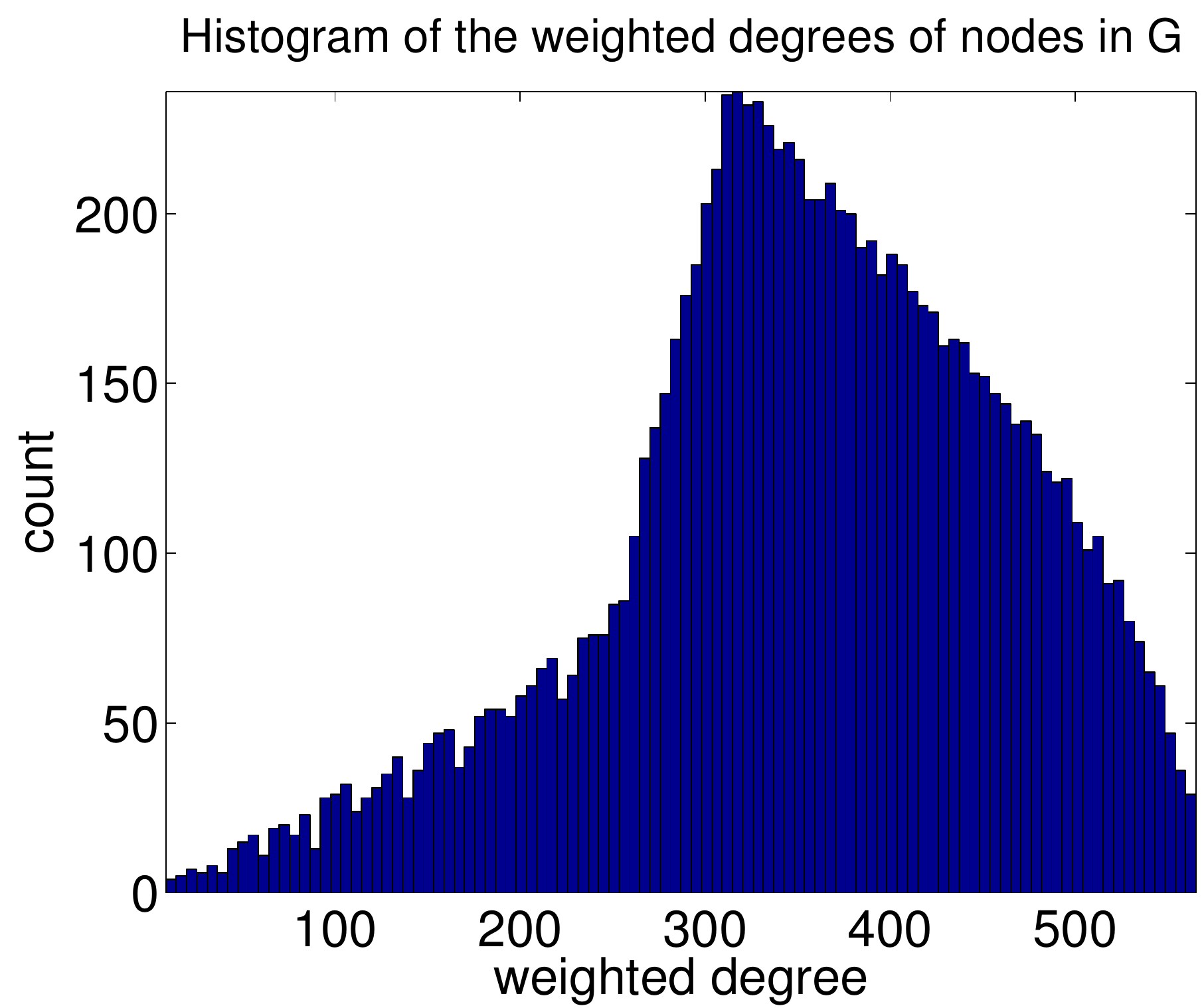}
\raise 3.7cm \hbox{\hbox{(d)}}
\hskip -3mm
\includegraphics[width=0.23\columnwidth]{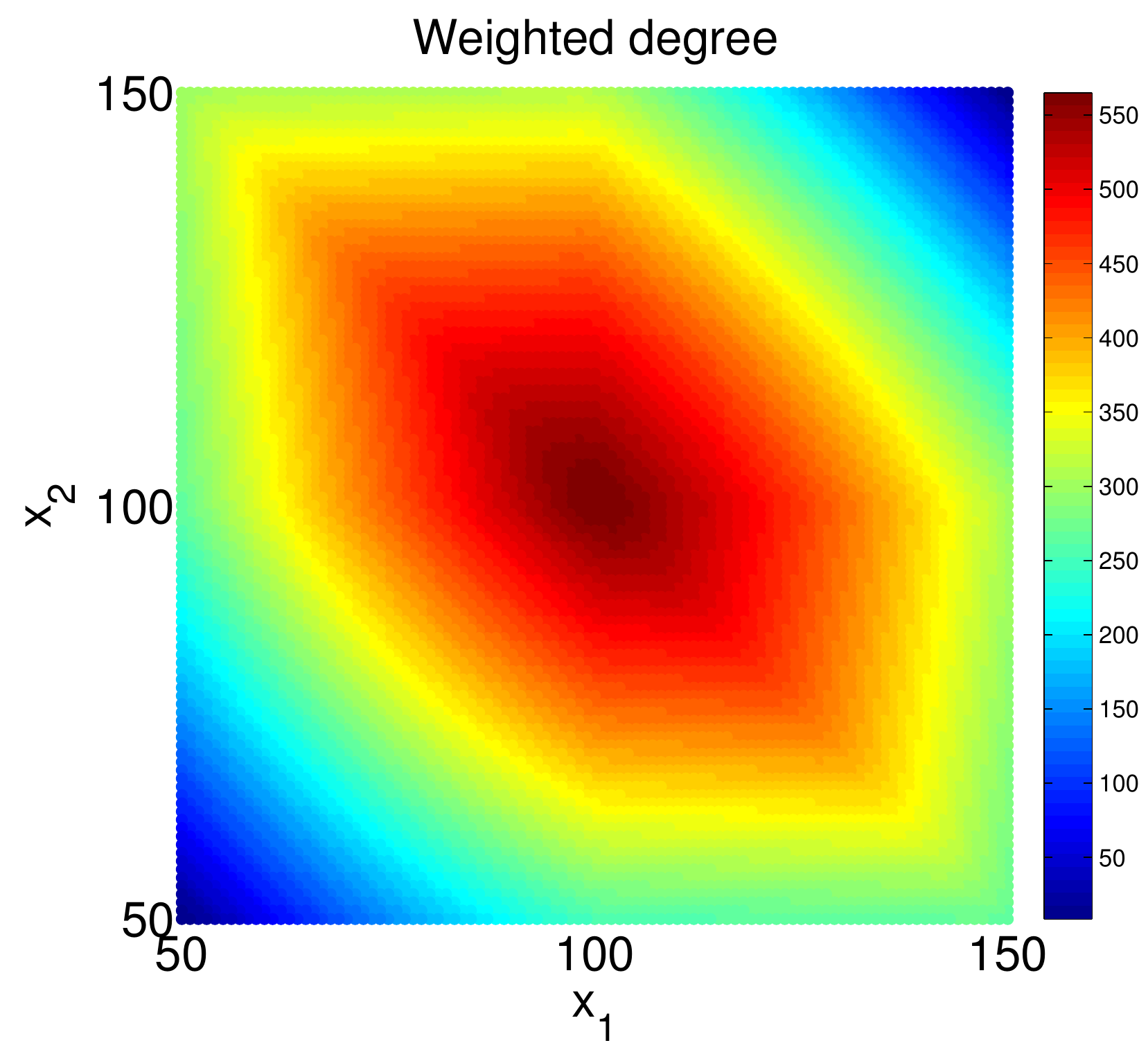}
}
\caption{{\it Illustrative Example} CS-I.
{\rm (a)} {\it The top $500$ eigenvalues of the associated combinatorial 
Laplacian, i.e. $(1-\lambda_i)$ for $i=1,2,\dots,500$.}
{\rm (b)} {\it The coloring of the nodes of $G$ (states of the observable 
space) according to their corresponding entry in the top 
eigenvector ${\boldsymbol \Phi}_1$ of $L$ given by $(\ref{ALap})$.}
{\rm (c)} {\it The weighted degree distribution of the ground state 
graph $G$.}
{\rm (d)} {\it A scatterplot of the states of the system, colored by 
their weighted degree.}} 
\label{figure3p1}
\end{figure}

\begin{figure}[t]
\centerline{
\hskip 3mm
\raise 3.7cm \hbox{\hbox{(a)}}
\hskip -3mm
\includegraphics[width=0.23\columnwidth]{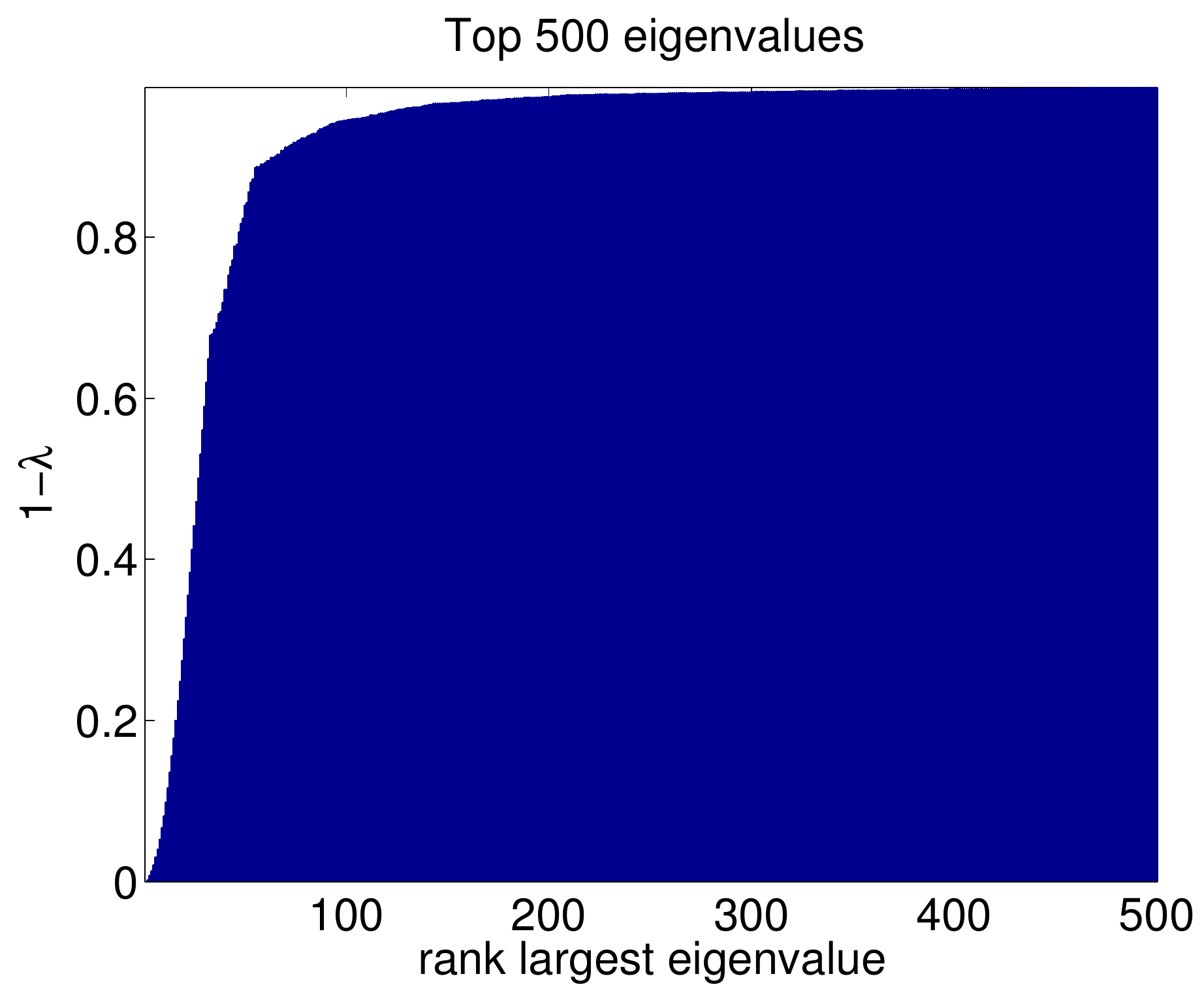}
\raise 3.7cm \hbox{\hbox{(b)}}
\hskip -3mm
\includegraphics[width=0.23\columnwidth]{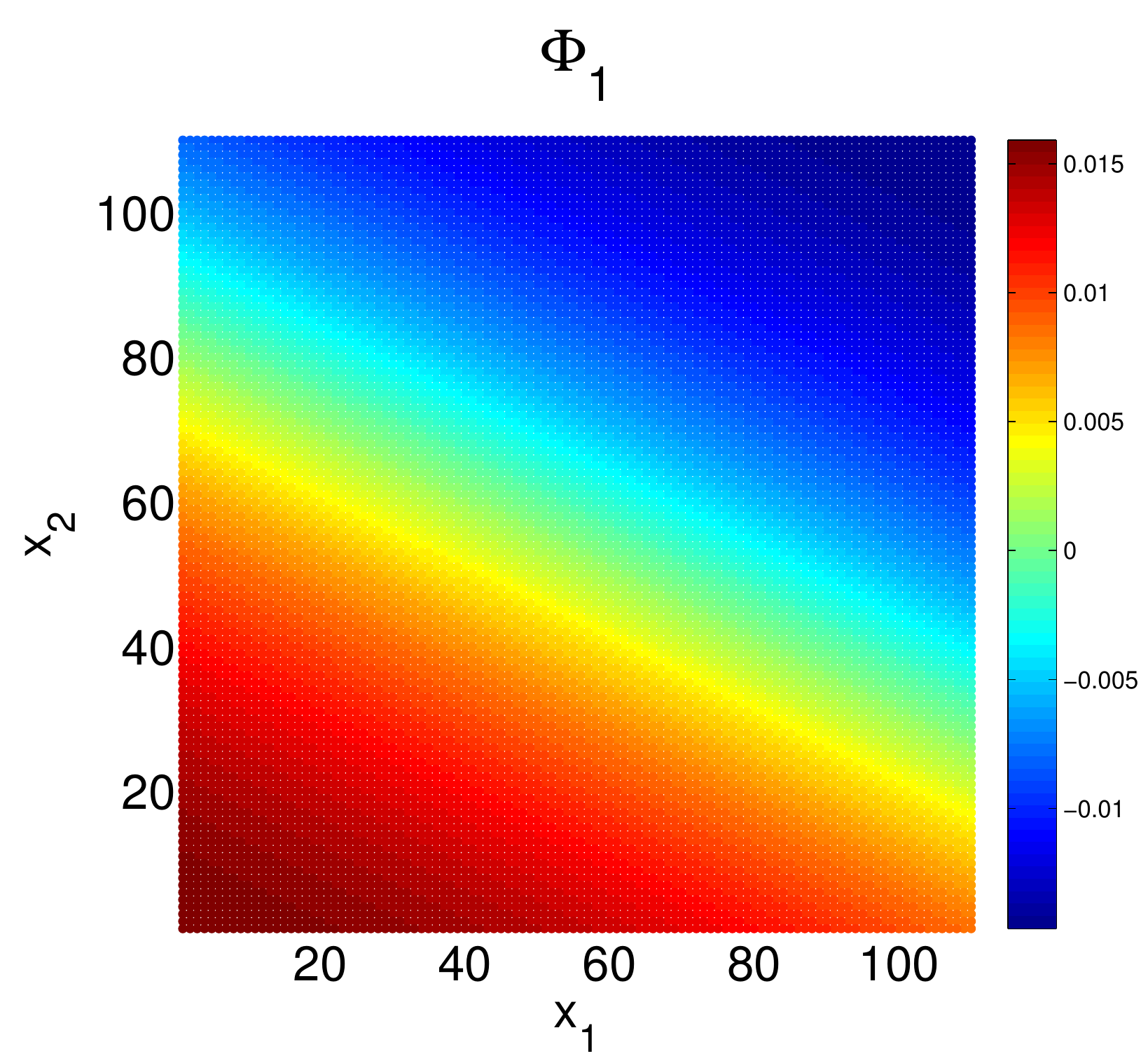}
\raise 3.7cm \hbox{\hbox{(c)}}
\hskip -3mm
\includegraphics[width=0.23\columnwidth]{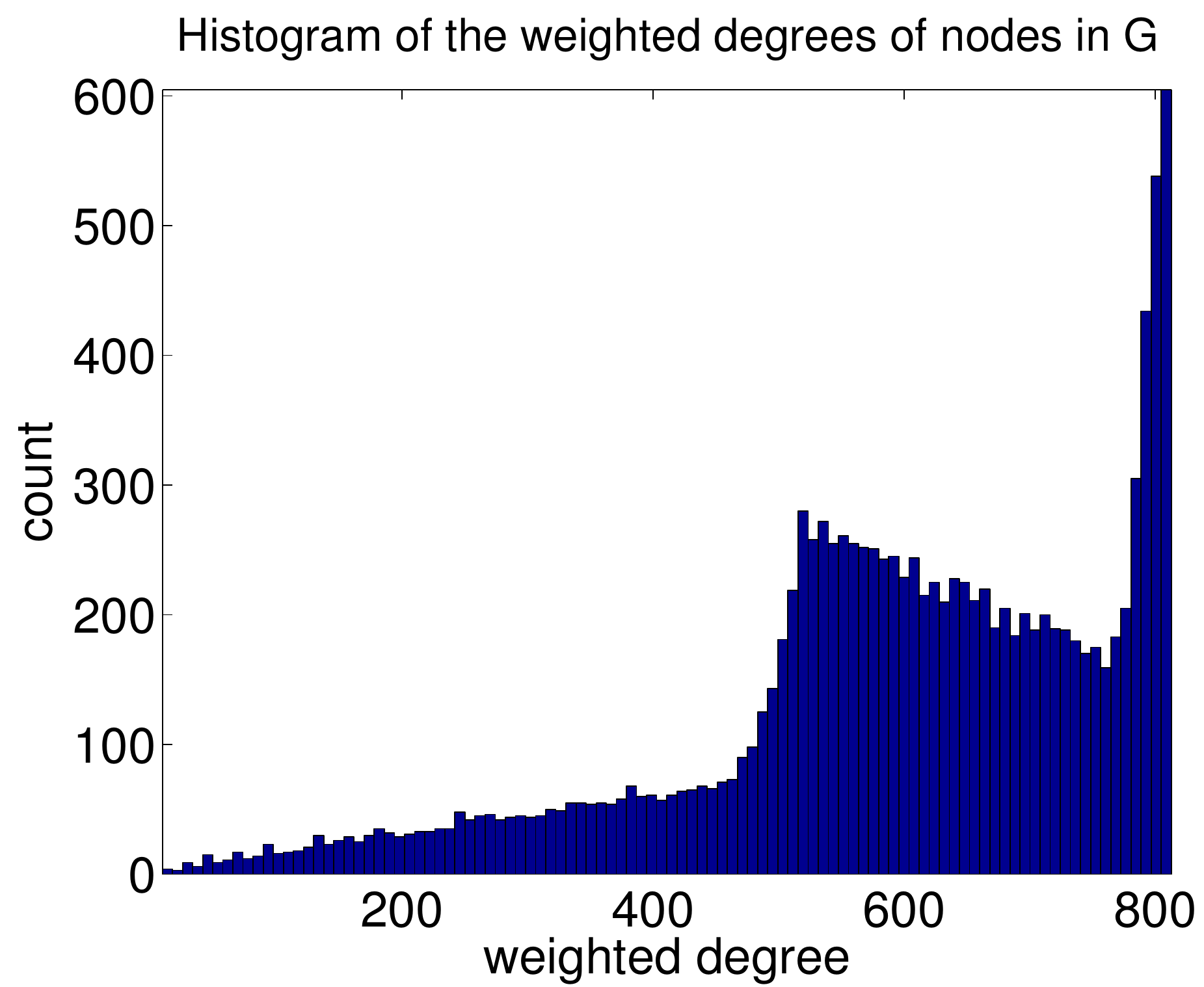}
\raise 3.7cm \hbox{\hbox{(d)}}
\hskip -3mm
\includegraphics[width=0.2331\columnwidth]{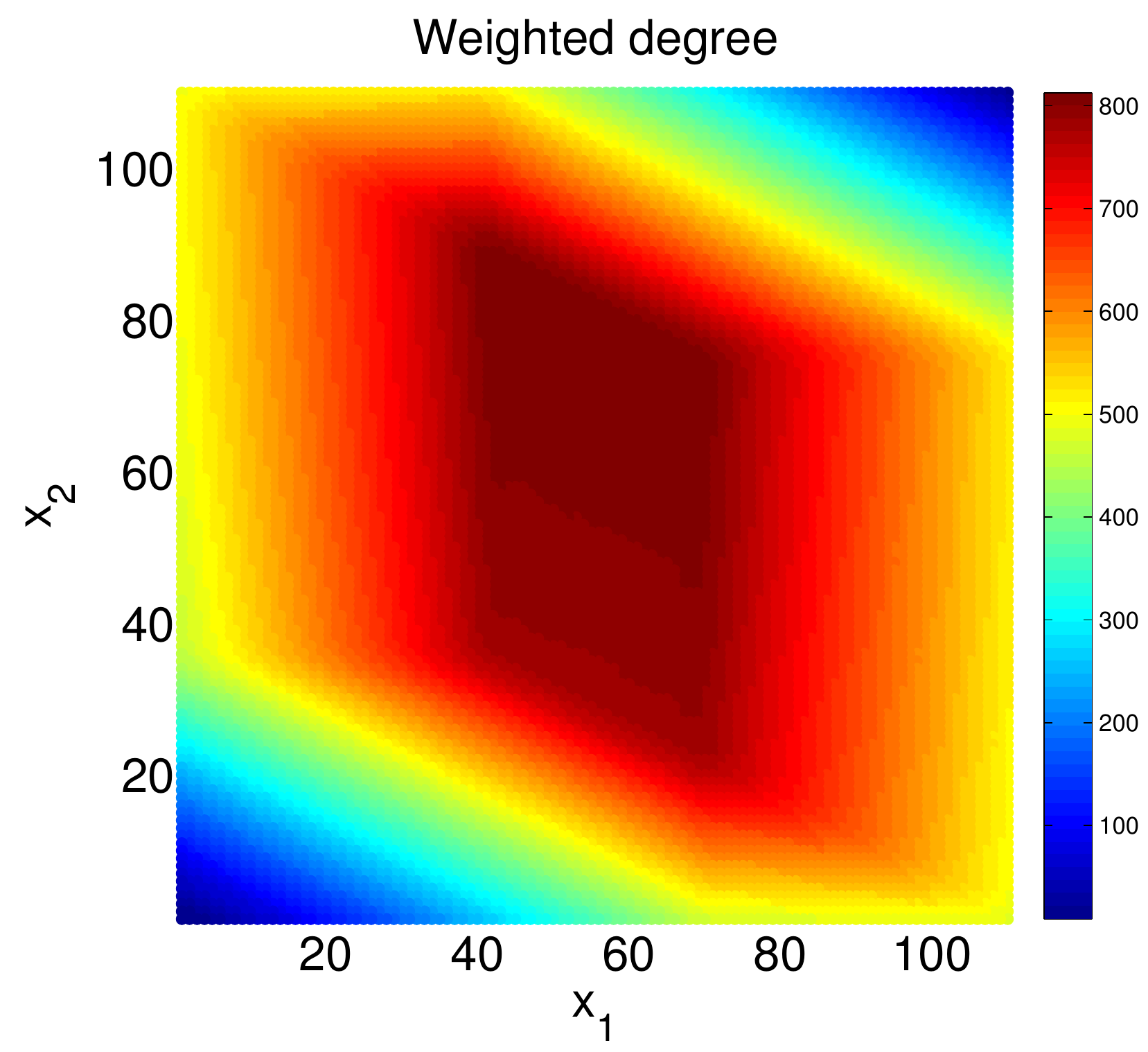}
}
\caption{{\it Illustrative Example} CS-II.
{\rm (a)} {\it The top $500$ eigenvalues of the associated combinatorial 
Laplacian, i.e. $(1-\lambda_i)$ for $i=1,2,\dots,500$.}
{\rm (b)} {\it The coloring of the nodes of $G$ (states of the observable 
space) according to their corresponding entry in the top 
eigenvector ${\boldsymbol \Phi}_1$ of $L$ given by $(\ref{ALap})$.}
{\rm (c)} {\it The weighted degree distribution of the ground state graph $G$.}
{\rm (d)} {\it A scatterplot of the states of the system, colored by 
their weighted degree.}} 
\label{figure3p2}
\end{figure}

Using the stochasticity of $L$, we can interpret it as a random 
walk matrix on the weighted graph $G = (V,E) $, where the set of 
nodes corresponds to the original observable states 
$(x_1,x_2)^{(i)},$ $i=1,2,\ldots,N$ (and implicitly to 
states $s^{(i)}$ of the slow variable), and there is an edge 
between nodes $i$ and $j$ if and only if $W_{ij}>0$. 
The associated combinatorial Laplacian is given by $\tilde{L} = D-W$.
Whenever the pair $(\lambda_i,{\boldsymbol \Phi}_i)$ is an 
eigenvalue-eigenvector solution to 
$L {\boldsymbol \Phi}_i = \lambda_i {\boldsymbol \Phi}_i$,
then so is $(1-\lambda_i,{\boldsymbol \Phi}_i)$ for 
the generalized eigenvalue problem 
$\tilde{L} {\boldsymbol \Phi}_i = \lambda_i D {\boldsymbol \Phi}_i$.
We plot in Figures \ref{figure3p1}(a)
and \ref{figure3p2}(a) the spectrum of the combinatorial
Laplacian $\tilde{L} = D-W$, for the chemical systems CS-I and CS-II.
In Figures \ref{figure3p1}(b) and \ref{figure3p2}(b) we 
color the states of the network with the top non-trivial 
eigenvector ${\boldsymbol \Phi}_1$.

Before considering the top eigenvector of $L$ for determining 
the underlying slow variable and estimating its stationary 
distribution, we propose to use a sparse graph Laplacian 
which differs from the ADM method in~\cite{amitslowvars}, 
where the Laplacian matrix is associated 
to a complete weighted graph. However, using 
a complete graph leads to computing the $\Sigma$-dependent 
squared distance in equation (\ref{defDist}) for any pair of 
nodes, thus an $O(N^2)$ number of computations is used.
In light of the approximation (\ref{approxChain}), a pair of 
points which are far away in the observable space (i.e., for 
which $d_{\Sigma}^2 ( (x_1,x_2)^{(i)}, (x_1,x_2)^{(j)} )$ 
is large) denotes a pair of corresponding states of the slow 
variable which are also far away in the inaccessible space. 
Thus we do not have to do such computations, because points 
far away in the unobservable space will have an exponentially 
small similarity $W_{ij}$  close to 0.
The fact that the shape of the local point cloud is an ellipse 
provides some insight in this direction. Thus we 
will build a sparse graph of pairwise measurements and no longer 
compute the $\Sigma$-dependent distance  between all points of 
the data set, but only between a very small subset of the points.
The spectrum of the covariance matrix $\Sigma_{i}$, in particular  
the ratio $\tau$ of its two eigenvalues given by (\ref{taudef}), 
guides us in building locally at each point, a sparse ellipsoid-like
neighborhood graph.

\begin{figure}[t]
\begin{center}
\vskip -0.1cm
\includegraphics[width=0.35\columnwidth]{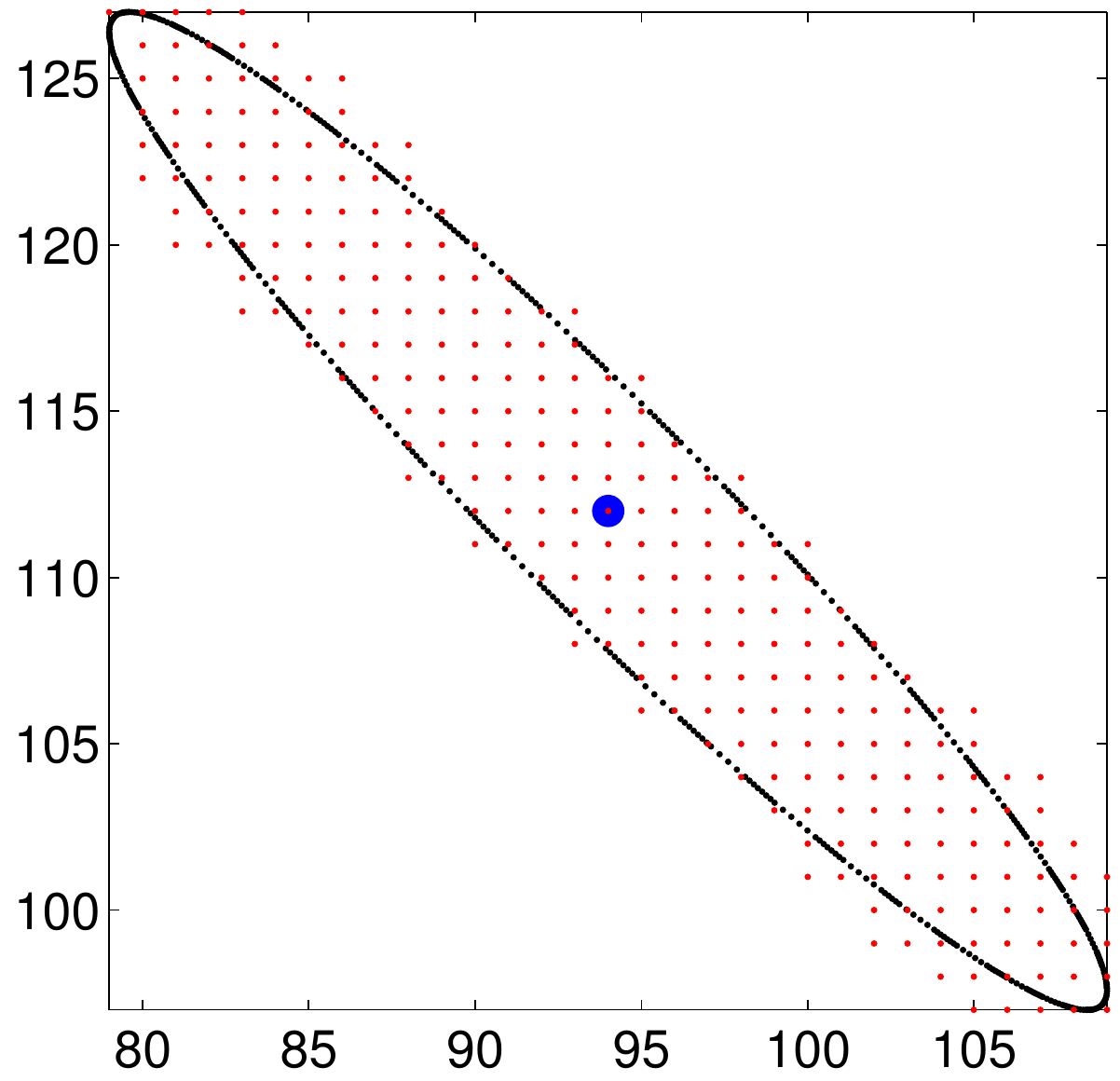}
\end{center}
\vskip -0.1cm
\caption{{\it The local neighborhood graph $G_i$ at a given 
node $(x_1,x_2)^{(i)}$; the shape is an ellipsoid whose axis 
ratio is given by the ratio of the eigenvalues $\tau$
of the local covariance matrix 
$d_{\Sigma}^2 ( (x_1,x_2)^{(i)}, (x_1,x_2)^{(j)} )$},
i.e. by $(\ref{taudef})$. The corresponding eigenvectors
are used to calculate the orientation of the ellipse.
}
\label{figure3p3}
\end{figure}

For each observable state $(x_1, x_2)^{(i)}$, we build 
a local adjacency graph, denoted by $G_i$,  in the shape of 
an ellipse pointing in the direction of the fast dynamics, and 
whose small axis points in the direction of the slow dynamics. 
Figure \ref{figure3p3} shows an example of such a 
local 1-hop neighborhood graph $G_i$, where the central node 
$(x_1, x_2)^{(i)}$ is connected to all points contained within the 
boundaries of an appropriately scaled ellipse centered at 
$(x_1, x_2)^{(i)}$. 
Finally, we define the sparse graph $G$ of size $N\times N$ 
associated to the entire network as the union of all locally 
defined ellipsoid-like neighborhood graphs
$G = \bigcup_{i=1}^{N} G_i$.
Note that the union graph $G$ is still a simple graph, with no 
self edges and no multiple edges connecting the same pair of 
nodes. We compute the distance $d_{\Sigma}$ by (\ref{defDist})
(and thus the similarity $W_{ij}$) between a pair of nodes 
$(x_1, x_2)^{(i)}$ and $(x_1, x_2)^{(j)}$ if 
and only if the corresponding edge $(i,j)$ exists in $G$.

We plot in Figures \ref{figure3p1}(c)
and \ref{figure3p2}(c) the 
histogram of the weighted degrees of the nodes in the weighted 
graph $W$ defined in (\ref{defWij}), while 
Figures \ref{figure3p1}(d)
and \ref{figure3p2}(d)
show a scatterplot of the states of the system, where each state 
$i$ is colored by its weighted degree, i.e., the sum of all its 
outgoing weighted edges $W_{ij}, j=1,2,\ldots,n$. 
Throughout the computational examples in this paper,  
the smoothing parameter $\varepsilon$ which appears in 
(\ref{defWij}) was set to $\varepsilon = 0.1$.
In contrast to the approach in \cite{amitslowvars} which computes 
all $O(N^2)$ pairwise similarities, the sparsity of $G$ (and thus 
of the associated graph Laplacian $L$) in our approach only requires 
the computation of a much smaller number of distances, as low as 
linear, depending on the discretization of the domain, and makes 
it computationally feasible to solve problems with thousands or 
even tens of thousands of nodes. 

\section{A robust mapping from the observable space $\mathcal{O}$ to 
the ``dynamically meaningful" inaccessible space $\mathcal{H}$} 
\label{sec:EigBinning}

\begin{figure}[t]
\centerline{
\hskip 3mm
\raise 4.2cm \hbox{\hbox{(a)}}
\hskip -3mm
\includegraphics[width=0.28\columnwidth]{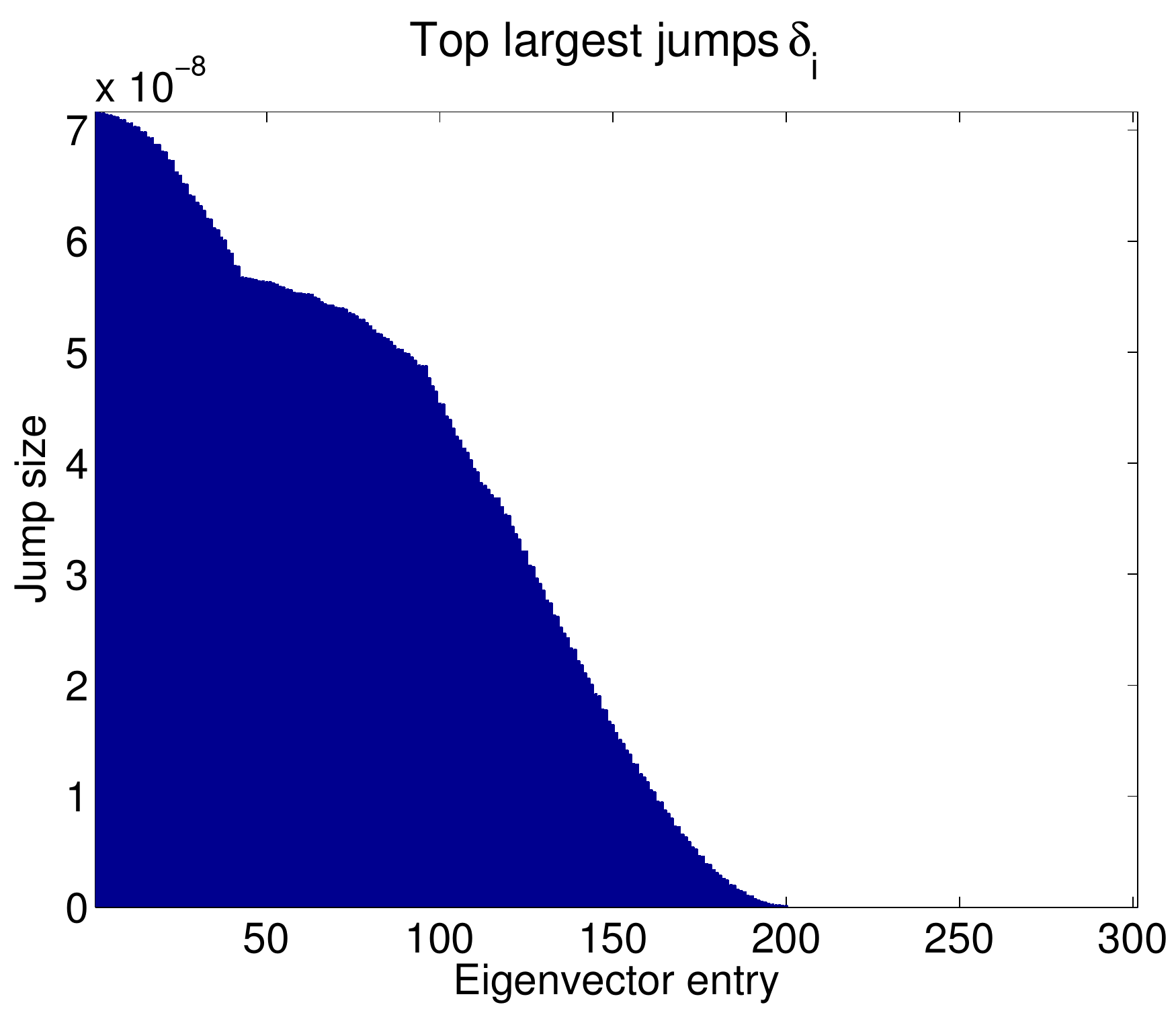}
\hskip 3mm
\raise 4.2cm \hbox{\hbox{(b)}}
\hskip -3mm
\includegraphics[width=0.29\columnwidth]{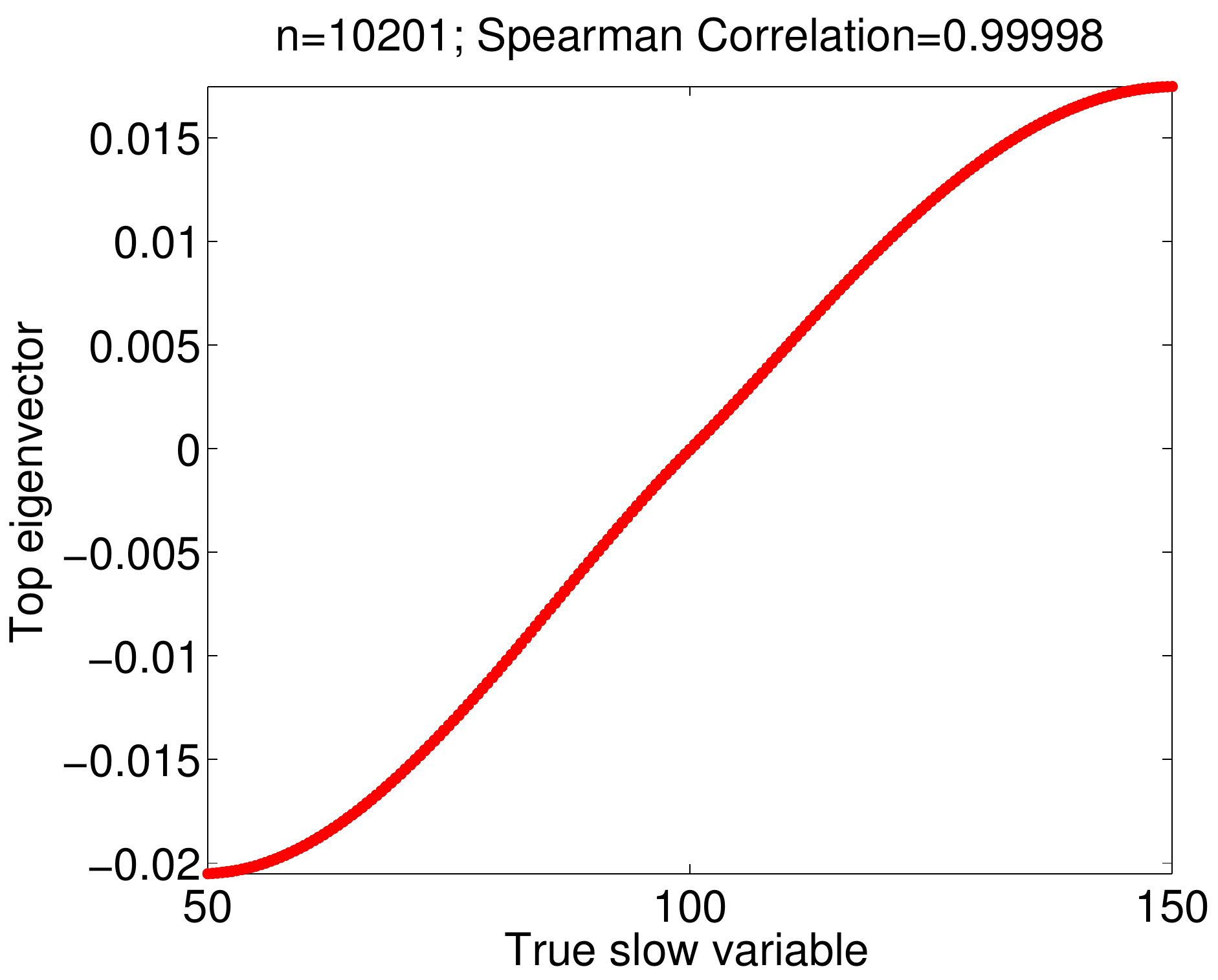}
\hskip 3mm
\raise 4.2cm \hbox{\hbox{(c)}}
\hskip -3mm
\includegraphics[width=0.29\columnwidth]{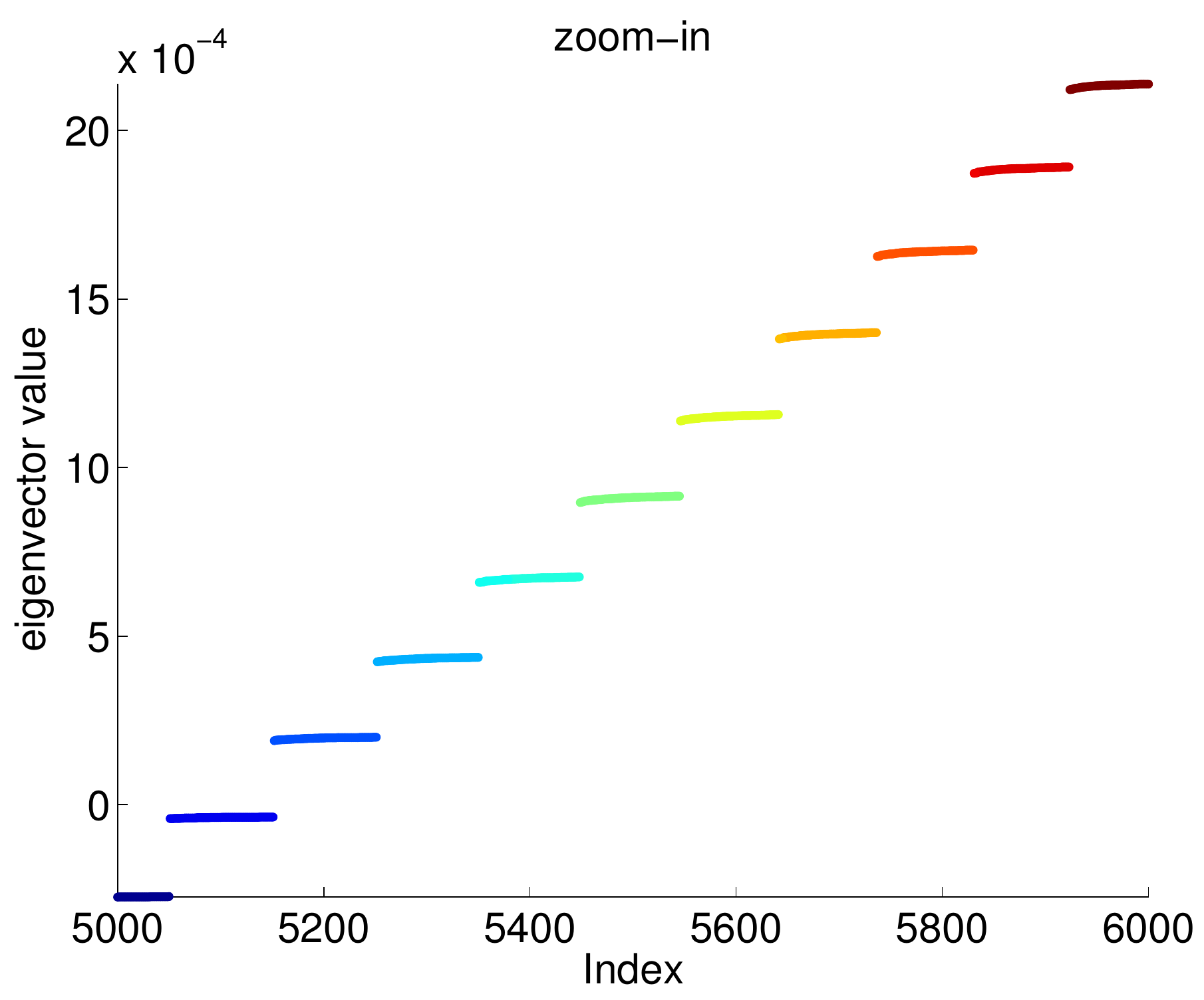}
}
\caption{{\it Illustrative example} {\rm CS-I.} 
{\rm (a)}
{\it Jump sizes $(\ref{deltInc})$ of the sorted eigenvector 
${\boldsymbol \Phi}_1$ of the sparse anisotropic graph Laplacian $L$.}
{\rm (b)}
{\it The correlation of ${\boldsymbol \Phi}_1$ with the ground truth slow variable $s=(x_1+x_2)/2$.}
{\rm (c)} 
{\it Zoom-in on the sorted top eigenvector $\bar{\boldsymbol \Phi}_1$  (the colors denote the corresponding slow variable)
showing that $\bar{\boldsymbol \Phi}_1$ is almost piece-wise 
constant on the bins that correspond to distinct slow variable states.
The kernel scale is set to $ \varepsilon = 0.1$.}
}
\label{figure4p1}
\end{figure}

\begin{figure}[t]
\centerline{
\hskip 3mm
\raise 4.2cm \hbox{\hbox{(a)}}
\hskip -3mm
\includegraphics[width=0.28\columnwidth]{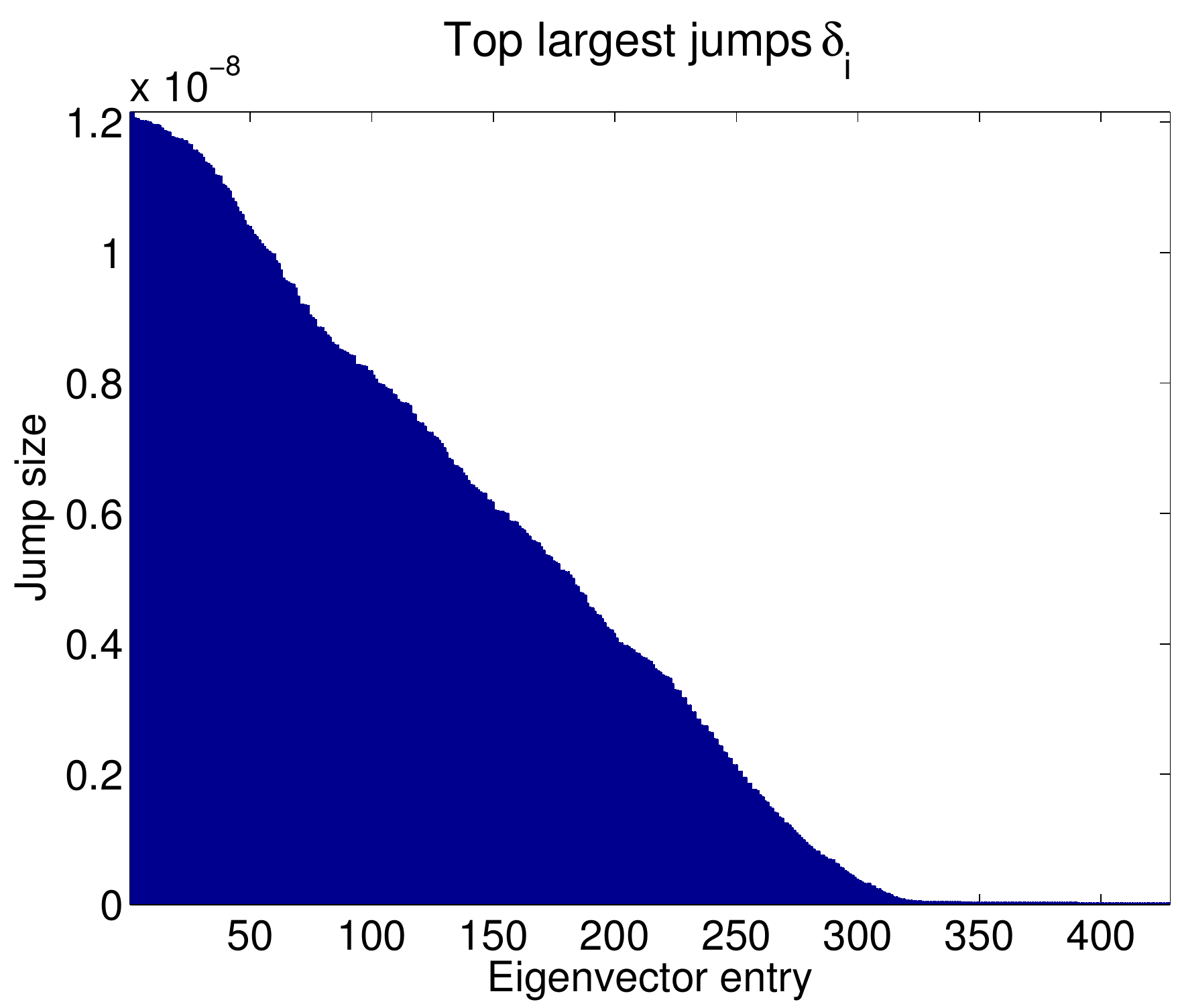}
\hskip 3mm
\raise 4.2cm \hbox{\hbox{(b)}}
\hskip -3mm
\includegraphics[width=0.29\columnwidth]{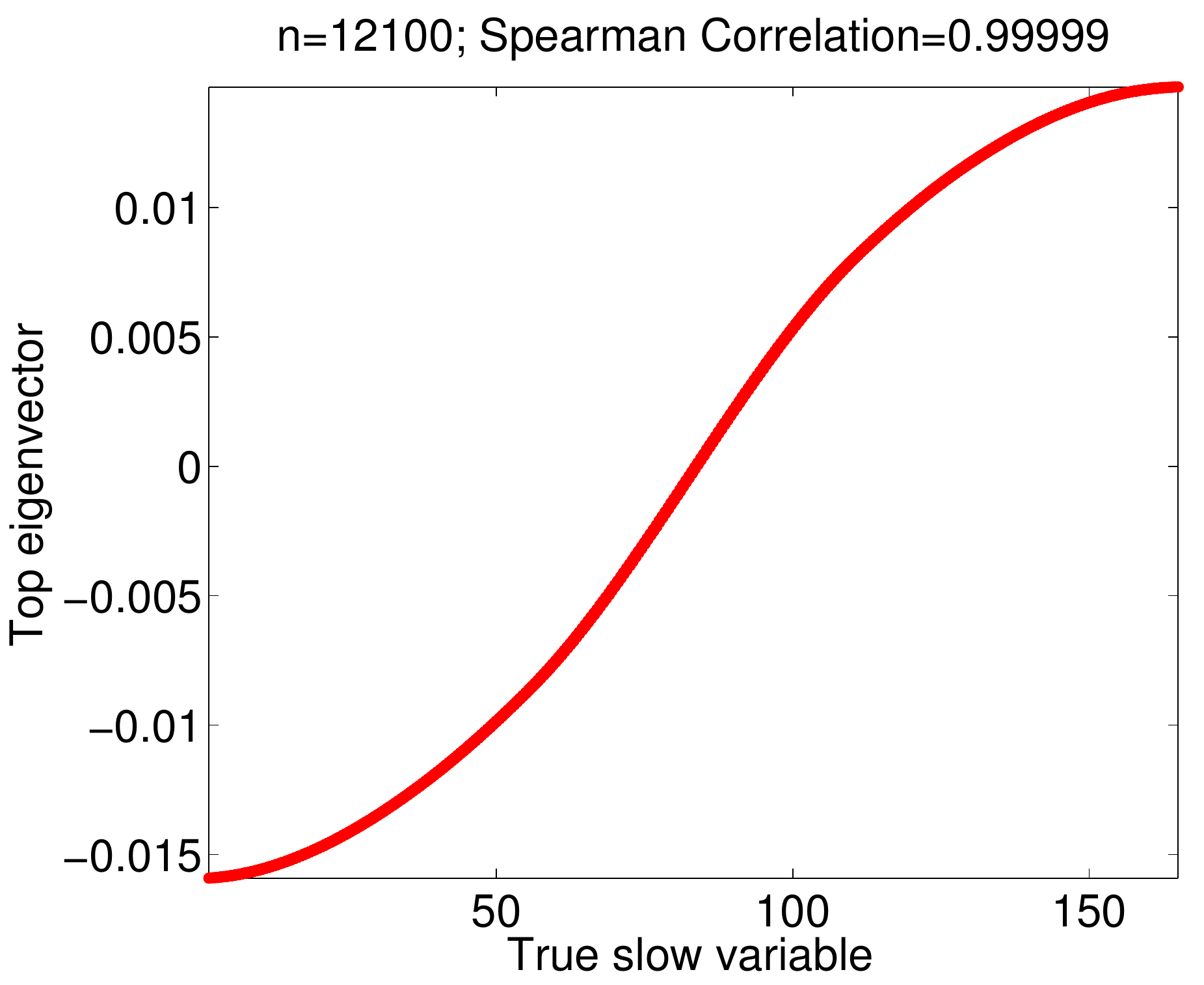}
\hskip 3mm
\raise 4.2cm \hbox{\hbox{(c)}}
\hskip -3mm
\includegraphics[width=0.29\columnwidth]{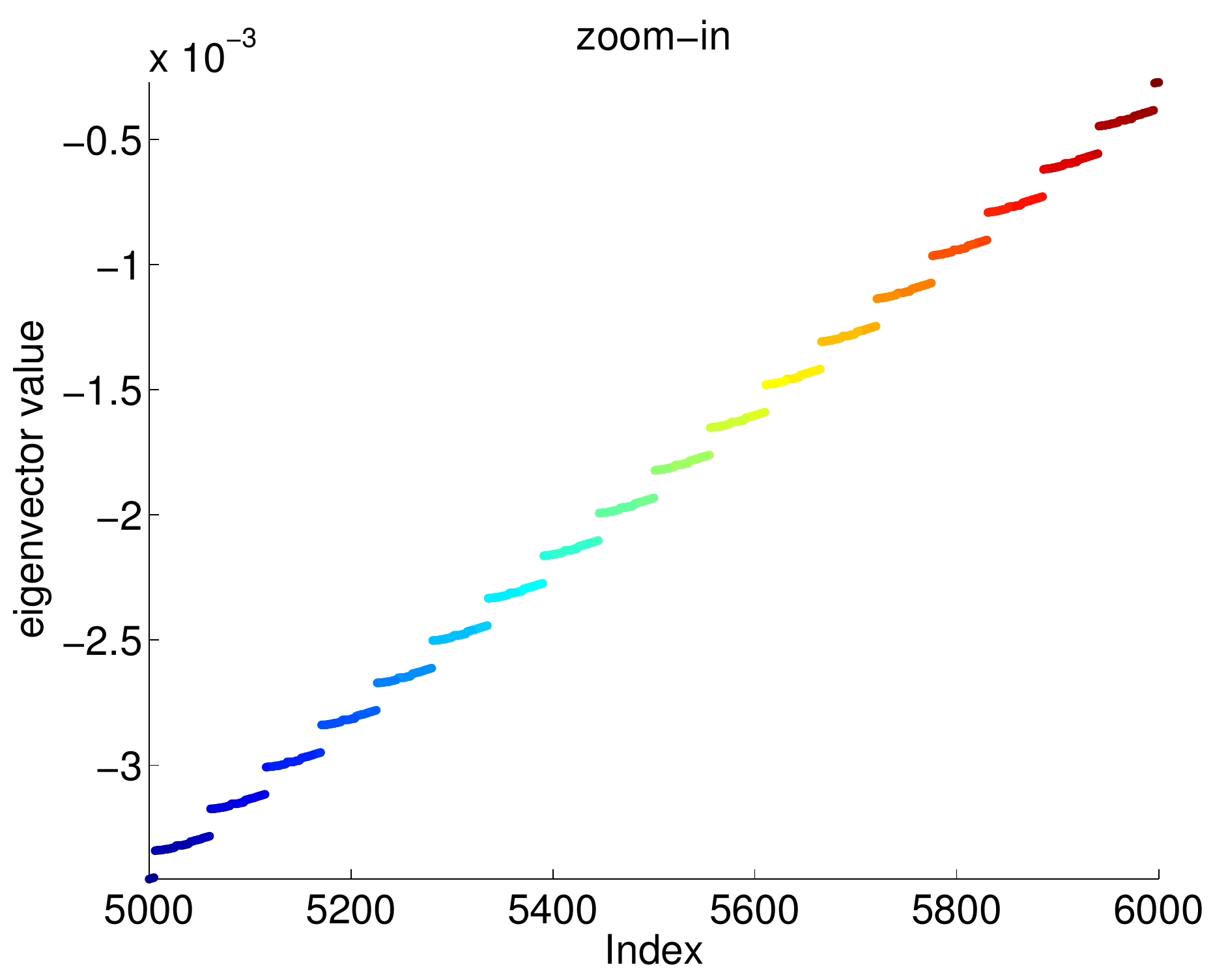}
}
\caption{ {\it Illustrative example} {\rm CS-II.} 
{\rm (a)}
{\it  Jump sizes $(\ref{deltInc})$ of the sorted eigenvector 
${\boldsymbol \Phi}_1$ of the sparse anisotropic graph Laplacian $L$.}
{\rm (b)}
{\it The correlation of ${\boldsymbol \Phi}_1$ with the ground truth slow variable $s=x_1+2 x_2$.}
{\rm (c)} 
{\it Zoom-in on the sorted top eigenvector $\bar{\boldsymbol \Phi}_1$  
(the colors denote the corresponding slow variable)
showing that $\bar{\boldsymbol \Phi}_1$ is almost piece-wise 
constant on the bins that correspond to distinct slow variable states.
The kernel scale is set to $ \varepsilon = 0.1$.}
}
\label{figure4p2}
\end{figure}

As a first step towards partitioning the nodes of the original graph 
$G$ and detecting the associated slow variable, we sort the entries 
of the top eigenvector ${\boldsymbol \Phi}_1$, which we then denote by 
$\bar{{\boldsymbol \Phi}}_1$ with 
$\bar{\Phi}_1(1) \geq \bar{\Phi}_1(2) \geq \ldots \geq \bar{\Phi}_1(N)$. 
This sorting process defines permutation $\sigma$ of the original 
index set $i=1, 2, \dots, N$ so that $\bar{\Phi}_1(\sigma(i)) = \Phi_1(i).$ 
We consider the increments between two consecutive (sorted) values 
\begin{equation}
\delta_{i} = \bar{\Phi}_1(i) - \bar{\Phi}_1(i+1), 
\qquad i=1,2,\ldots, N-1.
\label{deltInc}
\end{equation}
Next, we sort the vector of such increments, denote its entries 
by $\bar{\delta}_{1} \geq  \bar{\delta}_{2} \geq \ldots \bar{\delta}_{N-1}$, 
and show in Figure \ref{figure4p1}(a) (resp. Figure \ref{figure4p2}(a)) 
the top 300 (resp. top 420) largest such increments $\bar{\delta}_{i}$
for illustrative example CS-I (resp. CS-II). Note that this already give 
us an idea about the number of distinct slow states in the system, a 
set which we denote by $\mathcal{S}$. 
Ideally, the difference $\Phi_1(i) - \Phi_1(j) $ in the entries of 
the top eigenvector corresponding to two observable states 
$(x_1,x_2)^{(i)}$ and $(x_1,x_2)^{(j)}$ that belong 
to the same slow variable $s$ (i.e., 
$x_1^{(i)} + 2 x_2^{(i)} = x_1^{(j)} + 2 x_2^{(j)} = s$
for illustrative example CS-II) 
should be zero or close to zero, in which case we expect that 
only approximately  $|\mathcal{S}|$ of the $N-1$   
increments $\delta_i$ are significantly larger than zero, 
while the remaining majority are zero or close to zero.

In Figures \ref{figure4p1}(b) and \ref{figure4p2}(b) we highlight 
the correlation between the entries of the top non-trivial
eigenvector $\Phi_1$ and the corresponding slow variable $S$. 
In Figures \ref{figure4p1}(c) and \ref{figure4p2}(c), we zoom 
on a subset of states 
to make the point that the eigenvector $\Phi_1$ is almost 
constant on the $\mathcal{O}$-states that correspond to the same 
value of the slow variable.
The plots in Figures \ref{figure3p1}(b) and \ref{figure3p2}(b)
show a coloring of the networks generated by the two chemical systems
CS-I and CS-II, based on the first nontrivial 
eigenvector of the associated sparse Laplacian $L$. Note that the 
eigenvector looks almost piecewise constant along the lines that 
point to the evolution of the fast variable,
for a given value of the slow variable 
($S=(X_1+X_2)/2$ for CS-I and $S=X_1+2X_2$ for CS-II), 
yet nowhere along the way we have input this
information into the method. In the next step we use this top 
eigenvector to identify all nodes of 
the graph (original states of the chemical system) that correspond 
to the same value of the underlying slow variable.
In other words, all nodes whose corresponding eigenvector entries 
are between an appropriately
chosen interval (that we shall refer to a \textit{bin}) will be 
labeled as belonging to the same slow variable S.
In other words, we seek a partition of the observable states 
in $\mathcal{O}$, i.e., of the nodes of $G$, such that all 
original states $(x_1,x_2)^{(i)}$ with the same value of the 
corresponding slow variable $s((x_1,x_2)^{(i)})$ end up in the 
same bin. Our goal is to find a partition  
$\mathcal{P} =\{\mathcal{P}_1, \mathcal{P}_2, \ldots, \mathcal{P}_k \}$ 
of  $\mathcal{O}$ such that 
\begin{equation}
\mathcal{P}_j =
\{ (x_1, x_2)^{(i)} \in \mathcal{O} \mid    s(x_1, x_2) = q_j \}  
\quad \mbox{ and } \quad 
\bigcup_{j=1}^{k} \mathcal{P}_j 
= \mathcal{O},
\label{partP}
\end{equation}
where $k$ denotes the number of distinct values $q_j$,
$j=1,2,\dots,k$, of the slow variable $S$. As an example, in 
the case of CS-I given by (\ref{ex1system}), the partition 
$\mathcal{P}_{j} = \{ (1,99), (2,98), \ldots, (99,1) \}$ 
corresponds to all nodes in the graph for which the value 
of the associated slow variable is constant $q_j=50$. The key 
observation we exploit here is that the top eigenvector of 
the Laplacian matrix is almost piecewise constant on the 
bins that partition $\mathcal{O}$, since the nodes of $G$ 
that correspond to the same value of the slow variable 
have a very high pairwise similarity, with $W_{ij}$ 
very close to 1.

One may also interpret the above problem as a clustering problem, 
where the similarity between pairs of points is given by 
(\ref{defWij}), and is such that nodes that belong to the 
same bin have a much higher similarity compared to nodes 
that belong to two different bins, an effect due to the 
strong separation of scales. In the case of illustrative example CS-I, 
the clusters correspond to lines in the two-dimensional plane such 
that $(x_1 + x_2)/2 = c$,  for a constant $c$. We point 
out the interested reader 
to the work of~\cite{MeilaShi}, where the top eigenvectors of the random 
walk Laplacian are used for clustering. While in practice 
one uses the top several eigenvectors as the reduced 
eigenspace where clustering is subsequently performed, in 
our case the top eigenvector alone suffices to capture the 
many different clusters (i.e., bins), a fact we attribute 
to the strong separation of scales exhibited by the 
illustrative chemical systems CS-I and CS-II. If several
eigenvectors were considered then one could use a clustering algorithm, 
such as k-means or spectral 
clustering~\cite{ShiMalik00_NCut,luxburg05_survey},
to obtain the partitioning (\ref{partP}).
However, a simpler method has been successfully used for the 1-dimensional 
eigenspace in both examples we considered. It is described as follows.
Recall the sorted vector of increments 
$\bar{\delta}_{1} \geq  \bar{\delta}_{2} 
\geq \ldots \geq \bar{\delta}_{N-1}$ 
defined in equation (\ref{deltInc}), and consider the set 
of the $k-1$ largest such increments
$\{ \bar{\delta}_{1}, \bar{\delta}_{2}, \ldots, \bar{\delta}_{k-1} \}$
where $\bar{\delta}_{1} \geq  \bar{\delta}_{2} \geq \ldots \geq \bar{\delta}_{k-1}.$
Next, from the sorted eigenvector $\bar{{\boldsymbol \Phi}}_1$ 
we extract the position 
of the entries whose associated increment (with respect to its right-next 
neighbor index) belongs to 
$\{ \bar{\delta}_{1}, \bar{\delta}_{2}, \ldots, \bar{\delta}_{k-1} \}$. 
In others words, we compute
\begin{equation}
b_t = \underset{i=1,2,\ldots,N-1}{\operatorname{arg}}  
\bar{\Phi}_{1}(i) - \bar{\Phi}_{1}(i+1) 
= \bar{\delta}_{t}, \qquad \hbox{where} \;\;\; t=1,2,\ldots,k-1,
\label{bdelimiters}
\end{equation}
and $b_0=0$ and $b_k=N.$
Finally, we compute an estimated partition $\hat{\mathcal{P}}$ 
of $\mathcal{O}$ by $\hat{\mathcal{P}}_q 
= \{ i \in \mathcal{O} \mid \sigma(i) \in (b_{q-1}, b_q] \},$ 
where $q=1,2,\ldots,k,$ and $\sigma$ is the permutation 
of the original index set $i =1,2, \dots, N,$ given in the
definition of $\bar{{\boldsymbol \Phi}}_1$, i.e.
$\bar{\Phi}_1(\sigma(i)) = \Phi_1(i).$ 

\begin{figure}[t]
\centerline{
\hskip 3mm
\raise 3.5cm \hbox{\hbox{(a)}}
\hskip -3mm
\includegraphics[width=0.23\columnwidth]{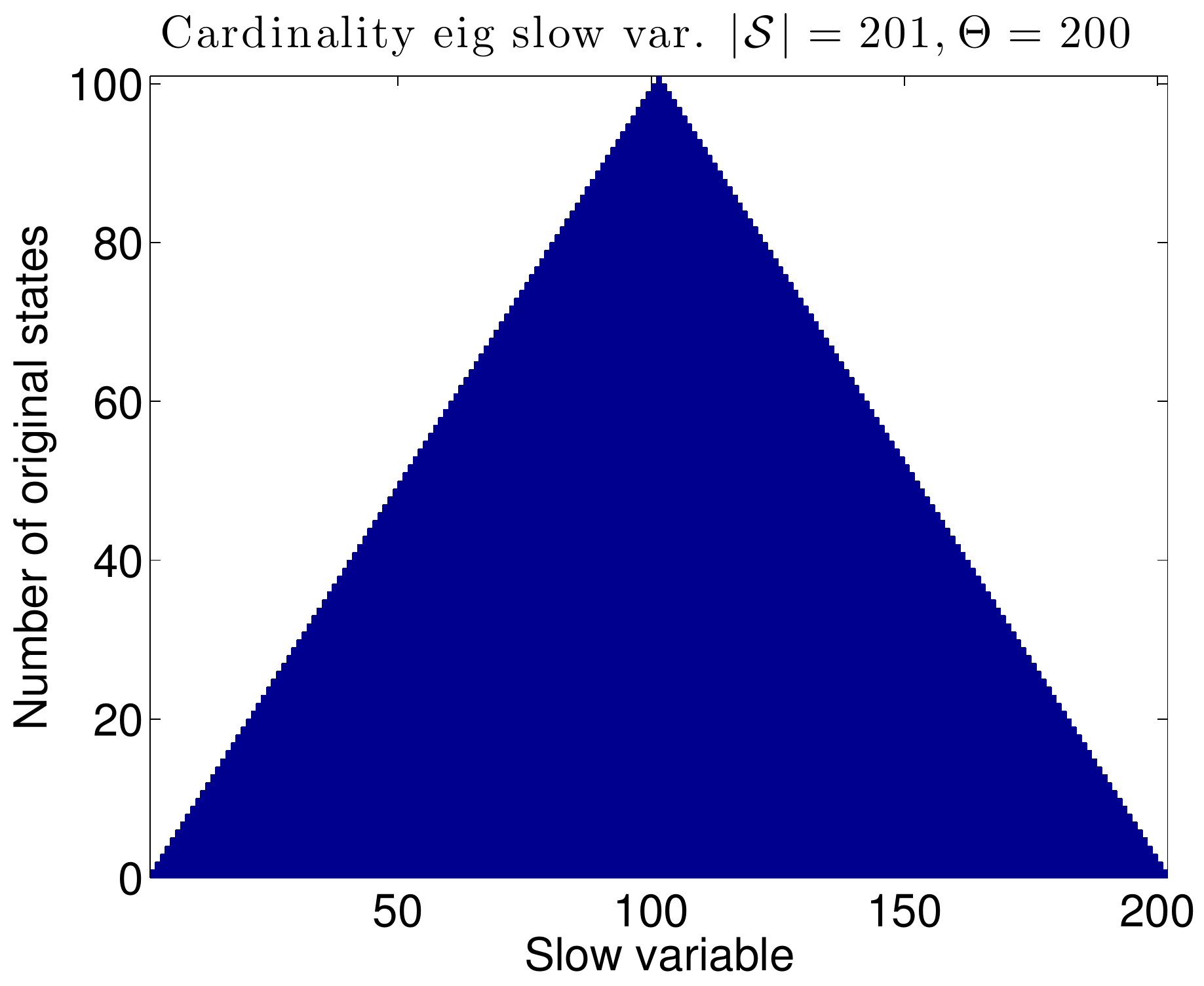}
\raise 3.5cm \hbox{\hbox{(b)}}
\hskip -3mm
\includegraphics[width=0.23\columnwidth]{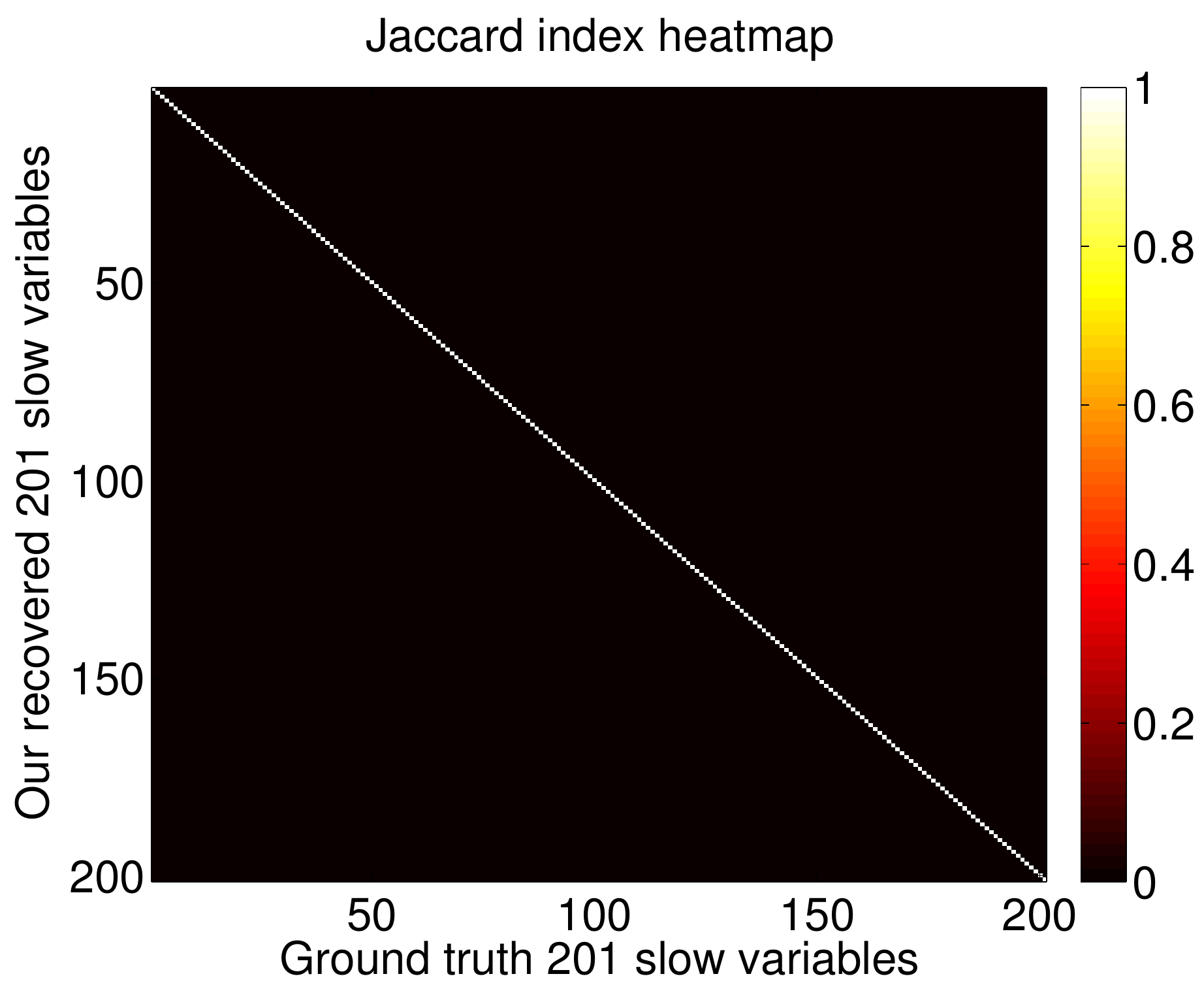}
\raise 3.5cm \hbox{\hbox{(c)}}
\hskip -3mm
\includegraphics[width=0.23\columnwidth]{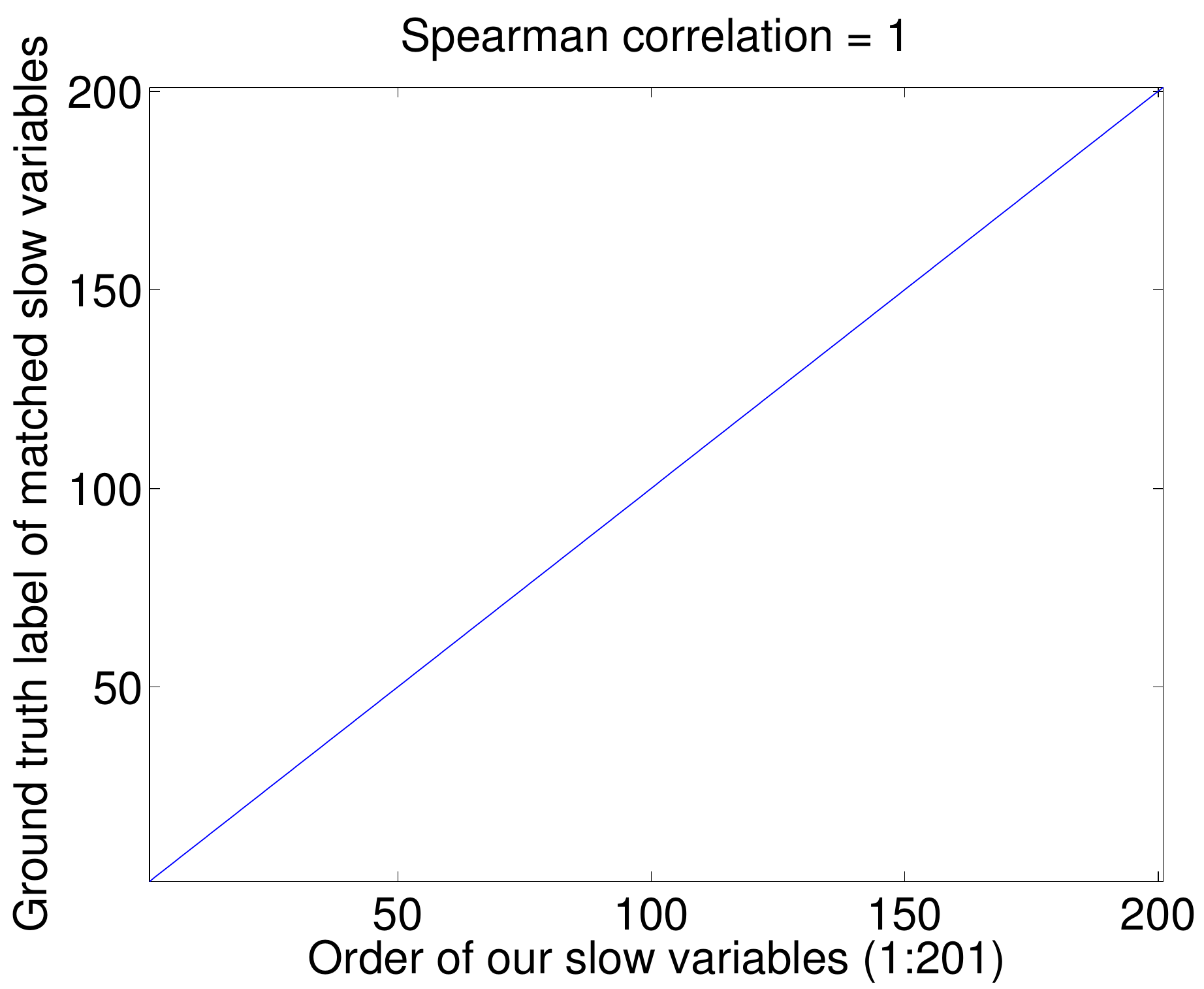}
\raise 3.5cm \hbox{\hbox{(d)}}
\hskip -3mm
\includegraphics[width=0.23\columnwidth]{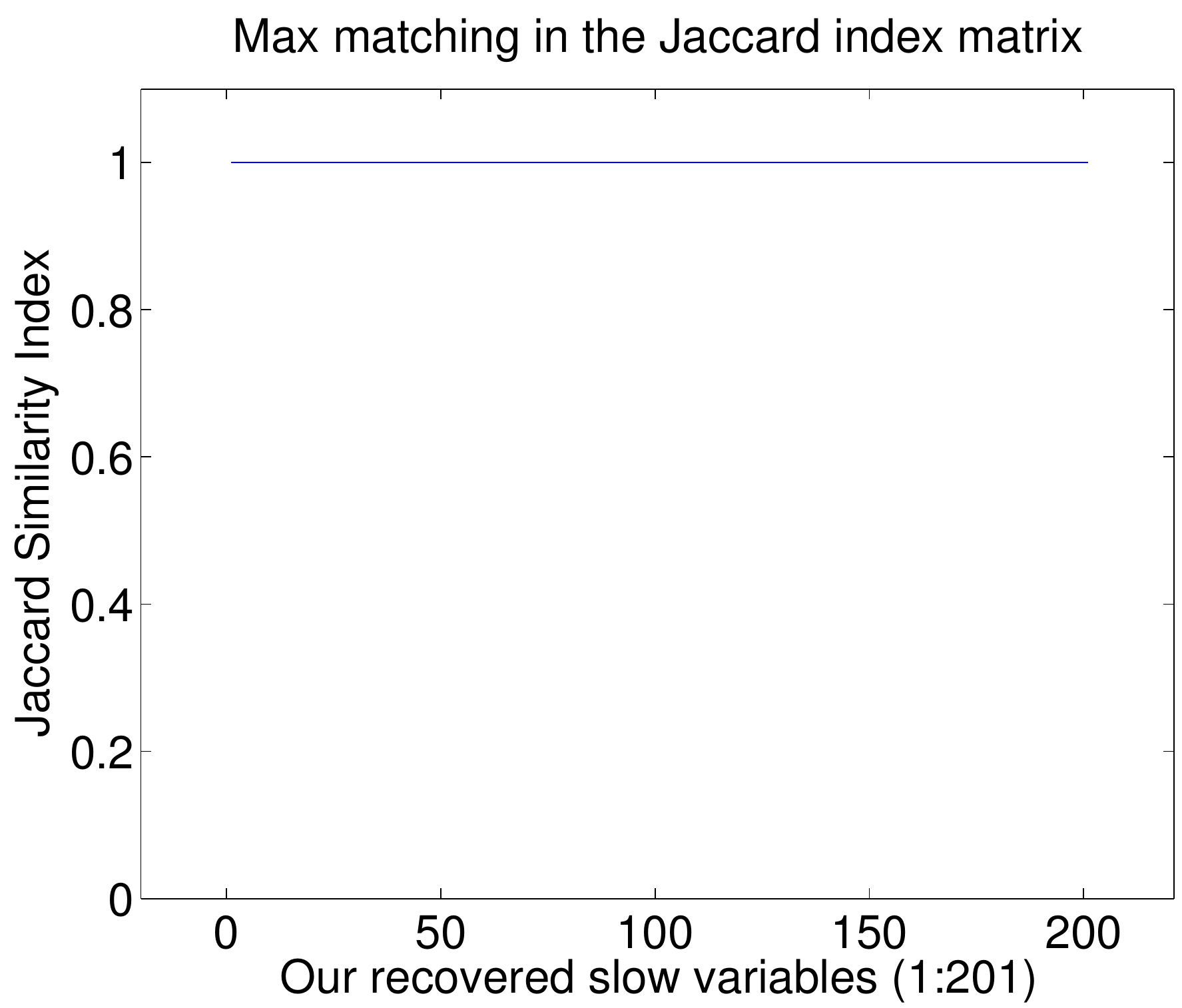}
}
\caption{{\it Illustrative example} {\rm CS-I.}
{\rm (a)} {\it The eigenvector-based slow variable cardinality. The Theta 
score $\Theta$ is the smoothness measure of the bin cardinalities, defined 
in $(\ref{def:thetaContScore}).$
The algorithm perfectly recovers the ground truth partition. 
}
{\rm (b)} 
{\it The heatmap of the pairwise Jaccard similarity matrix given 
by $(\ref{JacIndex})$.}
{\rm (c)}
{\it The correlation between the ordering of the ground truth slow variable 
and the eigenvector recovered slow variable.}
{\rm (d)}
{\it The Jaccard index of the pairwise matched bins (from the maximum
matching).}
}
\label{figure4p3}
\end{figure}

\begin{figure}[t]
\centerline{
\hskip 3mm
\raise 4.2cm \hbox{\hbox{(a)}}
\hskip -3mm
\includegraphics[width=0.28\columnwidth]{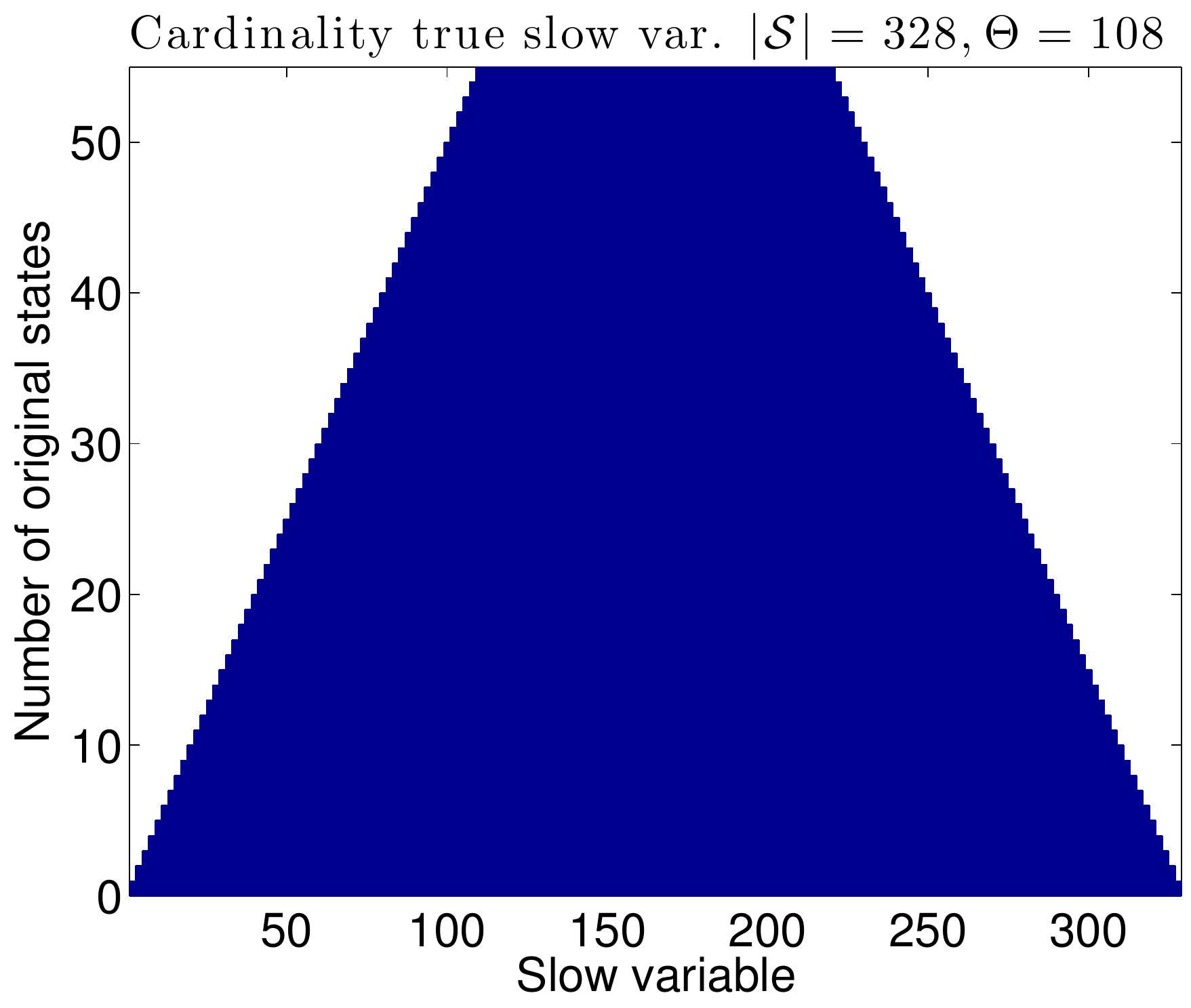}
\hskip 3mm
\raise 4.2cm \hbox{\hbox{(b)}}
\hskip -3mm
\includegraphics[width=0.28\columnwidth]{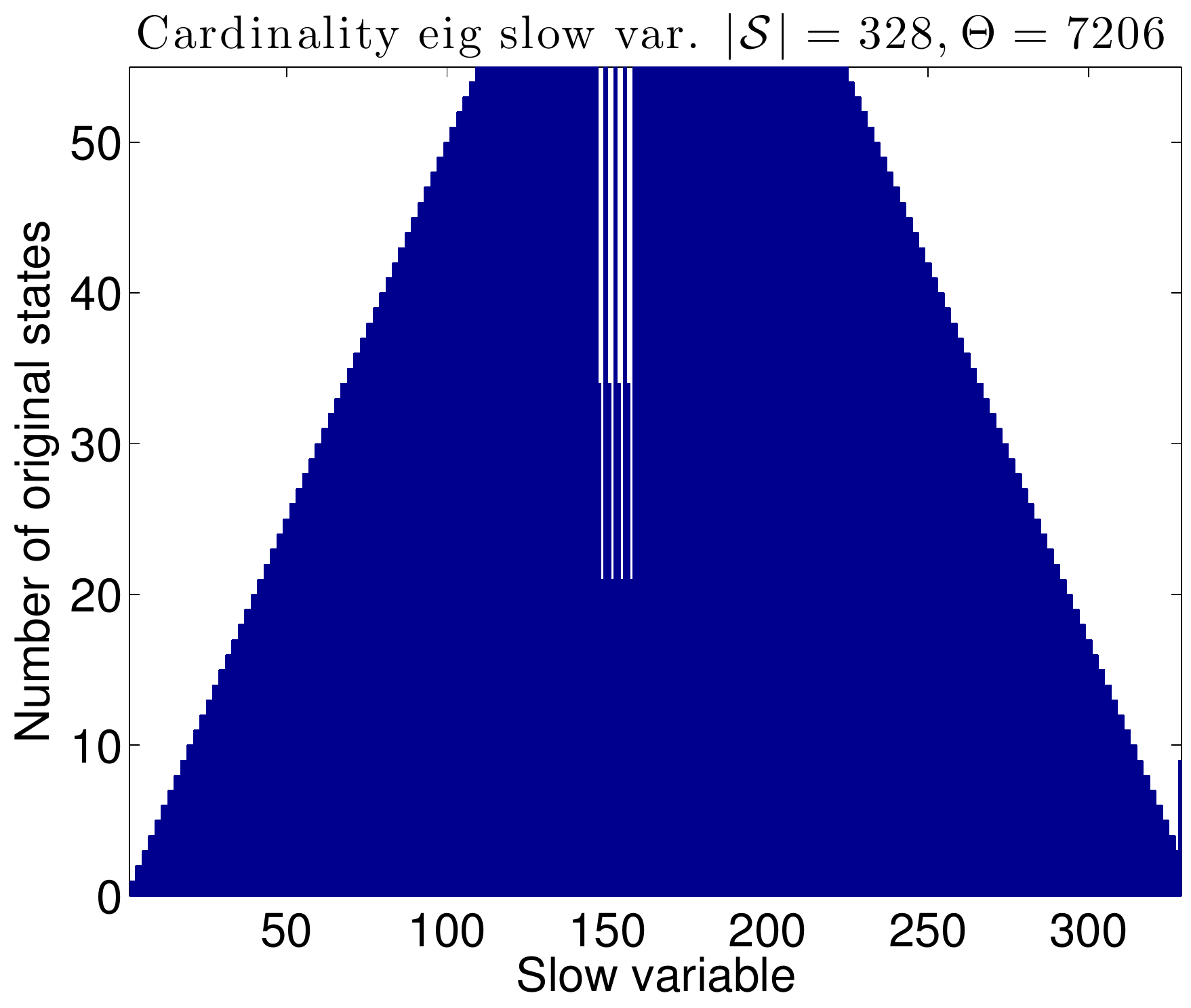}
\hskip 3mm
\raise 4.2cm \hbox{\hbox{(c)}}
\hskip -3mm
\includegraphics[width=0.28\columnwidth]{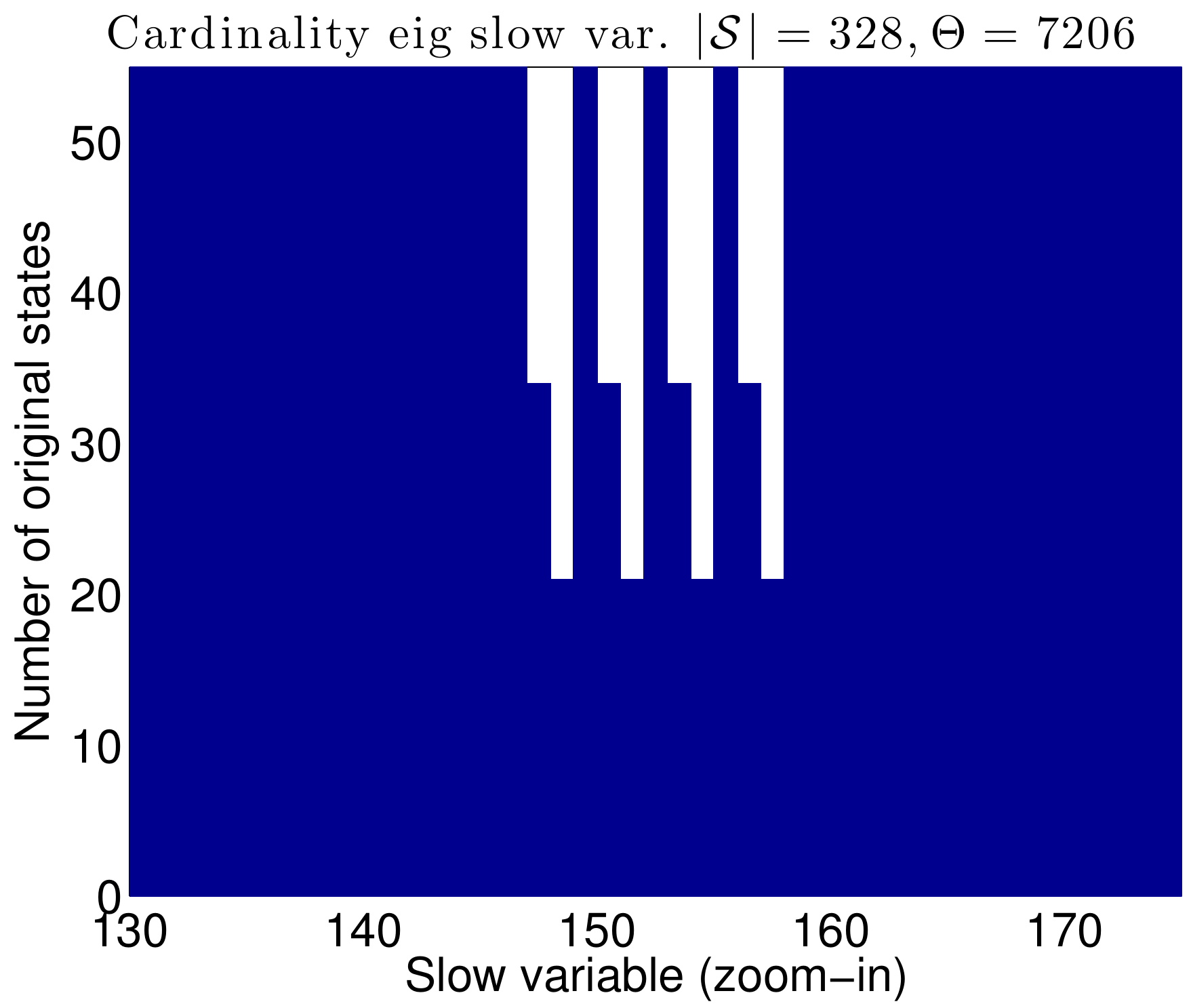}
}
\caption{{\it Illustrative example} {\rm CS-II.} 
{\rm (a)}
{\it The ground truth slow variable cardinality.}
{\rm (b)} 
{\it The eigenvector-based slow variable cardinality. $\Theta$ captures the smoothness 
of the bin cardinalities, as introduced in $(\ref{def:thetaContScore})$.}
{\rm (c)}
{\it Plot of the cardinalities of a subset of bins, showing the erroneous 
bin assignments in the  eigenvector-based partition. This 
is a zoomed-in version of panel} {\rm (b)}. 
}
\label{figure4p4}
\end{figure}

To illustrate the correctness of our proposed technique, we compute 
the Jaccard index between each proposed partition set $\hat{\mathcal{P}}_j$,
$j=1,2,\dots,k$ and each ground truth partition set 
$\mathcal{P}_i$, $i=1,2,\dots,|\mathcal{S}|$:
\begin{equation}
J_{ij} =  
\frac{|\mathcal{P}_i \cap \hat{\mathcal{P}}_j|}{|\mathcal{P}_i \cup \hat{\mathcal{P}}_j|},
\qquad
\mbox{where} \quad i=1,2,\dots,|\mathcal{S}|, \; \; \; j=1,2,\dots,k,
\label{JacIndex}
\end{equation}
and show a heatmap of this matrix in Figure \ref{figure4p3}(b). 
Since we are interested not only in the partition, but also in recovering 
the ordering of the slow variable, we show in  Figure \ref{figure4p3}(c) 
the correlation between the ground truth ordering of the slow variable 
and our recovered ordering. Note that we can only recover the ordering 
up to a global sign, since $-\Psi_1$ is also an eigenvector of $L$. 
Finally, we compute the maximum weight matching (using, for example, 
the Hungarian method \cite{Kuhn1955}) in the bipartite graph with 
node set $\mathcal{P} \cup \hat{\mathcal{P}}$ and edges across the two 
sets given by matrix $J$ in (\ref{JacIndex}). 
In Figure \ref{figure4p3}(d) we plot 
the Jaccard index of the matched partitions. For the first 
chemical system CS-I, note that the algorithm perfectly recovers the ground 
truth partition. In Figure \ref{figure4p4} we present the outcome of the binning algorithm
for the illustrative example CS-II, which is no longer satisfactorily by 
itself and requires further improvement. Though the bin cardinalities in 
the initial solution visually resemble the ground truth, there are 
numerous mistakes being made. To illustrate this, for a given 
partition $\mathcal{P}$, we compute the following measure of continuity 
of the recovered bin cardinalities 
\begin{equation}
\Theta_{\mathcal{P}}  
= 
\Theta( \mathcal{P}_1,\ldots,\mathcal{P}_S) 
=  \sum_{i=1}^{S-1} ( | \mathcal{P}_i| - | \mathcal{P}_{i+1} | )^2.
\label{def:thetaContScore}
\end{equation}
In other words, $ \Theta_{\mathcal{P}} $ captures the squared difference 
in the cardinalities of two consecutive bins. 
For the chemical system CS-II, the ground truth yields a score 
$\Theta = 108$, while for the eigenvector-recovered solution 
$\Theta = 7206$, thus indicating already that numerous 
misclassifications are being made, without even computing 
the Jaccard similarity matrix (\ref{JacIndex}) between the two partitions. 
To this end, we introduce in the next subsection a heuristic 
denoising technique followed by a truncation of the domain, 
which altogether lead to a better partitioning of 
$\mathcal{O}$ into groups of states that correspond 
to the same slow variable. 

\subsection{A bin denoising scheme} 
\label{sec:subBinDenoising}

While the eigenvector-based partition procedure detailed above yields 
accurate results for the CS-I in  (\ref{ex1system}), this procedure 
alone is not sufficient for obtaining a satisfactory partition for the 
more complex CS-II  considered in (\ref{ex2system}), as illustrated 
by the high corresponding $\Theta$-score ($\Theta = 7206$) shown 
in Figure \ref{figure4p4}(b). In Figure 
\ref{figure4p4}(c) we zoom into some of the recovered bins, 
showing that the eigenvector-based reconstruction splits some 
of the inner bins, which explains the high associated $\Theta$-score. 
In other words, states/bins which in the ground truth solution 
correspond to the same values of the slow state variable, are divided 
into two adjacent bins, and mistaken for two distinct states 
of the slow variable.

\begin{algorithm}[t]
\caption{Bin-merging algorithm:} \label{alg:binMerging}
\begin{algorithmic}[1]
\State Initialize FLAG = TRUE
\While{FLAG is TRUE}
\State Compute 
 $ \alpha_i :=  
 \Theta \left( \mathcal{P}_1,\ldots, \mathcal{P}_i \cup  \mathcal{P}_{i+1},
  \ldots \mathcal{P}_{|{\mathcal S}|} \right), 
 \qquad \forall i = 1,2,\ldots |\mathcal{S}|-1 $ 
	using definition (\ref{def:thetaContScore}) 
    \If  
  {$ \underset{i=1,2,\ldots,|{\mathcal S}|-1}{\operatorname{min}}  
  \alpha_i <  
    \Theta( \mathcal{P}_1,\ldots,\mathcal{P}_{|{\mathcal S}|}) $  }
	\State  
$ q = 
\underset{i=1,2,\ldots,|{\mathcal S}|-1}{\operatorname{argmin}}  \alpha_i $ 
        \State 
$\mathcal{P} : =  \mathcal{P}_1,\ldots, \mathcal{P}_q \cup 
 \mathcal{P}_{q+1}, \ldots \mathcal{P}_ {|{\mathcal S}|}$ 
	\State $|{\mathcal S}| = |{\mathcal S}|-1$
    \Else 
	\State FLAG = FALSE
    \EndIf
\EndWhile
\end{algorithmic}
\end{algorithm}

\begin{figure}[t]
\centerline{
\hskip 3mm
\raise 3.5cm \hbox{\hbox{(a)}}
\hskip -3mm
\includegraphics[width=0.2331\columnwidth]{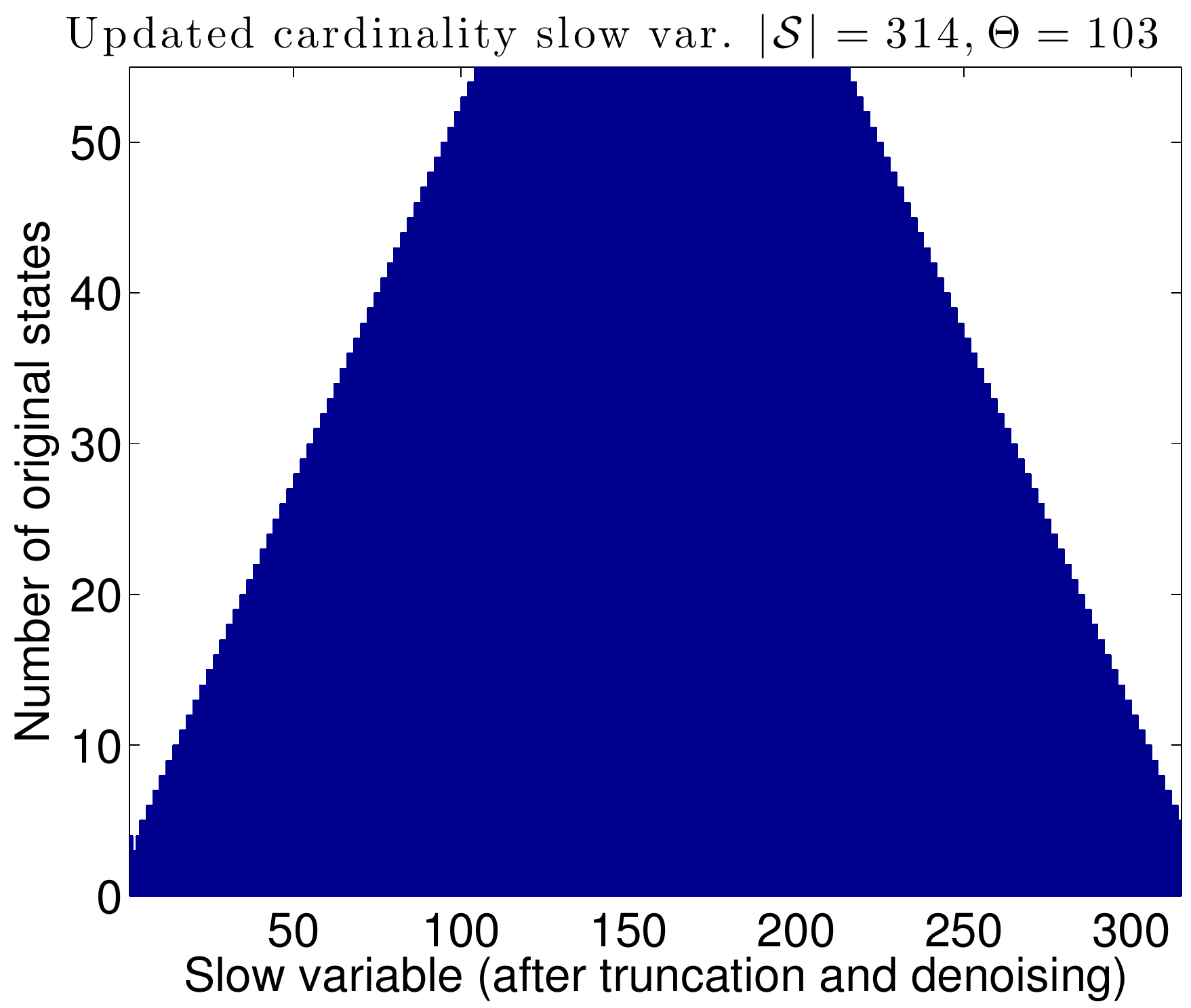}
\raise 3.5cm \hbox{\hbox{(b)}}
\hskip -3mm
\includegraphics[width=0.2331\columnwidth]{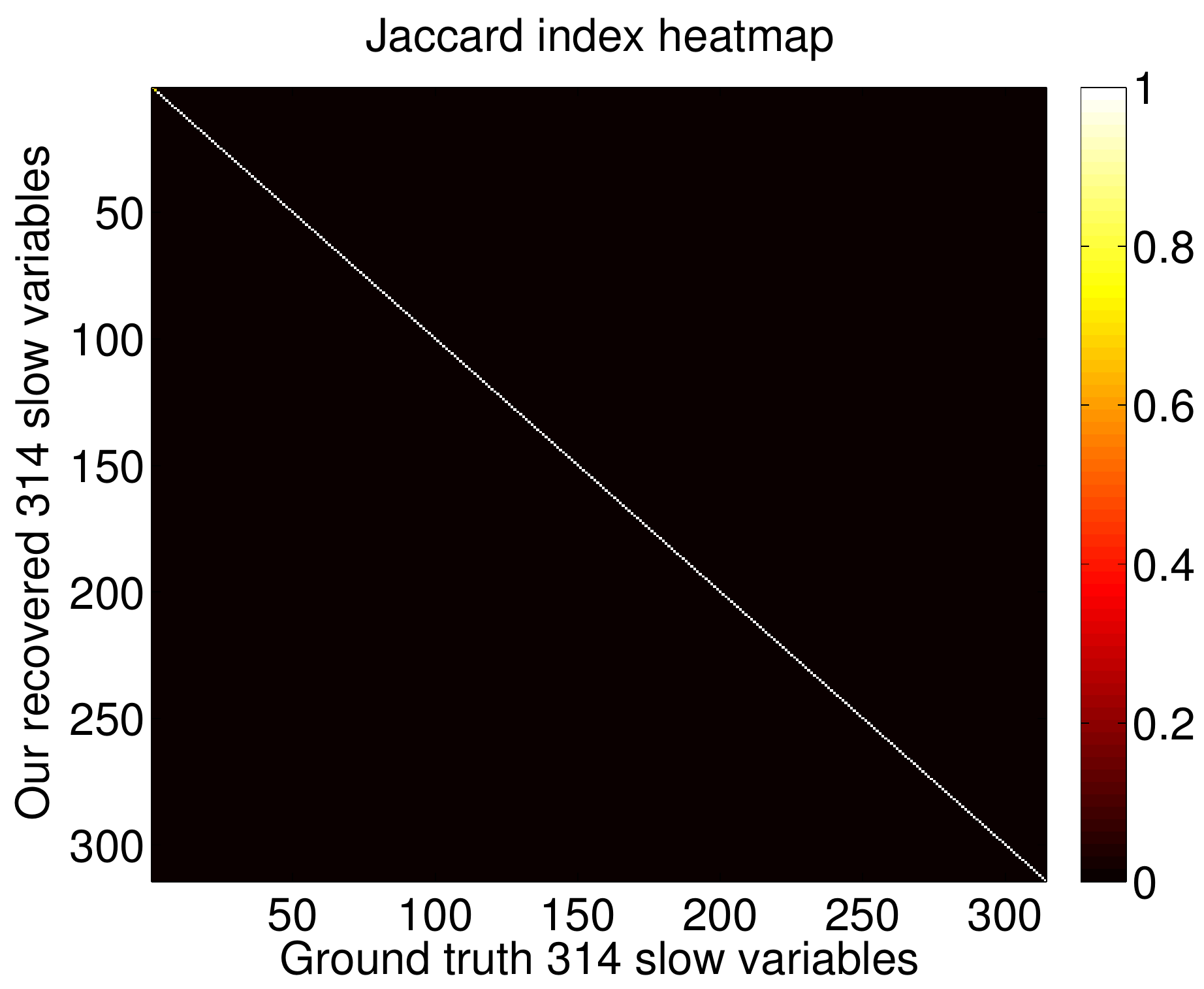}
\raise 3.5cm \hbox{\hbox{(c)}}
\hskip -3mm
\includegraphics[width=0.2331\columnwidth]{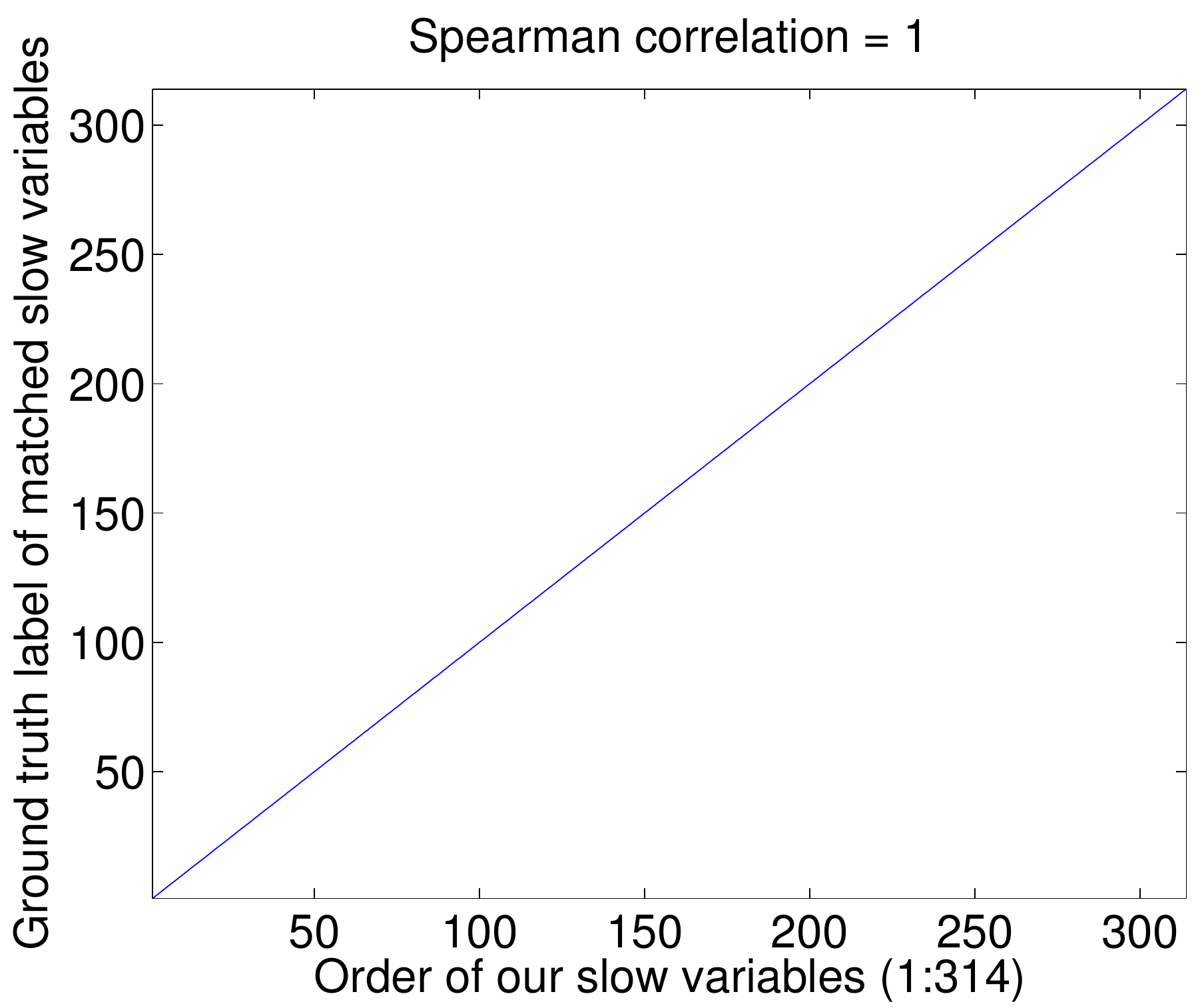}
\raise 3.5cm \hbox{\hbox{(d)}}
\hskip -3mm
\includegraphics[width=0.2331\columnwidth]{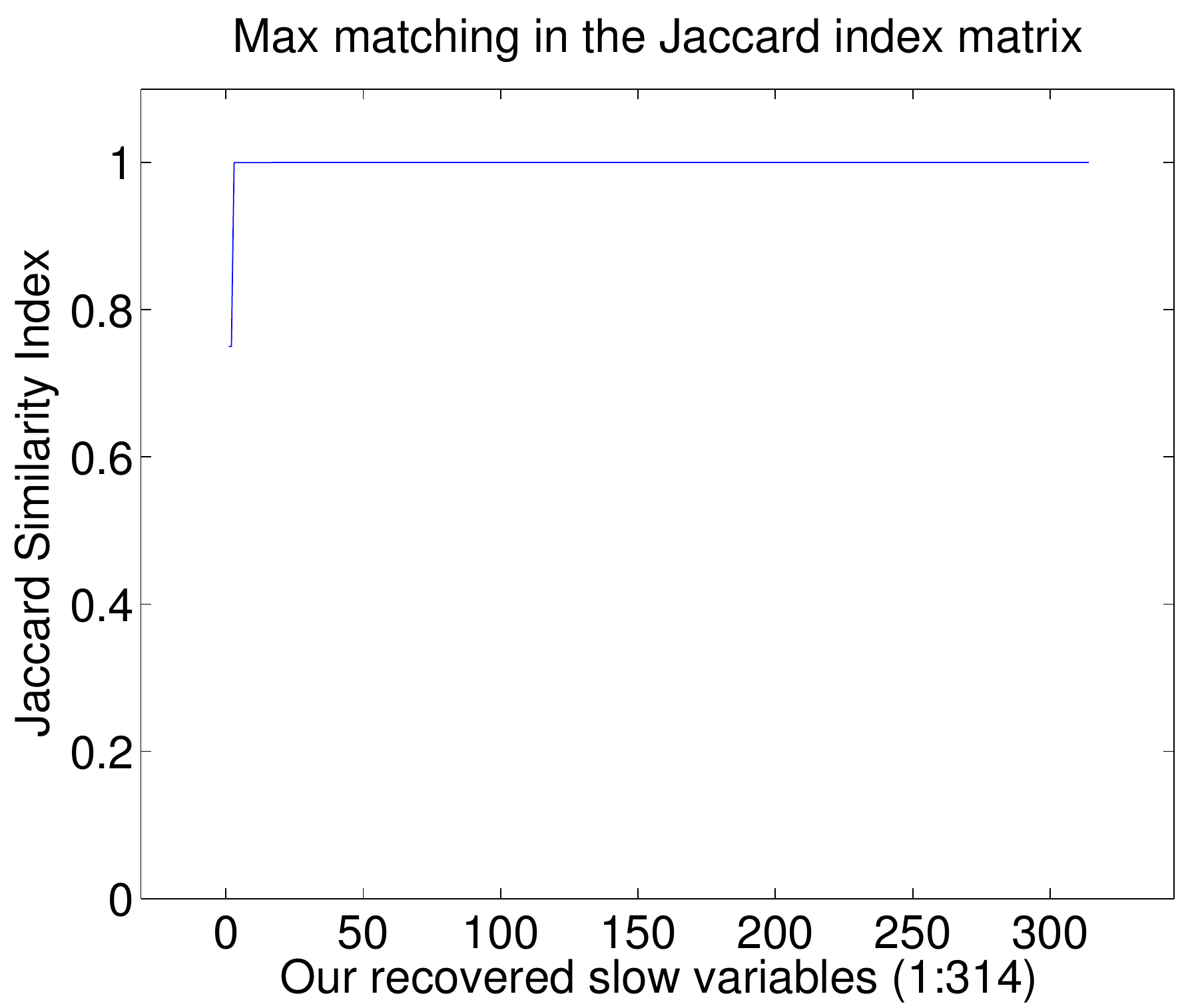}
}
\caption{{\it Illustrative example} {\rm CS-II.}
{\rm (a)} {\it The eigenvector-based slow variable cardinality after 
truncating and bin denoising. The Theta score $\Theta$ is the 
smoothness measure of the bin cardinalities, defined in $(\ref{def:thetaContScore}).$
}
{\rm (b)} 
{\it The heatmap of the pairwise Jaccard similarity matrix given
by $(\ref{JacIndex})$.}
{\rm (c)}
{\it The correlation between the ordering of the ground truth 
slow variable and the eigenvector recovered slow variable.}
{\rm (d)}
{\it The Jaccard index of the pairwise matched bins 
(from the maximum matching).}
}
\label{figure4p5}
\end{figure}

To solve this issue, we propose a bin-denoising heuristic that 
robustly assigns data points to their respective bins. In hindsight, 
the continuity of the eigenfunctions of the Laplacian should 
be reflected in the continuity of the histogram of state 
counts in bins corresponding to adjacent intervals. We detail 
in Algorithm \ref{alg:binMerging} an iterative heuristic procedure
which, at each step, merges two adjacent bins such that the 
resulting $\Theta$-score is minimized across all possible pairs 
of adjacent bins that can be merged. We show in Figure 
\ref{figure4p5}(a) the resulting bin cardinalities after 
the bin-merging heuristic and after truncating at the boundary 
of the slow variable. Note that the new denoised partition 
yields $\Theta = 103$, and the number of bins (states of the 
slow variable) decreases from $|{\mathcal S}|=328$ 
to $|{\mathcal S}|=314$. Furthermore, 
in Figure \ref{figure4p5}(b) we compute the Jaccard 
similarity matrix between the ground truth and the newly 
obtained partition, showing in Figure \ref{figure4p5}(d) that 
we almost perfectly recover the structure of the ground truth bins.

\section{A Markov approach for computing the steady distribution 
of the slow variable} 
\label{sec:MarkovSteady}

In this section, we focus on the final step of the ADM-CLE approach, 
of estimating the stationary distribution of the slow variable, 
without any prior knowledge of what the slow variable actually is. 
One of the ingredients needed along the way is an estimation of 
the  conditional distribution $\mathbb{P}(F| S = s)$ of the fast 
variable $F$ given a value $s$ of the slow variable $S$, 
which we compute via two approaches. 
As the first approach, we consider the Conditional Stochastic 
Simulation Algorithm (CSSA) \cite{cssa} which is given
in Algorithm~\ref{algoCSSA}. It samples from the distribution 
of the fast variable conditioned on the slow variable. 
The second approach is entirely analytic and free of any 
stochastic simulations, and amounts to analytically solving 
the CME for each set in the 
partition $\mathcal{P} = \{ \mathcal{P}_1, \mathcal{P}_2,
\ldots, \mathcal{P}_k \}$. 
We then compare our results to the Constrained Multiscale Algorithm (CMA) 
introduced in \cite{cssa}, which approximates the effective 
dynamics of the slow variable as a SDE, after estimating the 
effective drift and diffusion using the CSSA (Algorithm~\ref{algoCSSA}).

\subsection{A stochastic simulation algorithm for estimating 
the conditional probability (CSSA)}
\label{sec:StochConditional}

Our next task is to estimate the conditional distribution 
$\mathbb{P}(F|S=s)$ of the fast variable $F$ given a value 
$s$ of the slow variable $S$. One possible approach for doing 
this relies on the CSSA algorithm to globally integrate  
the effective dynamics of the slow variable. One iteration 
of the CSSA is given in Algorithm \ref{algoCSSA}.
Ideally, one repeats steps 1--6 of Algorithm \ref{algoCSSA}
and samples values of $F$ until the distribution $\mathbb{P}(F|S=s)$ converges. 
In practice, we run Algorithm \ref{algoCSSA} until $L_c$ changes 
of the slow variable $S$ occur. 
This computation is done for each value in the range of the 
slow variable 
$ \mathcal{S} 
= \{ s_1, s_2, \ldots, s_{|{\mathcal{S}}|} \}
$.

\begin{algorithm}[t]
\caption{One iteration of the CSSA for computing the 
conditional distribution $\mathbb{P}(F| S=s)$ of the fast variable $F$ given 
a value $s$ of the slow variable $S$} \label{algoCSSA}
\begin{algorithmic}[1]
\State Compute the propensity functions $\alpha_i (t)$, for 
$i=1,2,\ldots,m$, and their sum 
$\alpha_0(t) = \displaystyle \sum_{i=1}^m \alpha_i(t)$.
\vspace{-3mm}
\State Generate $r_1$ and $r_2$, two uniformly distributed random 
numbers in $(0,1)$.
\State Compute the next reaction time as $t + \tau$ where
$\tau = - \log(r_1) / \alpha_0(t)$. 
\State Use $r_2$ to select reaction $R_j$ which occurs at
time $t+\tau$ \hfill\break 
(each reaction $R_i$, $i=1,2,\dots,m$
occurs with probability $\alpha_i/\alpha_0$).
\State If the slow variable $S$ changes its current state from $s$ 
to $s' \neq s$ due to reaction $R_j$ occurring, reset 
$S=s$ to its previous value, while not changing the value 
of the fast variable $F$.
\State If any of the variables $X_i$ goes outside the boundary of
the considered domain, then revert to the state of the system 
in Step 4 before reaction $R_j$ occurred.
\end{algorithmic}
\end{algorithm}

\begin{table}[t]
\begin{center}
\begin{tabular}{|l|l|l  ||  l|l|l  ||   l|l|l| }
\hline
$X_1$ & $X_2$&    $S$ &  $X_1'$&  $X_2'$&   $S'$ &  $R_j$   & $\alpha_j/\alpha_0$ & $\alpha_j$ \\
\hline
    \textbf{1} &  \textbf{3} & \textbf{7} &  \textbf{2} & \textbf{3} &  \textbf{8} &   \textbf{1} &   $\mathbf{7.8 \times 10^{-3}}$  & 96\\
    1 &     3 &     7 &     0 &     3 &     6 &     2 &     $  7.8 \times 10^{-5} $  & 0.96 \\
\textbf{1} & \textbf{3} & \textbf{7}& \textbf{2} & \textbf{3} & \textbf{8} & \textbf{3} & $\mathbf{1.5 \times 10^{-2}}$  & 184.38 \\
    1 &     3 &     7 &     0 &     3 &     6 &     4 &     $1.6 \times 10^{-3}$ & 19.75 \\
    1 &     3 &     7 &     3 &     2 &     7 &     6 &     $ 9.7 \times 10^{-1}$ &  12000 \\
\hline \hline
\textbf{3} & \textbf{2} & \textbf{7} & \textbf{4} & \textbf{2} & \textbf{8} & \textbf{1} &   $\mathbf{7.3 \times 10^{-3}}$ &  64 \\
    3 &     2 &     7 &     2 &     2 &     6 &     2 &    $2.0 \times 10^{-4}$  &  1.92 \\
\textbf{3} & \textbf{2} & \textbf{7} & \textbf{4} & \textbf{2} & \textbf{8} & \textbf{3} & $ \mathbf{2.1 \times 10^{-2}} $  & 184.38 \\
    3 &     2 &     7 &     2 &     2 &     6 &     4 &    $6.7 \times 10^{-3}$  &  59.25 \\
    3 &     2 &     7 &     1 &     3 &     7 &     5 &    $5.4 \times 10^{-2}$   &  480 \\
    3 &     2 &     7 &     5 &     1 &     7 &     6 &    $9.1 \times 10{-1}$  &  8000 \\
\hline \hline
\textbf{5} & \textbf{1} & \textbf{7} & \textbf{6} & \textbf{1} & \textbf{8} & \textbf{1} & $\mathbf{5.4 \times 10^{-3}}$  &  32  \\
    5 &     1 &     7 &     4 &     1 &     6 &     2 &     $3.0 \times 10^{-4}$   & 1.6  \\
\textbf{5} & \textbf{1} & \textbf{7} & \textbf{6} & \textbf{1} & \textbf{8} & \textbf{3} & $\mathbf{3.1 \times 10^{-2} }$   & 184.38  \\
    5 &     1 &     7 &     4 &     1 &     6 &     4 &   $ 1.6 \times 10^{-2} $   &  98.75 \\
    5 &     1 &     7 &     3 &     2 &     7 &     5 &    $2.7 \times 10^{-1} $  &  1600 \\
    5 &     1 &     7 &     7 &     0 &     7 &     6 &   $6.7 \times 10^{-1} $ &  4000 \\
\hline \hline
\textbf{7} & \textbf{0} & \textbf{7} & \textbf{8 }& \textbf{0 }& \textbf{8} & \textbf{3} & $ \mathbf{5.0} \times 10^{-2} $  & 184.38 \\
    7 &          0 &    7 &     6 &        0 &    6 &     4 &     $ 3.7 \times 10^{-2} $  &  138.25 \\
    7 &          0 &    7 &     5 &        1 &    7 &     5 &     $ 9.1 \times 10^{-1} $  &  3360  \\
\hline
\end{tabular}
\end{center}
\caption{{\it Illustrative example} {\rm CS-II}. 
{\it The set of all ground states of the system $(x_1,x_2)$ 
corresponding to the slow variable $S=X_1 + 2 X_2 = 7$.
We denote by $(x_1',x_2')$ the states reachable from $(x_1,x_2)$ 
in one transition step, and by $S'$ the associated corresponding slow 
variable such that $S' = X_1' + 2 X_2'$. $R_j$ denotes the reaction 
channel that takes the chemical system from state 
$(x_1,x_2)$ to $(x_1',x_2')$, 
with corresponding propensity $\alpha_j$. We highlight in bold letters 
the subset of all states via which the system can transition in one step 
from the slow variable $S=7$ to $S=8$.} 
}
\label{table5p1}
\end{table}

\subsection{An analytical derivation of the conditional distribution} 
\label{sec:AnalyticConditional}

\begin{figure}[t]
\centerline{
\hskip 3mm
\raise 5.1cm \hbox{\hbox{(a)}}
\hskip 2mm
\includegraphics[width=0.39 \columnwidth]{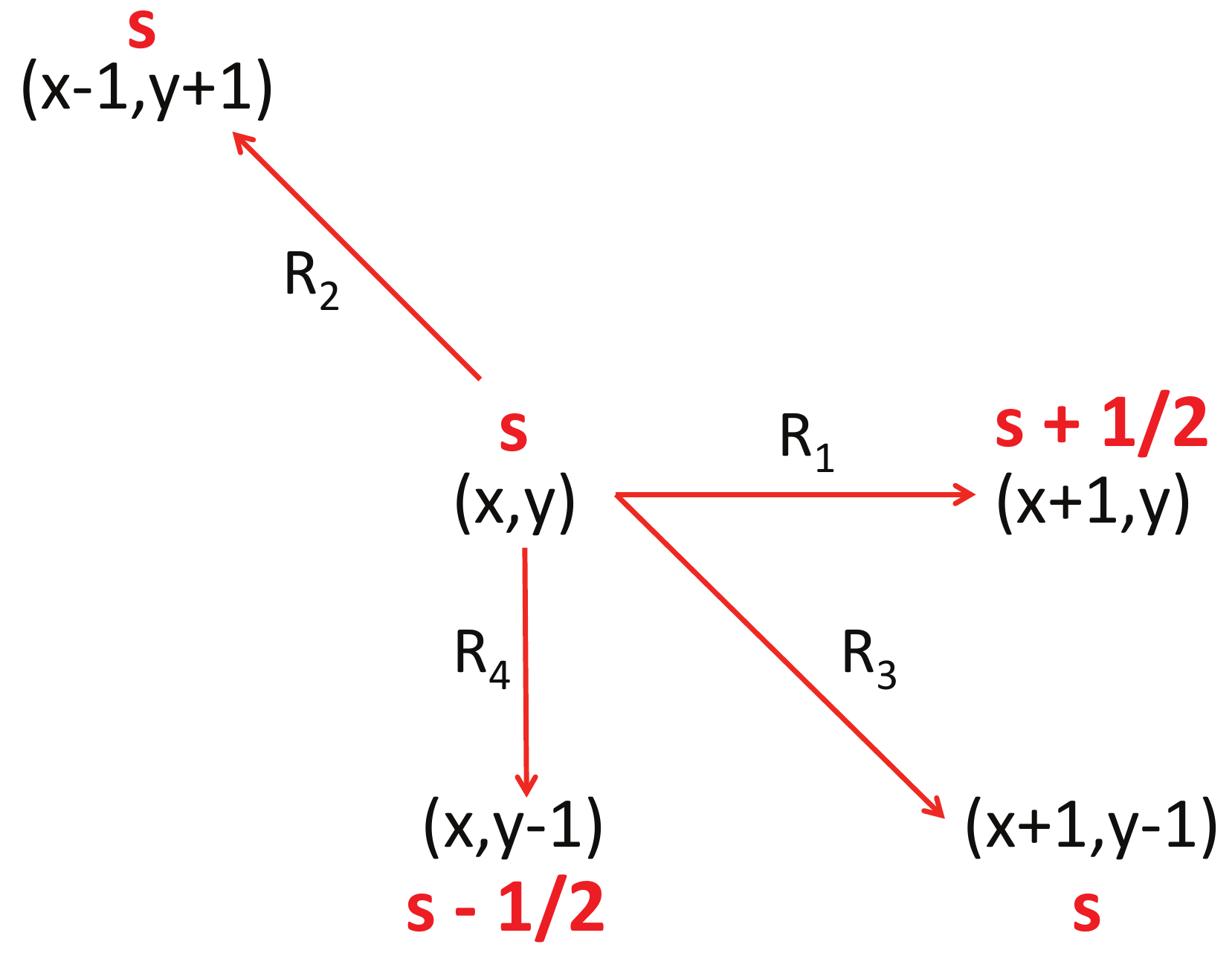}
\hskip 2mm
\raise 5.1cm \hbox{\hbox{(b)}}
\hskip 2mm
\includegraphics[width=0.58 \columnwidth]{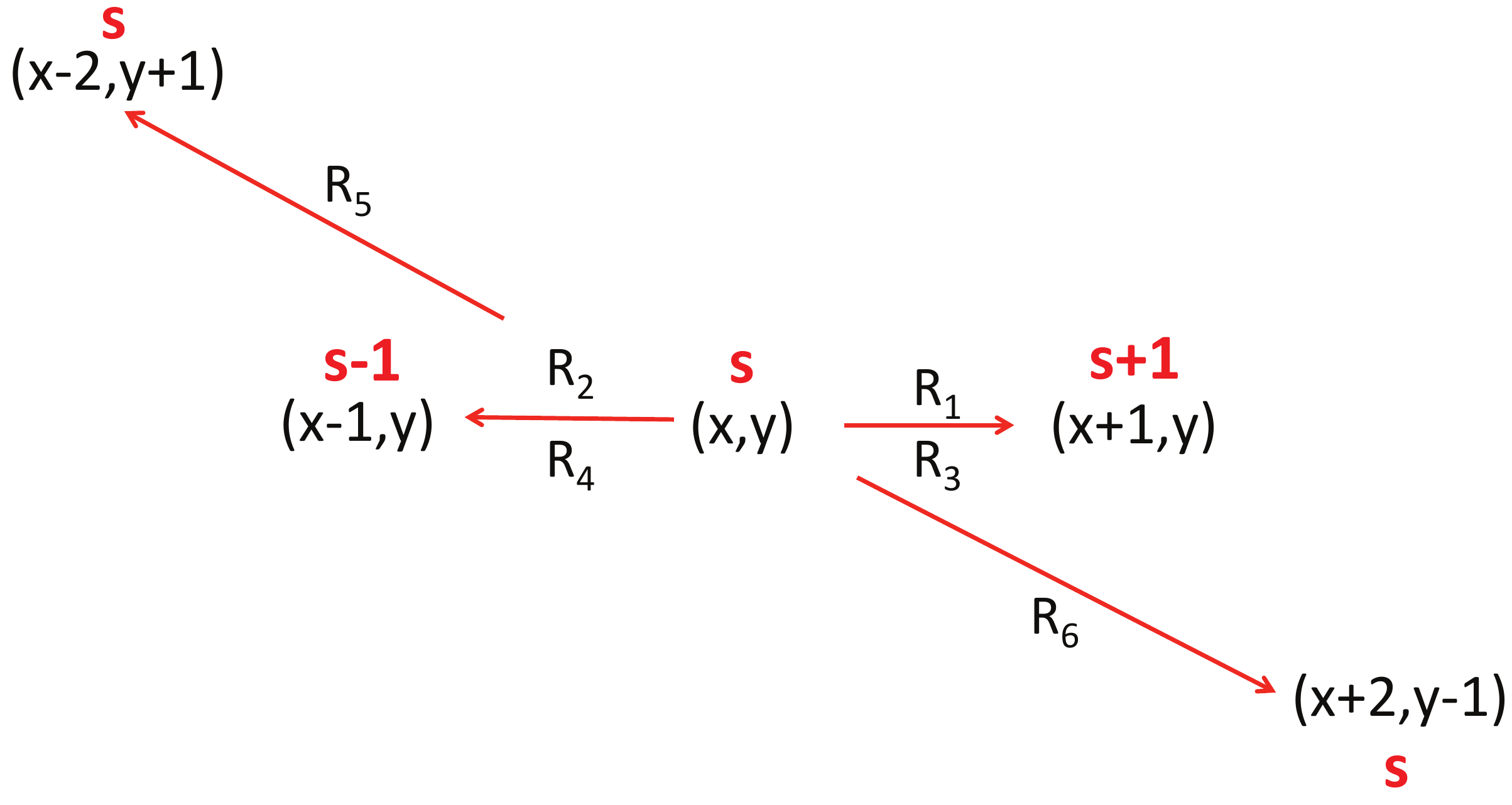}
}
\caption{{\it
Transition diagrams for the two chemical systems:}
{\rm (a) CS-I;} {\it and} {\rm (b) CS-II.}
}
\label{figure5p1}
\end{figure}

An alternative approach which we follow in this paper relies on an 
analytical computation of the conditional distribution 
$\mathbb{P}( F | S = s )$, 
thus eliminating the need for any expensive stochastic simulations.
We illustrate in Figure \ref{figure5p1} the transition diagrams 
for the two chemical systems we consider in this paper. For chemical 
system CS-II, 
the system can transition from a given state $(x_1,x_2)$ to four adjacent 
distinct $\mathcal{O}$-states: 
to $(x_1-2,x_2 +1)$ via channel $R_5$, to $(x_1+2,x_2-1)$ 
via channel $R_6$, to $(x_1 -1,x_2)$ via channels $R_2$ and $R_4$, 
and finally to 
state $(x_1+1,x_2)$ via channels $R_1$ and $R_3$. However, in terms of 
the underlying slow variable $S$, the system can  transition to only 
two adjacent states $S=s-1$ (via channels $R_2$ and $R_4$) and $S=s+1$ 
(via channels $R_1$ and $R_3$), or remain at the current state $S=s$, 
via channels $R_5$ and $R_6$. 
Considering the subsystem of fast reactions of CS-II and conditioning 
on the line $s = x_1 + 2 x_2$, the stationary CME takes the form
\begin{eqnarray*}
0 
&
=
&
k_5 (x_1+2) (x_1+1) \, \mathbb{P}(X_1=x_1+2, X_2=x_2-1)  
+ k_6 (x_2+1) \, \mathbb{P}(X_1=x_1-2, X_2=x_2+1) \\ 
&-& ( k_5 x_1 (x_1-1) + k_6 x_2) \, \mathbb{P}(X_1=x_1, X_2=x_2).
\end{eqnarray*}
Thus, the conditional distribution for CS-II is, for $0 \le x_1 \le s$, 
given by 
$$
\mathbb{P}(F=x_1 | S=s) 
=  
\frac{C}{x_1! \, x_2!} \left( \frac{k_5}{k_6} \right)^{x_2}  
= 
\frac{C}{x_1! \, ((s-x_1)/2)! } \left( \frac{k_5}{k_6} \right)^{(s-x_1)/2},
\qquad
\mbox{if} \; (s-x_1) \; \mbox{is an even number}.
$$
Here, $C$ is the normalization constants and 
$\mathbb{P}(F=x_1 | S=s) = 0$ if $(s-x_1)$ is 
an odd number. 

A similar argument can be used for CS-I. 
The stationary CME of the fast subsystem of CS-I is written as
\begin{eqnarray*}
0 
&
=
&
k_2 \, (x_1+1) \, \mathbb{P}(X_1 = x_1+1, X_2 = x_2-1)  
+ k_3 \, (x_2+1) \, \mathbb{P}(X_1=x_1-1, X_2 = x_2+1)  \\
&-& ( k_2 x_1 + k_3 x_2) \, \mathbb{P}(X_1=x_1, X=x_2).
\end{eqnarray*}
where $s = (x_1 + x_2)/2$. Thus, the conditional distribution for CS-I is, 
for $0 \le x_1 \le 2 s$, given by 
$$
\mathbb{P}(F=x_1 | S=s) 
= 
\frac{C}{x_1! x_2!}   \left(  \frac{k_2}{k_3} \right)^{x_2}
=
\frac{C}{x_1! (2 s- x_1)!} \left(  \frac{k_2}{k_3} \right)^{2 s-x_1},
$$
where $C$ is again the normalization constant.

\subsection{Aggregated transition rates and a Markov Chain 
on the state of slow variables} \label{sec:AggProbMarkovSlow}

\begin{figure}[t]
\begin{center}
\includegraphics[width=0.65\columnwidth]{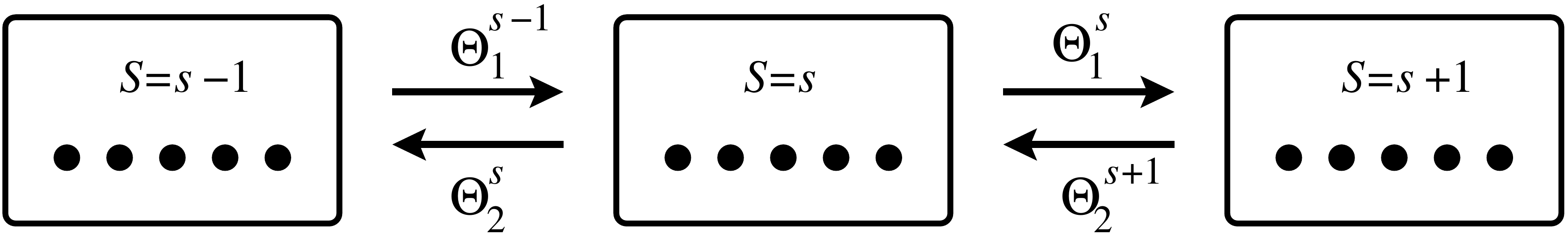}
\end{center}
\vskip -2.25cm
\hskip 3.9cm
$\mathcal{P}_{s-1}$
\hskip 3.65cm
$\mathcal{P}_{s}$
\hskip 3.65cm
$\mathcal{P}_{s+1}$
\vskip 1.8cm
\caption{{\it
The Markov chain on the slow variable state space, using the aggregated 
transition probabilities $(\ref{theta1})$--$(\ref{theta2})$
for the chemical system} {\rm CS-II.}}
\label{figure5p2}
\end{figure}

\begin{figure}[t]
\centerline{
\hskip 3mm
\raise 5.1cm \hbox{\hbox{(a)}}
\hskip -3mm
\includegraphics[width=0.32\columnwidth]{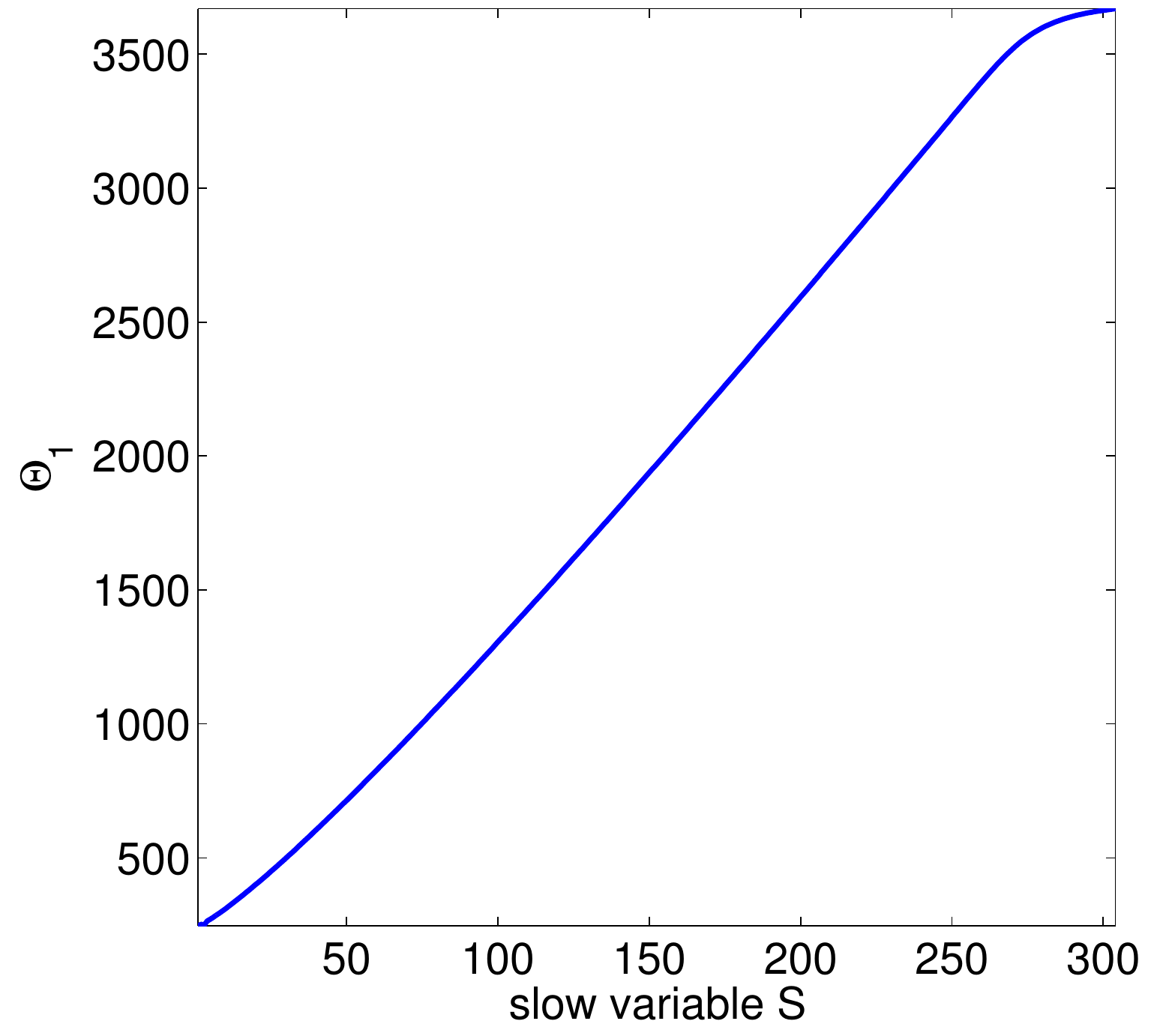}
\raise 5.1cm \hbox{\hbox{(b)}}
\hskip -3mm
\includegraphics[width=0.32\columnwidth]{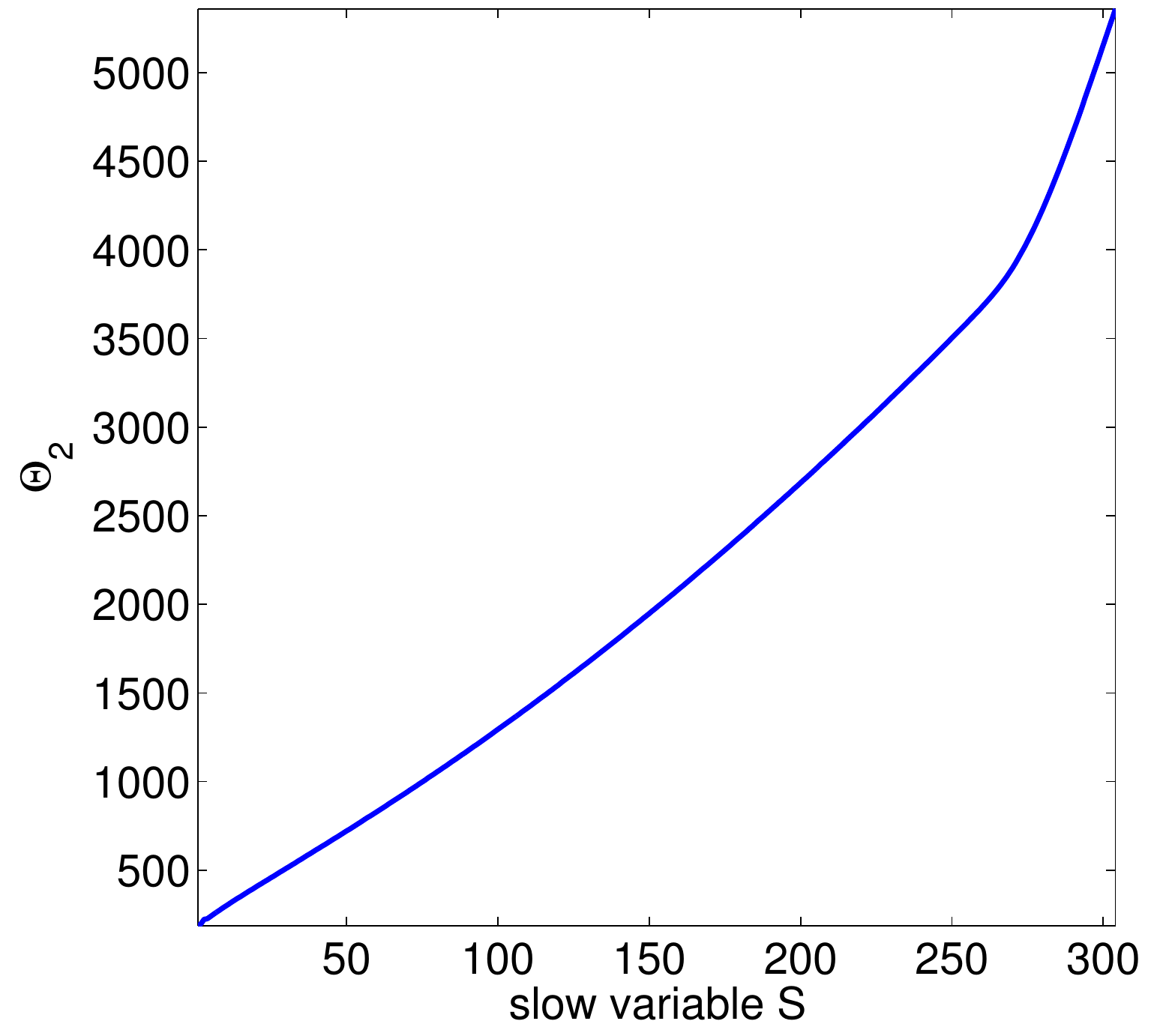}
\raise 5.1cm \hbox{\hbox{(c)}}
\hskip -3mm
\includegraphics[width=0.32\columnwidth]{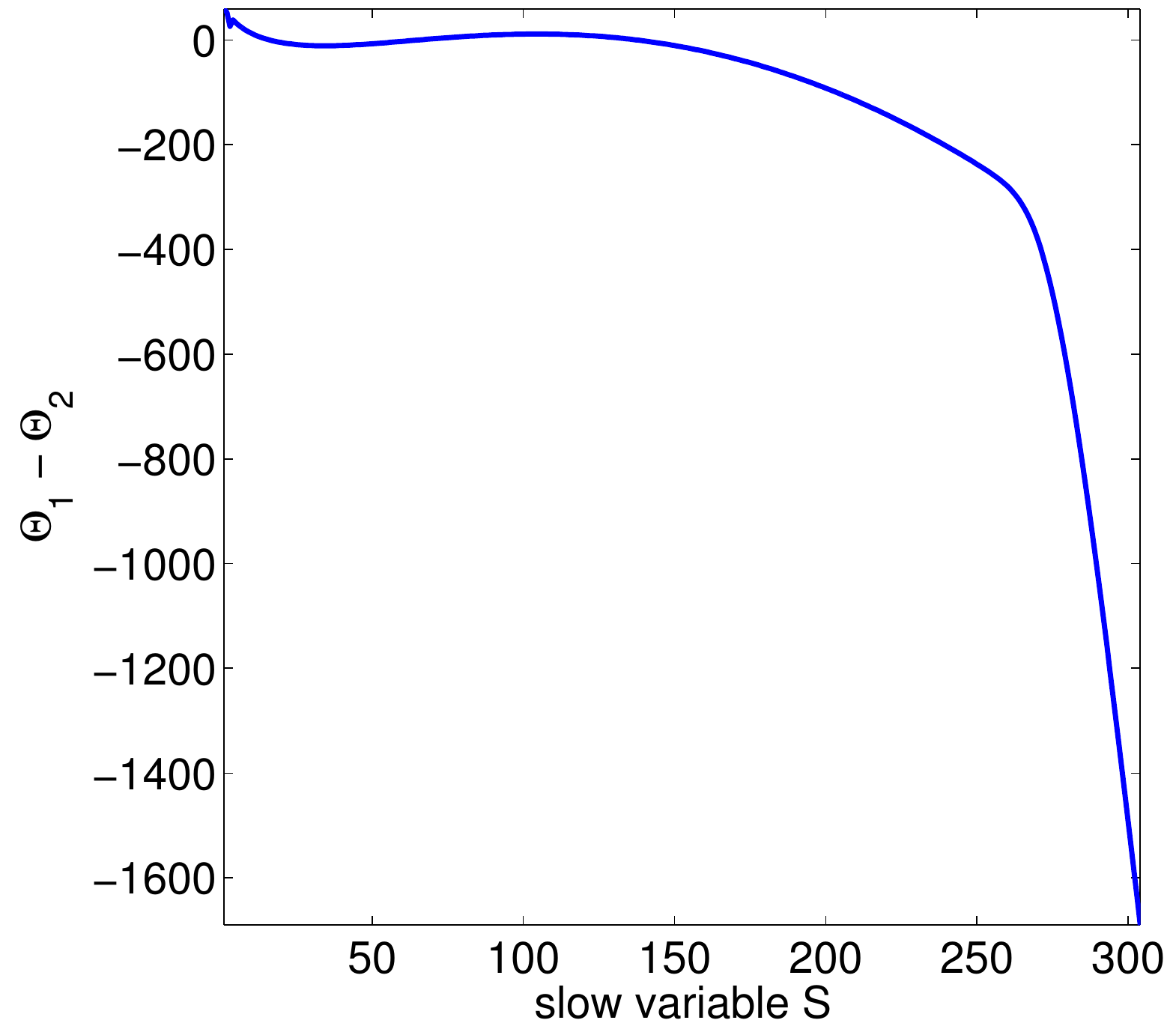}
}
\caption{{\it Plot of the aggregated transition rates for 
the illustrative example} {\rm CS-II: 
(a)} $\Theta_1$; {\rm (b)} $\Theta_2$; 
{\it and} {\rm (c)} $\Theta_1 - \Theta_2$.}
\label{figure5p3}
\end{figure}

In the final step of the ADM-CLE approach, we set up a Markov chain 
on the state space of slow variables with the end goal of estimating 
the stationary distribution of the slow variable. 
As illustrated in Figure \ref{figure5p1}(b), the system CS-II can 
can transition from a given state $S=s$ to 
two adjacent states $S=s-1$ (via reaction channels $R_2$ and $R_4$) and 
$S=s+1$ (via channels $R_1$ and $R_3$), or it can remain at the 
current state $S=s$, via channels $R_5$ and $R_6$. Consider now 
the set 
$\mathcal{P}_s = 
\{ {\mathbf x}^{(i)} = (x_1, x_2)^{(i)} | x_1 + 2 x_2 = s\} $, 
illustrated 
as the middle bin in Figure \ref{figure5p2}. To compute the transition 
rate between two adjacent bins $\mathcal{P}_s$ and $\mathcal{P}_{s+1}$, 
one has to aggregate over possible ways of getting from an observable state 
in bin $\mathcal{P}_s$ to an observable state in bin $\mathcal{P}_{s+1}$.
We compute $\Theta_1^{(s)}$ to be the aggregated transition 
rate from state $\mathcal{P}_s$ to state $\mathcal{P}_{s+1}$, 
over all possible states $({\mathbf x}^{(i)},{\mathbf x}^{(j)})$, 
such that ${\mathbf x}^{(i)} \in \mathcal{P}_S$ 
and ${\mathbf x}^{(j)} \in \mathcal{P}_{s+1}$, by
\begin{equation}
\Theta_{1}^{s} 
= \sum_{{\mathbf x}^{(i)} \in \mathcal{P}_s }   
\sum_{{\mathbf x}^{(j)} \in  \mathcal{P}_{s+1}}  
\mathbb{P}( F =x_1^{(i)} | S=s)   
\sum_{k=1}^{m}  \alpha_k({\mathbf x}^{(i)}) \,    
Q({\mathbf x}^{(i)},{\mathbf x}^{(j)} , R_k)
\label{theta1}
\end{equation}
where $Q({\mathbf x}^{(i)},{\mathbf x}^{(j)} , R_k)$ denotes 
the indicator functions of whether one can transition from 
the $\mathcal{O}$-state ${\mathbf x}^{(i)}$ to
$\mathcal{O}$-state ${\mathbf x}^{(j)}$ via reaction 
$R_k$. We define similarly the aggregated transition 
rate $\Theta_{2}^{s}$, that the chemical system transitions from the slow variable 
state $ \mathcal{P}_s$ to $ \mathcal{P}_{s-1}$ by
\begin{equation}
\Theta_{2}^{s} 
= \sum_{{\mathbf x}^{(i)} \in \mathcal{P}_s }   
\sum_{{\mathbf x}^{(j)} \in  \mathcal{P}_{s-1}}  
\mathbb{P}( F =x_1^{(i)} | S=s)   
\sum_{k=1}^{m}  \alpha_k({\mathbf x}^{(i)}) \, 
Q({\mathbf x}^{(i)},{\mathbf x}^{(j)} , R_k)
\label{theta2}
\end{equation}
We illustrate in Figure \ref{figure5p3} the aggregated transition  
rates between the slow state $S=s$ and its adjacent states 
$S=s-1$ and  $S=s+1$, for all values of the slow variable $S$.  
Note that in the derivations  
(\ref{theta1}) and (\ref{theta2}), we can either rely on the CSSA 
algorithm to sample from the conditional distribution of fast variables 
given values for the  slow variables, as shown in Section  \ref{sec:StochConditional}, or use the  analytic formulation which is
possible to derived for both CS-I and CS-II, see Section 
\ref{sec:AnalyticConditional}.

\begin{figure}[t]
\centerline{
\hskip 3mm
\raise 5.1cm \hbox{\hbox{(a)}}
\hskip -3mm
\includegraphics[width=0.3\columnwidth]{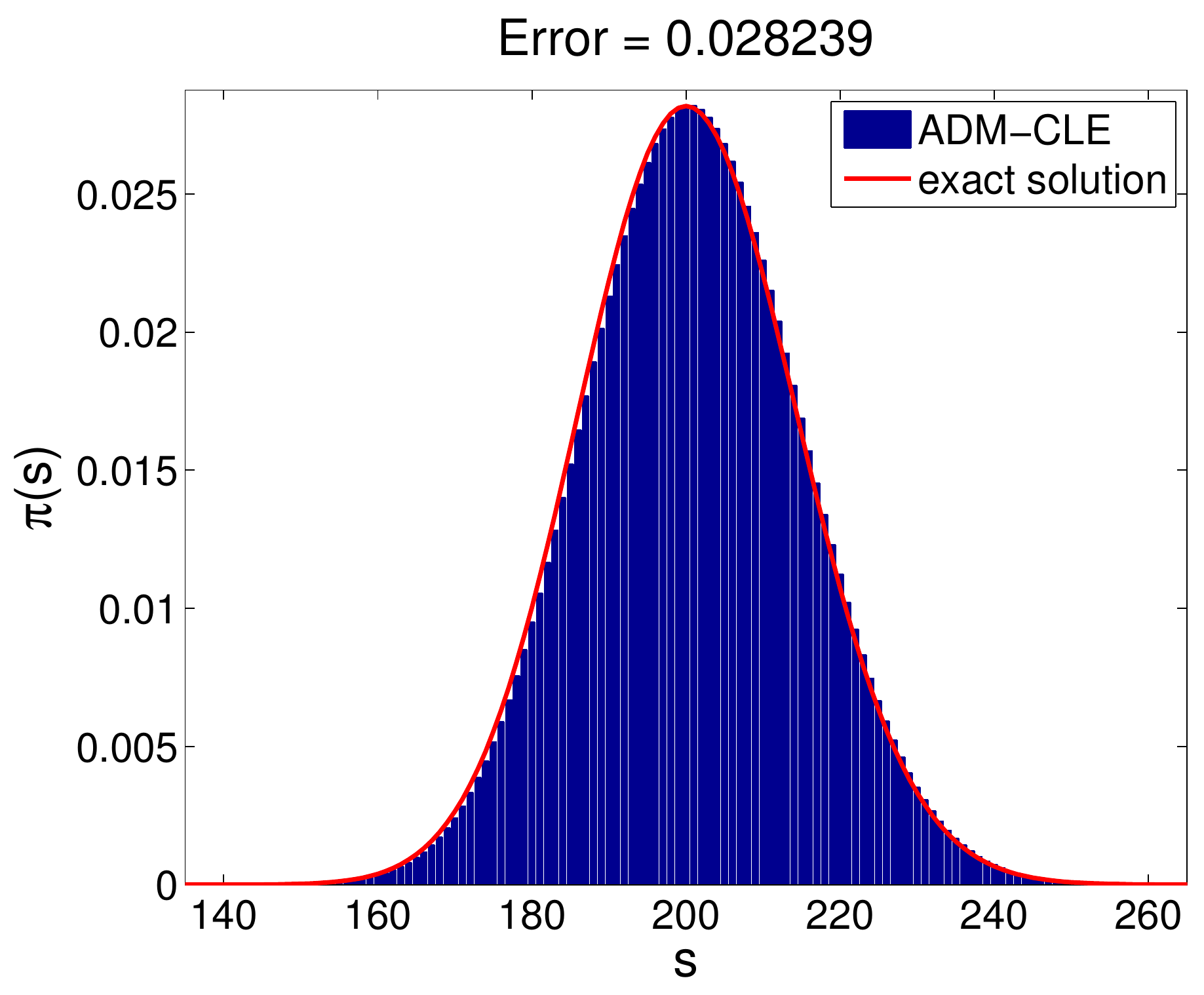} 
\hskip 3mm
\raise 5.1cm \hbox{\hbox{(b)}}
\hskip -3mm
\includegraphics[width=0.3\columnwidth]{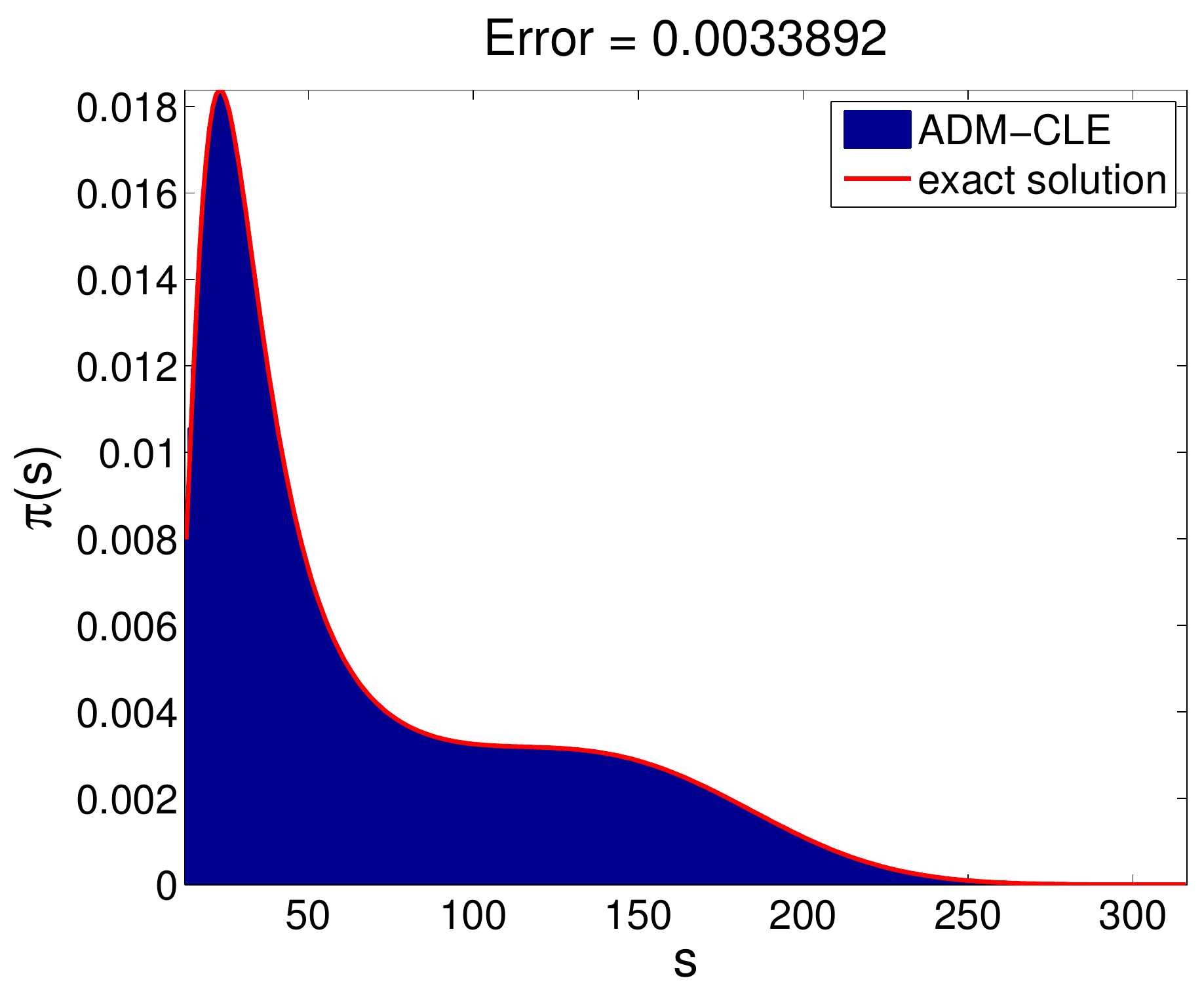} 
}
\caption{
{\it
The final estimated stationary distribution of the slow variable $S$ for the ADM-CLE approach, computed without knowledge of the slow variable, for}
(a) {\it the chemical system} {\rm CS-I;} {\it and} (b) 
{\it the chemical system} {\rm CS-II.} {\it (blue histograms).
Red solid lines are exact solutions computed by solving the CME
of the full model and using the corresponding definition of the
slow variable.}} 
\label{figure5p4}
\end{figure}

Finally, we compute the solution to the stationary CME associated 
to the system in Figure \ref{figure5p2} which can be written as
\begin{equation}
0 = \Theta_1^{s-1} \pi(s-1) 
+ 
\Theta_2^{s+1} \pi(s+1)
- 
\left(  \Theta_1^{s} +\Theta_2^{s} \right) \pi(s), 
\label{ex2CMESlow}
\end{equation}
where $\pi(s) \approx \mathbb{P}(S=s)$ is the probability that $S=s$ at 
time $t$. Assuming that $\pi(s)=0$ for all $s \not\in \mathcal{S}$
and using no-flux boundary conditions, 
we arrive at a linear system. The eigenvector of the resulting matrix, 
with associated eigenvalue $\lambda=0$, yields an approximate solution 
of the stationary CME, which we plot in blue in Figure \ref{figure5p4}.
Our result is visually indistinguishable from the exact solution
(plotted as the red solid line). 

\subsection{A comparison with the Constrained Multiscale Algorithm (CMA)}
\label{sec:CMASimon}

We compare the approach we introduced in the previous 
Section \ref{sec:AggProbMarkovSlow} with the CMA method proposed in~\cite{cssa}.
We compare the results of the two methods with 
the ground truth, and record the error defined as
\begin{equation}
\mbox{Error} \Big( \pi,\mathbb{P}(S=s) \Big) 
= 
\sum_{s \in \mathcal{S}} \Big| \pi(s) -  \mathbb{P}(S=s) \Big|, 
\label{errorDist}
\end{equation}
where $\mathbb{P}(S=s)$ denotes the ground truth probability 
distribution of the slow, and $\pi$ denotes the estimated solution,
either by the CMA and or the ADM-CLE. As Table \ref{tabpi} shows, 
the ADM-CLE approach yields lower errors compared to the CMA algorithm, 
even when we run the latter with the parameter $L_c$ as large 
as 20,000. Note that for the chemical system CS-I, the ground truth 
probability distribution of the slow variable $\mathbb{P}(S=s)$
can be easily computed using the multivariate Poisson 
distribution, as discussed in Section \ref{sec:subsectEx1Intro}. For the
second chemical system CS-II, we consider as ground truth the solution 
obtained by solving the associated CME of the full model 
in a large (truncated) domain.  

\begin{figure}[t]
\begin{center}
\subfigure[$ L=100$]{\includegraphics[width=0.234\columnwidth]{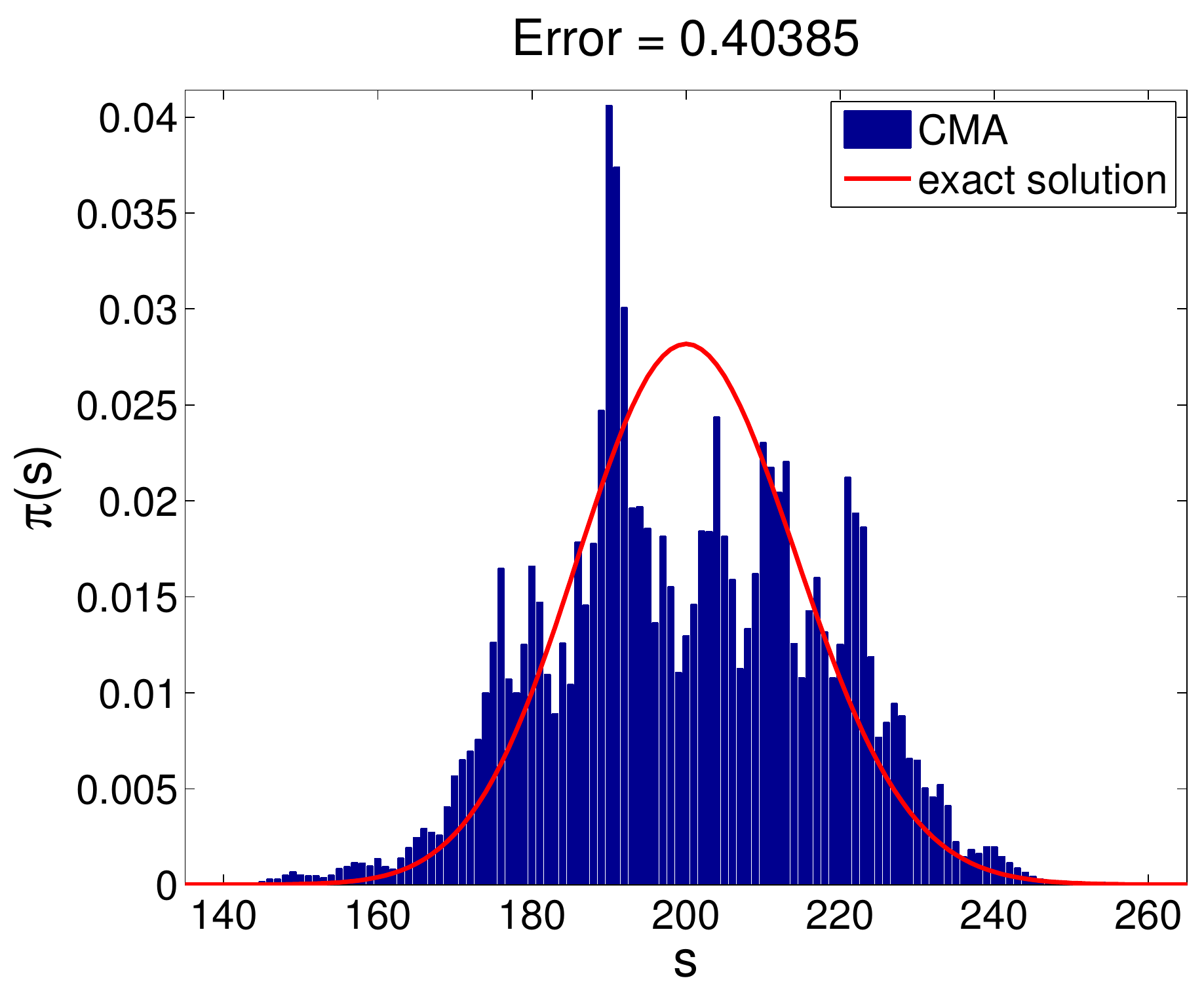}}
\subfigure[$ L=1,000$]{\includegraphics[width=0.234\columnwidth]{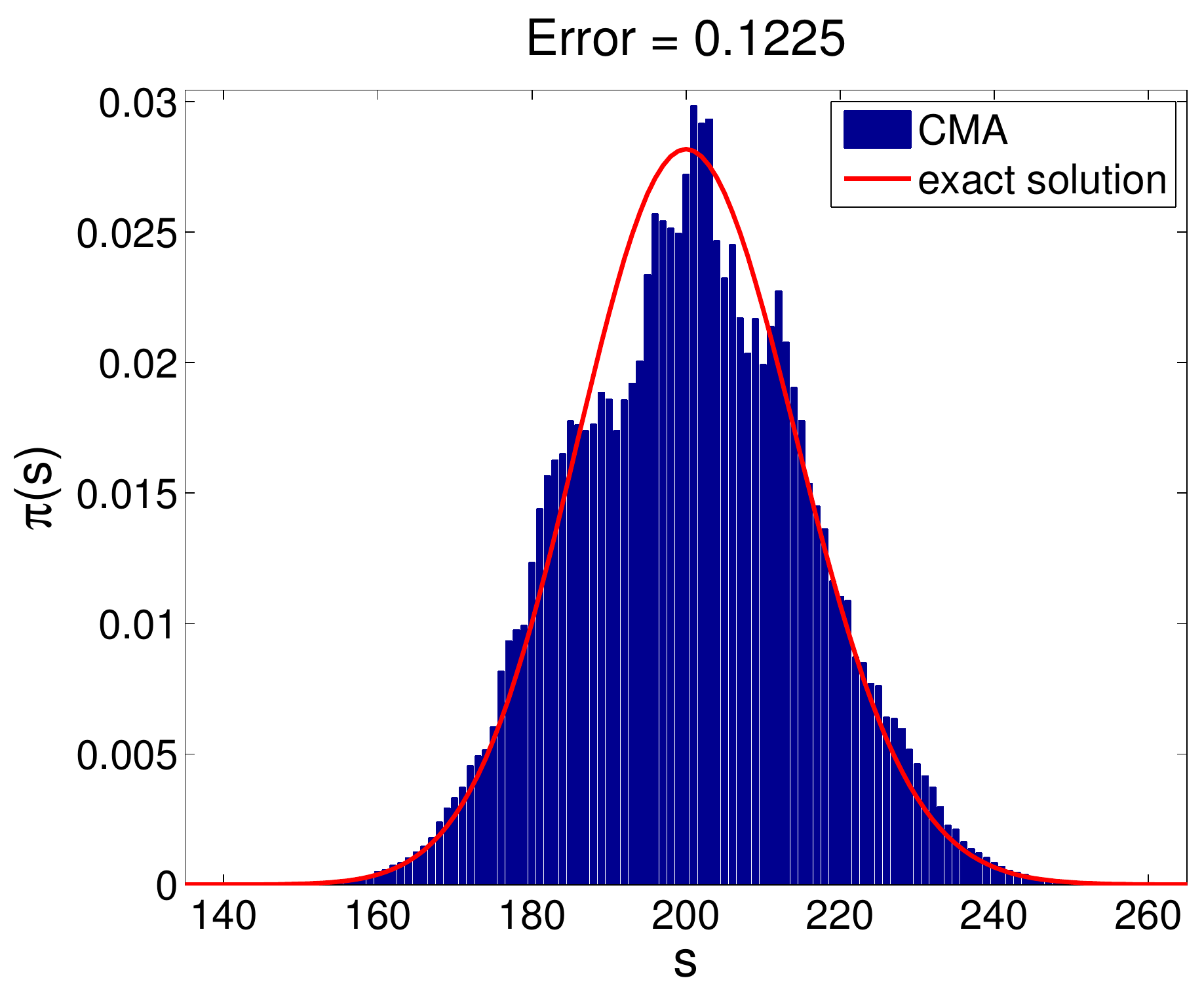}}
\subfigure[$ L=10,000$]{\includegraphics[width=0.234\columnwidth]{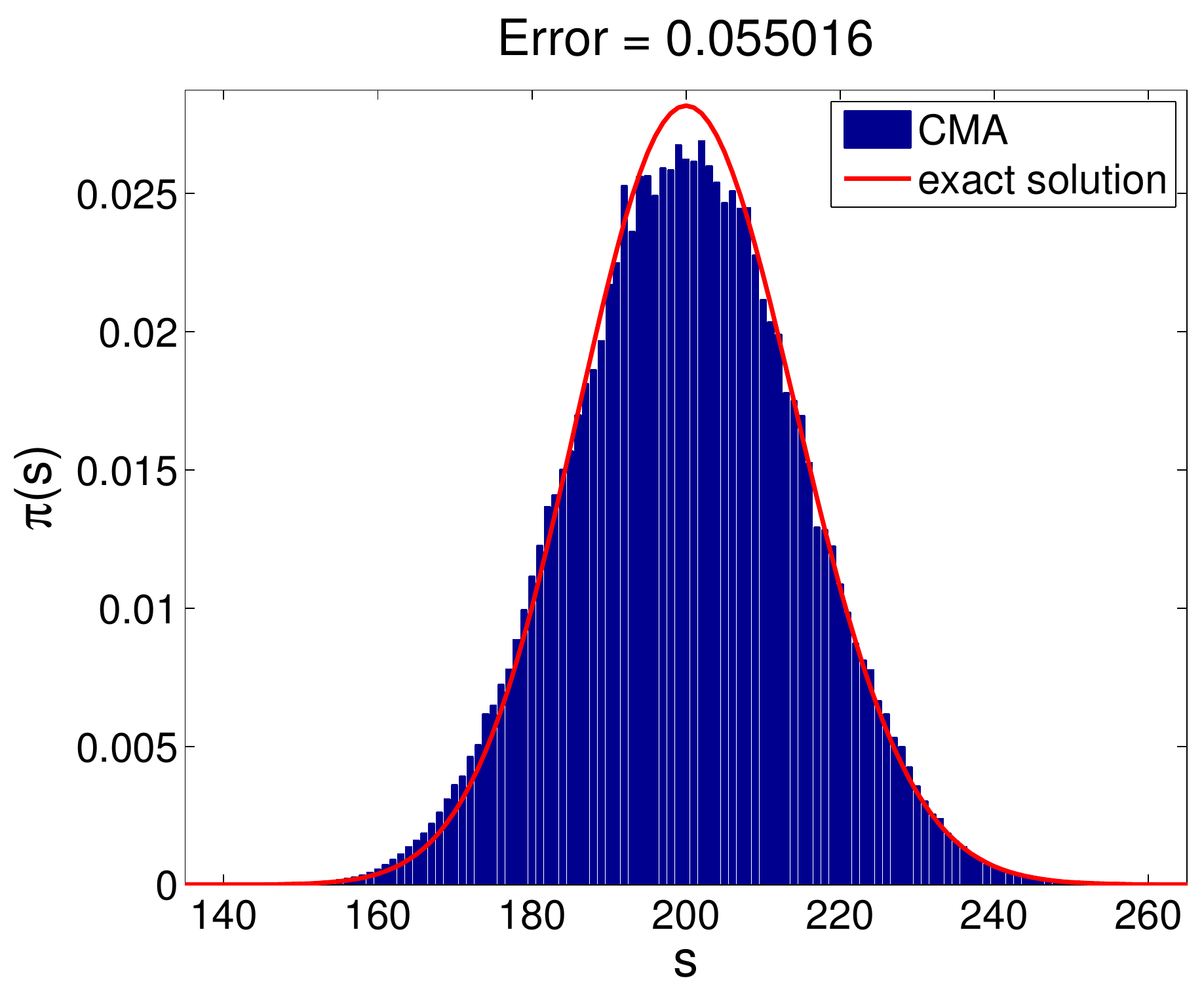}}
\subfigure[$ L=100,000$]{\includegraphics[width=0.234\columnwidth]{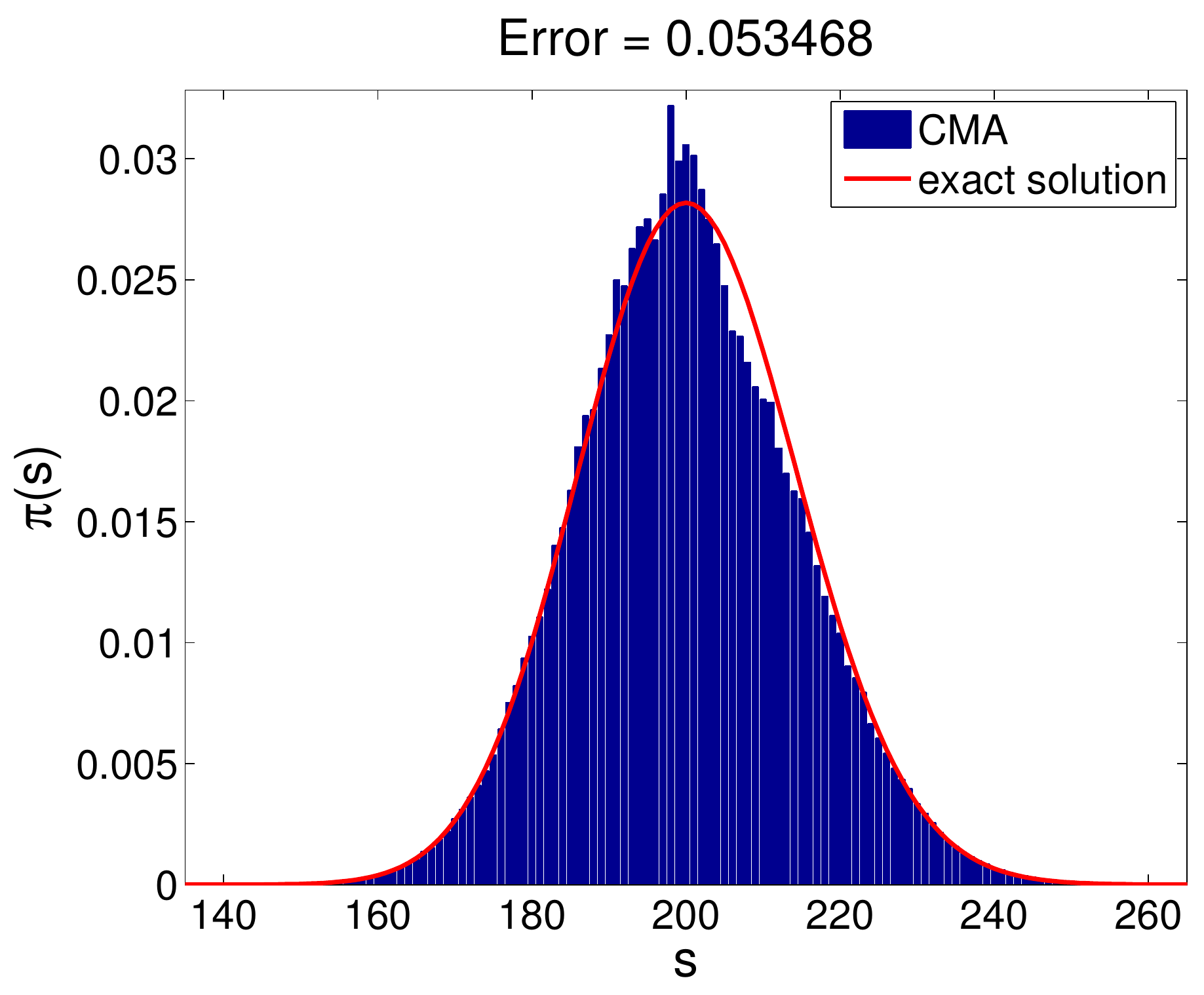}}
\subfigure[$ L=100$]{\includegraphics[width=0.234\columnwidth]{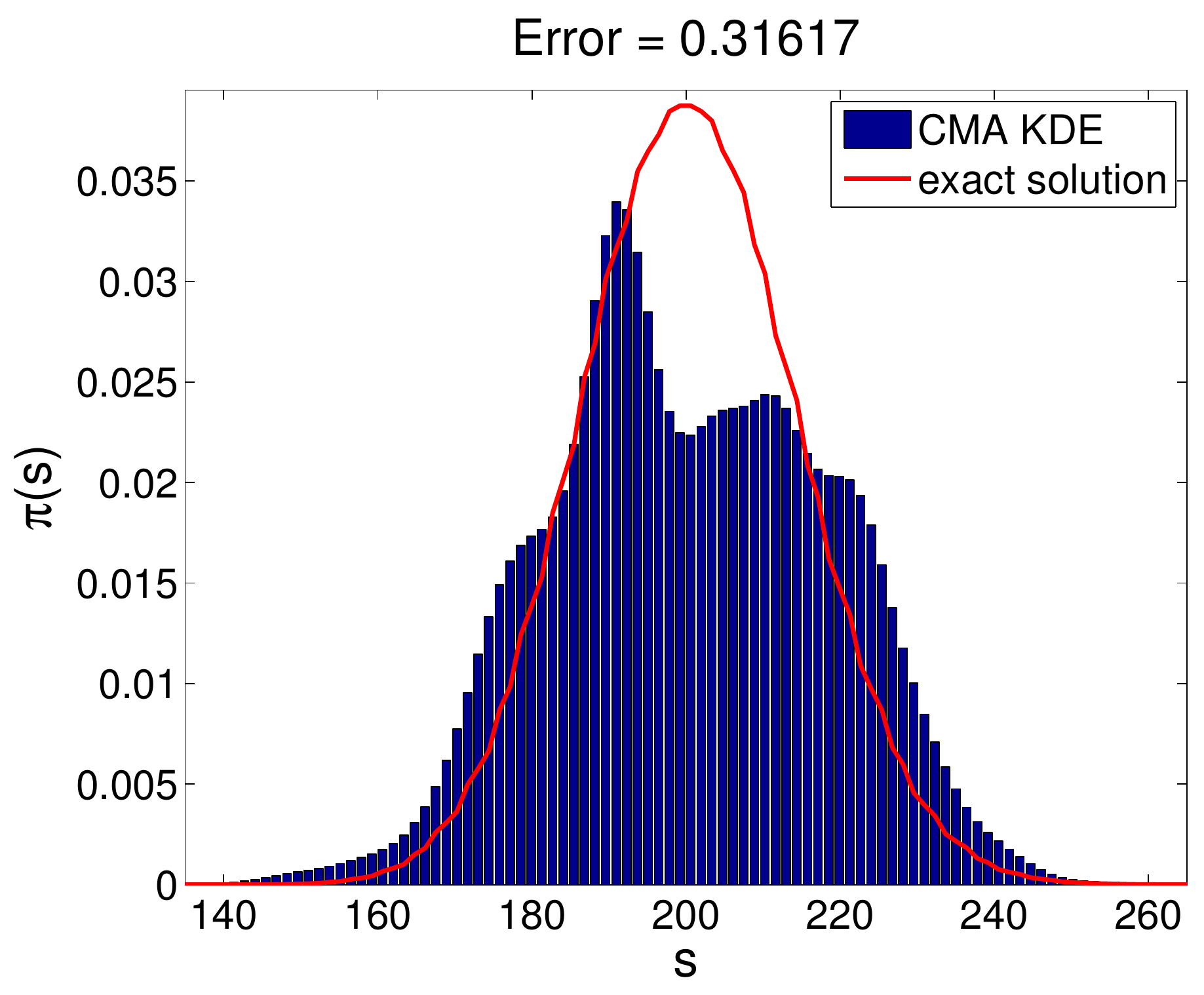}}
\subfigure[$ L=1,000$]{\includegraphics[width=0.234\columnwidth]{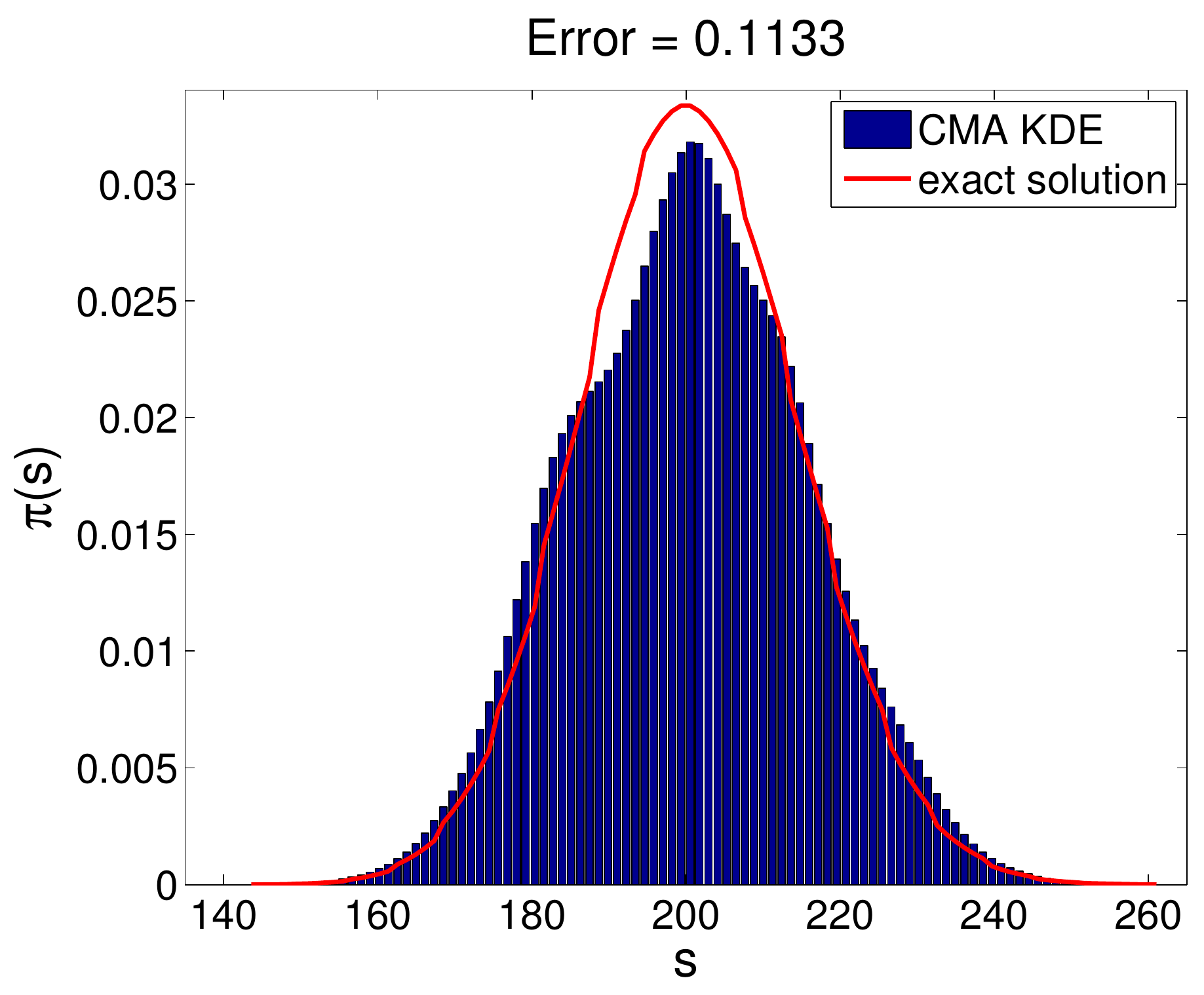}}
\subfigure[$ L=10,000$]{\includegraphics[width=0.234\columnwidth]{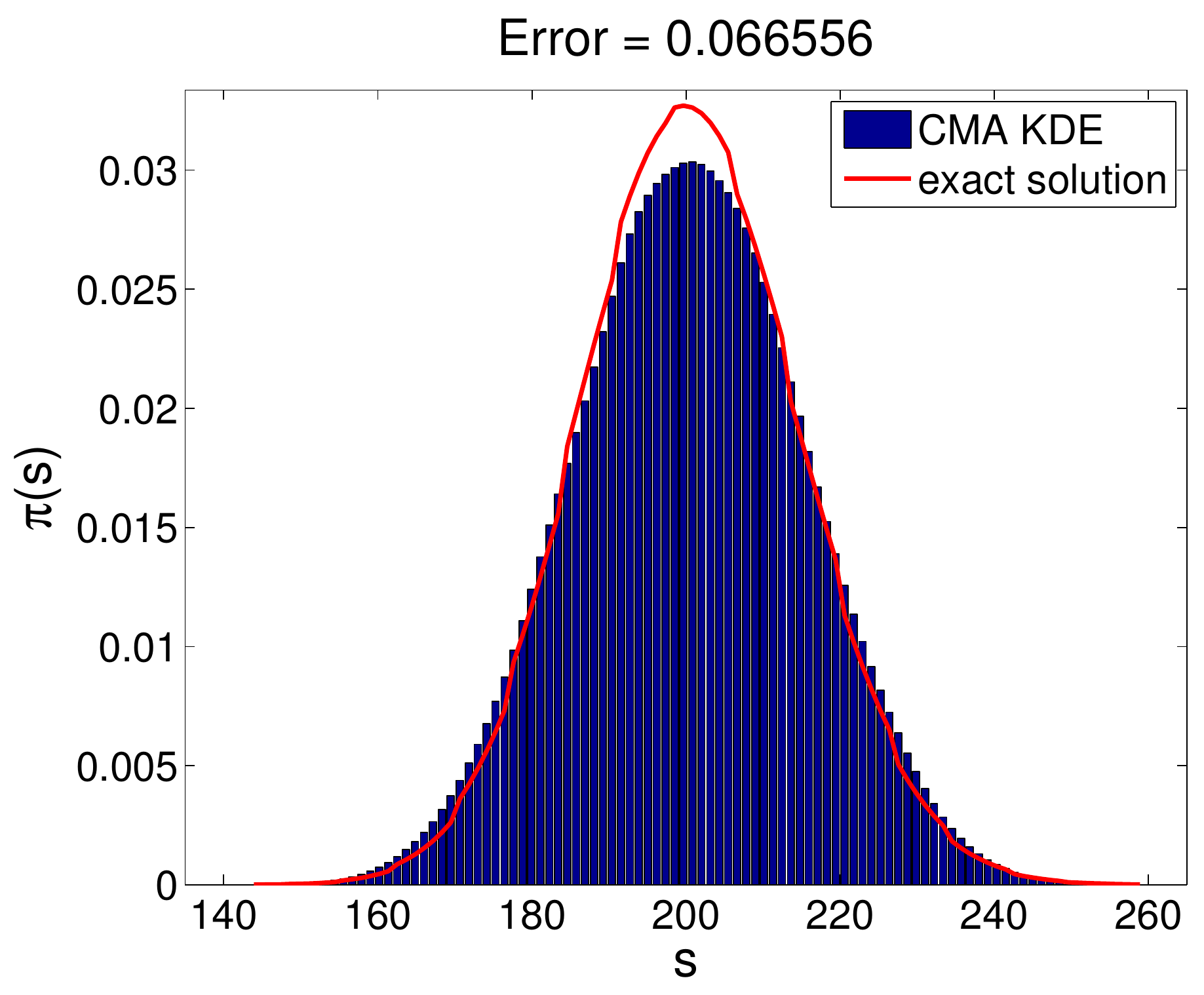}}
\subfigure[$ L=100,000$]{\includegraphics[width=0.234\columnwidth]{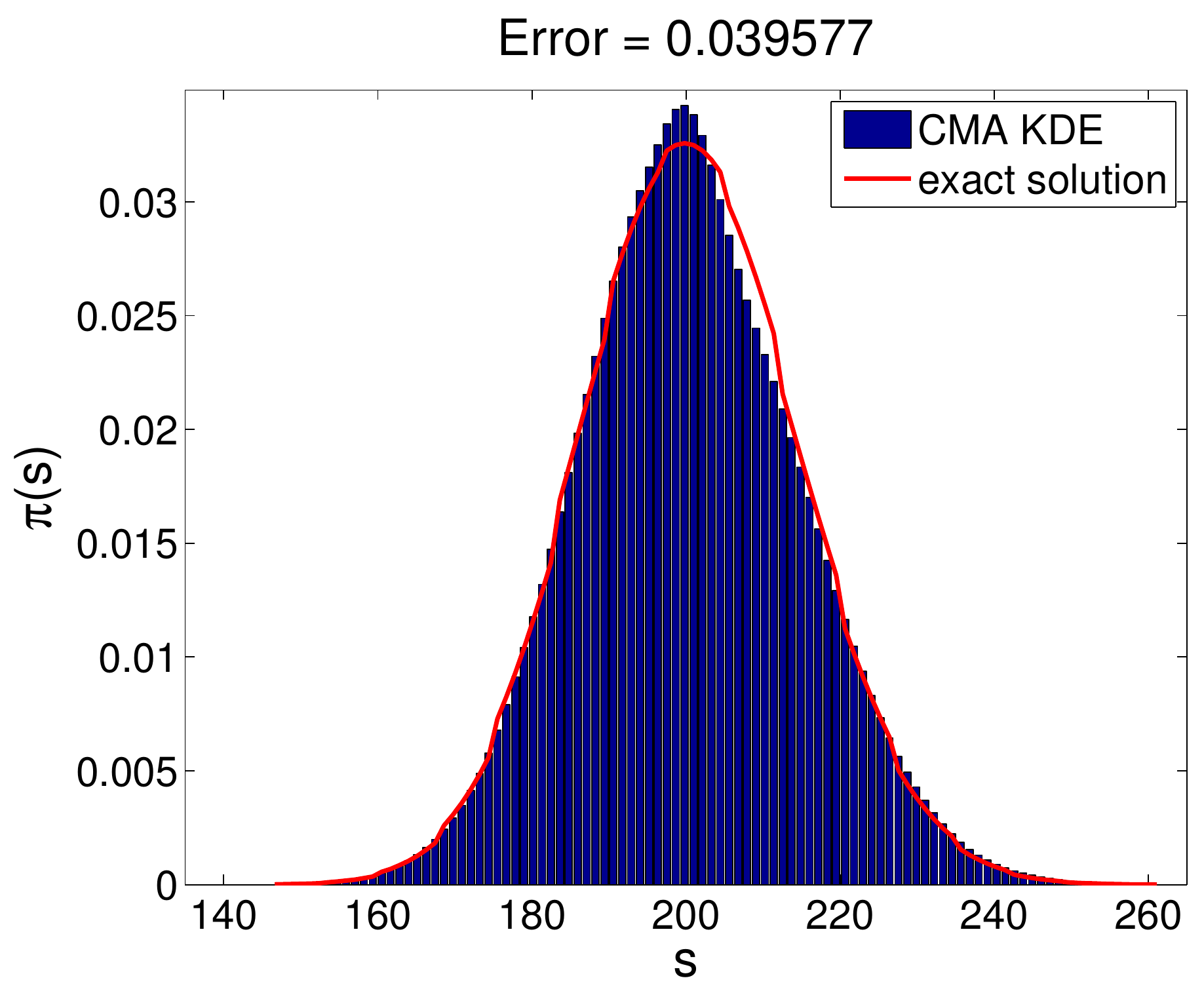}}
\end{center}
\caption{
{\rm (a)--(d)} {\it The stationary distribution of the slow variable computed by the
CMA for} {\rm CS-I,} {\it using knowledge of the slow variable (blue histograms).
The red solid line is the exact solution, $\mathbb{P}(S=s)$, obtained by 
solving the CME of the full system} {\rm CS-I.} 
{\it The CMA approach runs the CSSA algorithm for each value of 
the slow variable $S=s$ until $L_c=\{ 10^2, 10^3, 10^4, 10^5 \}$ changes of the 
slow variable occur. Panels}
{\rm (e)--(f)} {\it show the CMA-computed distribution after smoothing 
out by the Kernel Density Estimation procedure.}
}
\label{figure5p5}
\end{figure}

\begin{figure}[t]
\begin{center}
\subfigure[$ L=100$]{\includegraphics[width=0.234\columnwidth]{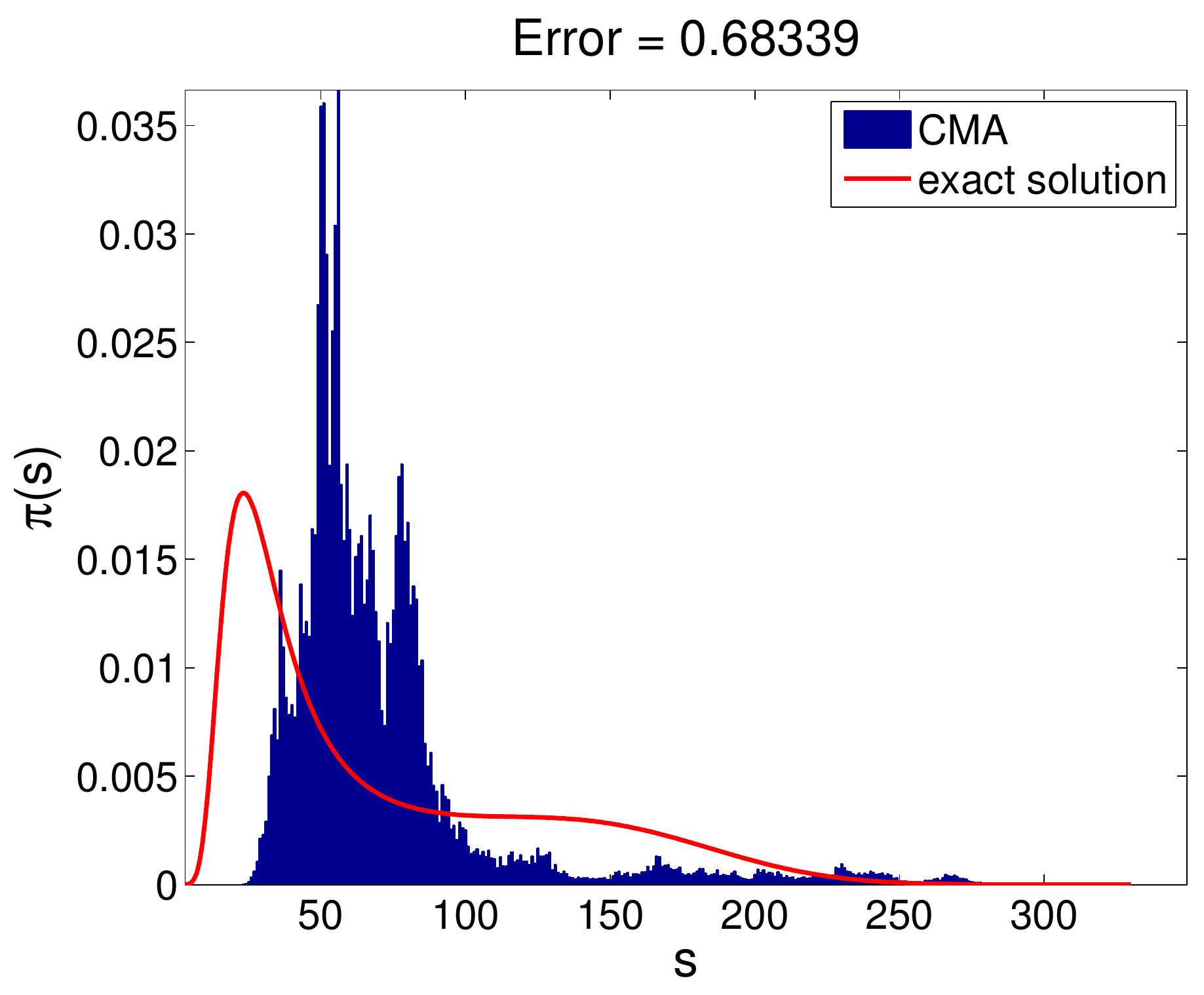}}
\subfigure[$ L=1,000$]{\includegraphics[width=0.234\columnwidth]{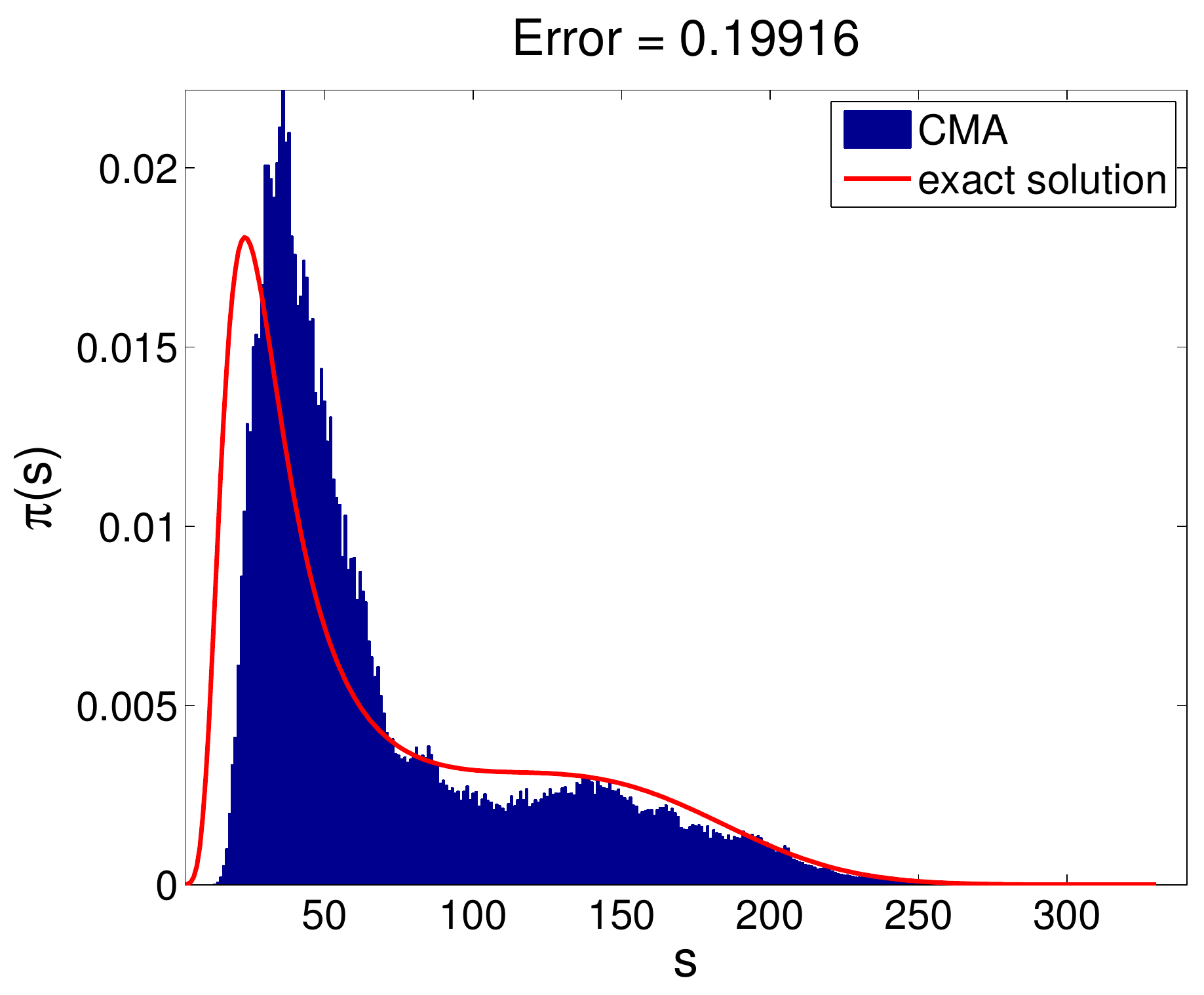}}
\subfigure[$ L=10,000$]{\includegraphics[width=0.234\columnwidth]{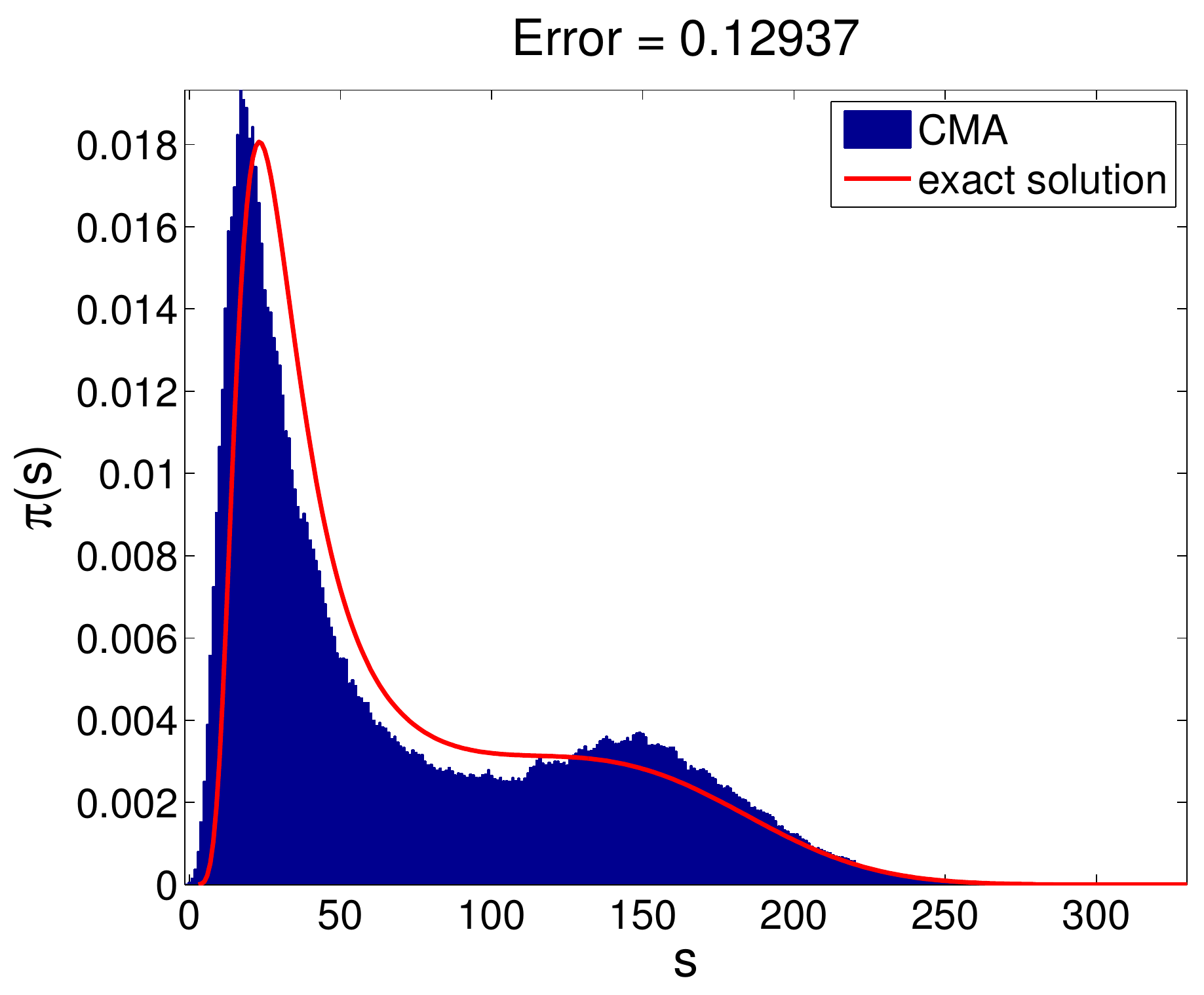}}
\subfigure[$ L=100,000$]{\includegraphics[width=0.234\columnwidth]{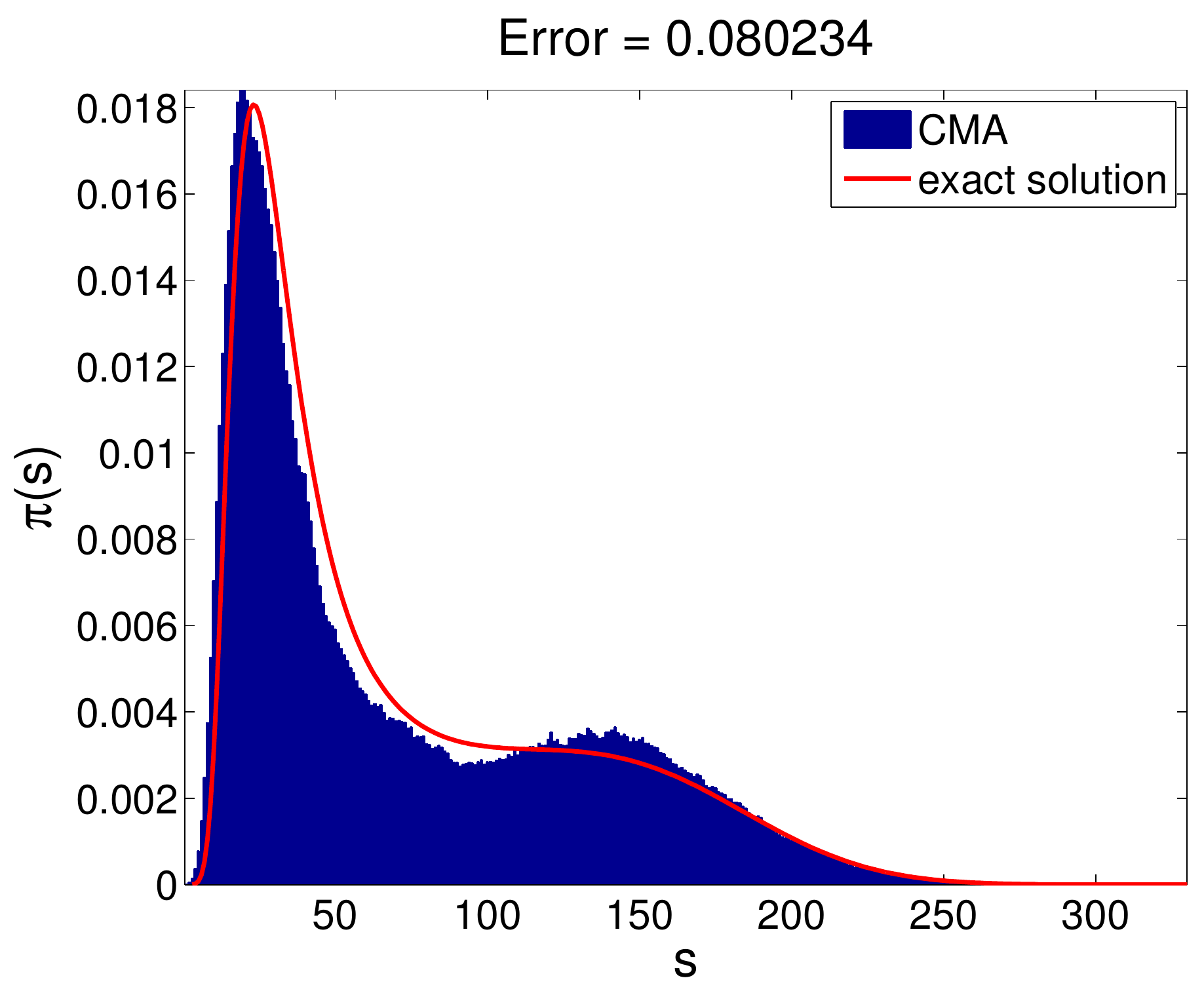}}
\subfigure[$ L=100$]{\includegraphics[width=0.234\columnwidth]{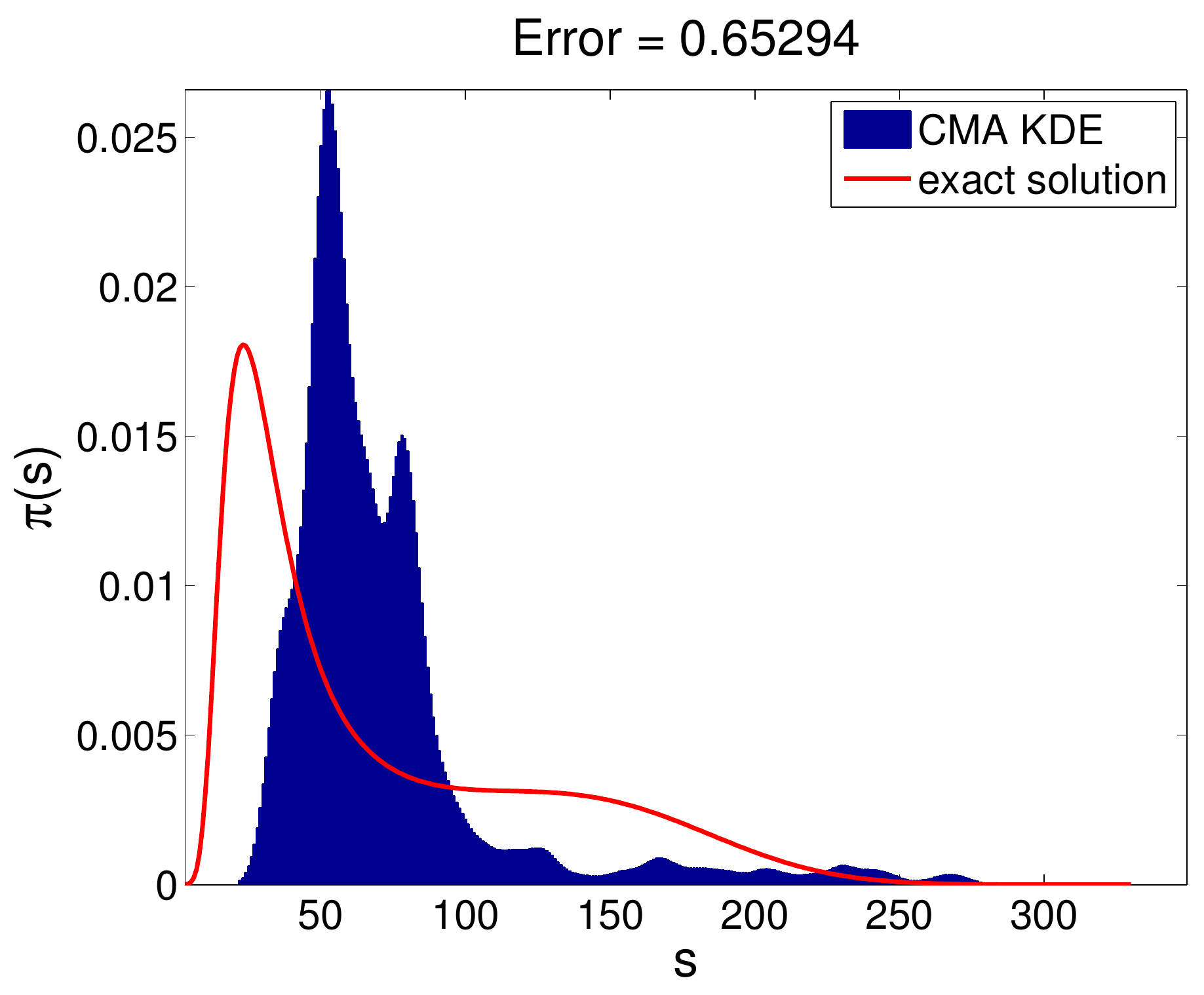}}
\subfigure[$ L=1,000$]{\includegraphics[width=0.234\columnwidth]{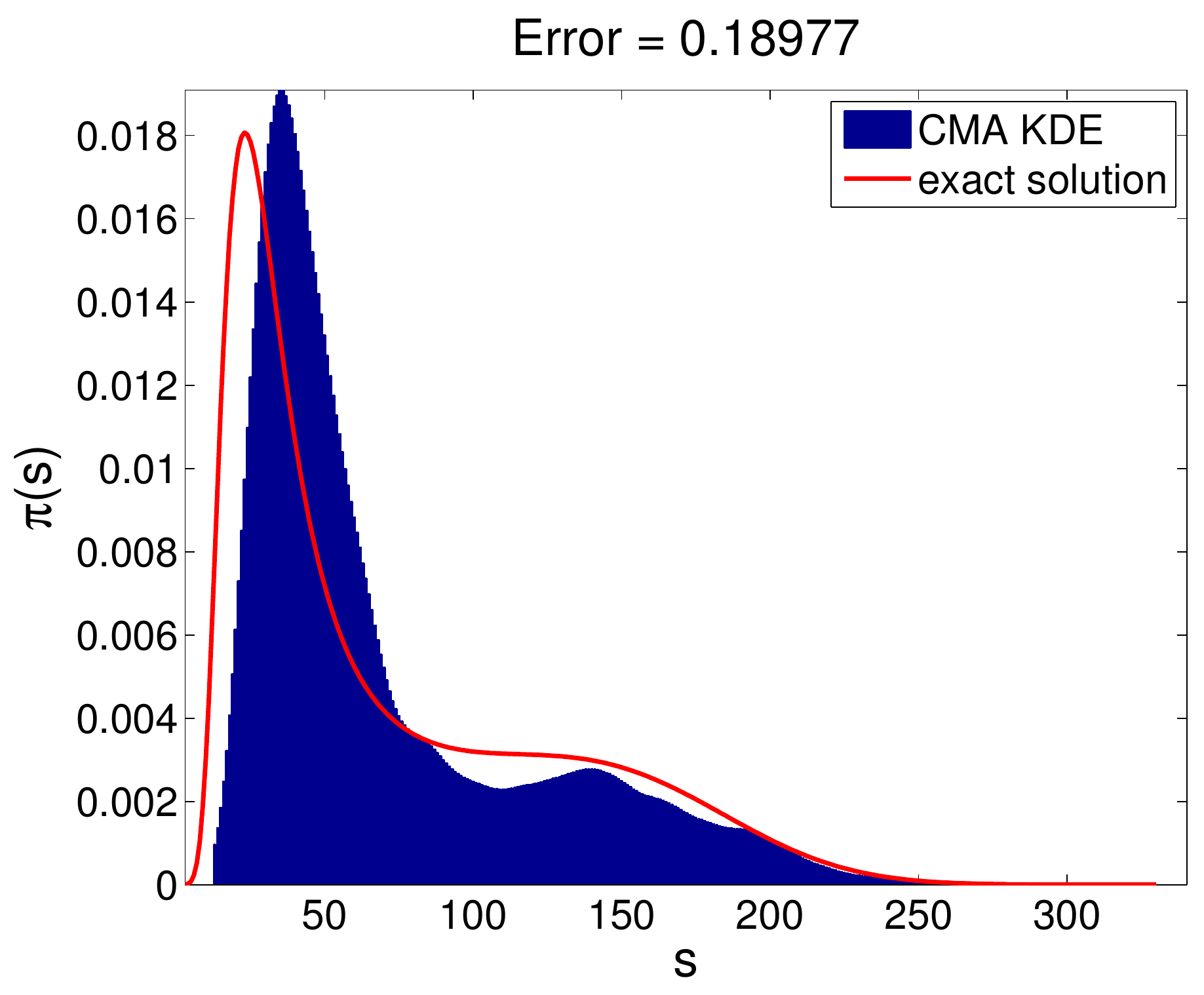}}
\subfigure[$ L=10,000$]{\includegraphics[width=0.234\columnwidth]{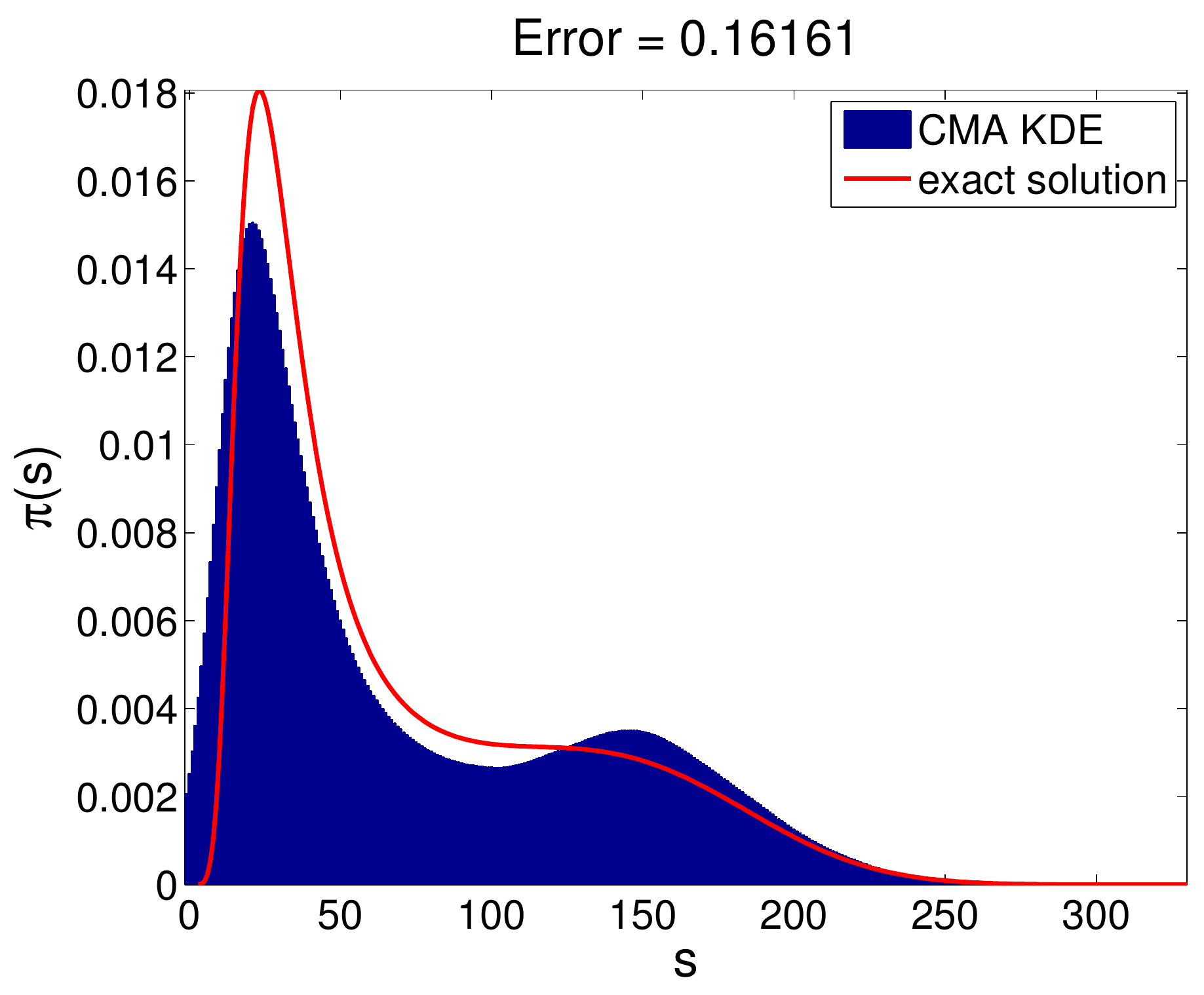}}
\subfigure[$ L=100,000$]{\includegraphics[width=0.234\columnwidth]{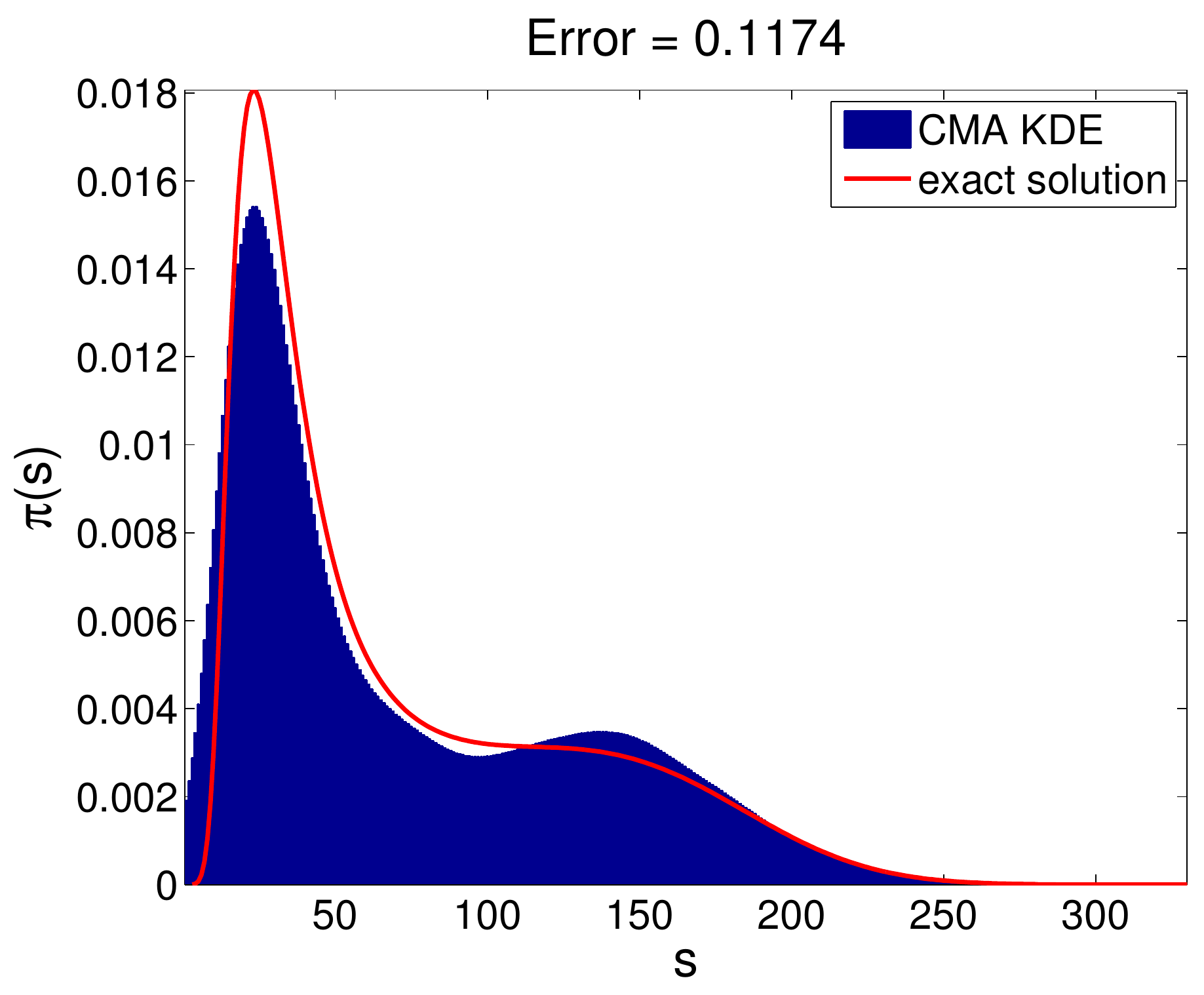}}
\end{center}
\caption{
{\rm (a)--(d)} {\it The stationary distribution of the slow variable computed by the
CMA for} {\rm CS-II,} {\it using knowledge of the slow variable (blue histograms).
The red solid line is the exact solution, $\mathbb{P}(S=s)$, obtained by 
solving the CME of the full system} {\rm CS-II.} 
{\it The CMA approach runs the CSSA algorithm for each value of 
the slow variable $S=s$ until $L_c=\{ 10^2, 10^3, 10^4, 10^5 \}$ changes of the 
slow variable occur. Panels}
{\rm (e)--(f)} {\it show the CMA-computed distribution after smoothing out 
by the Kernel Density Estimation procedure.}
}
\label{figure5p6}
\end{figure}

In Table \ref{tabpi} we show numerical results that highlight the accuracy 
improvement of the ADM-CLE approach compared to the CMA approach of \cite{cssa}. 
For the latter method, we run the CSSA algorithm \cite{cssa} for each value 
of the slow variable, until $L_c$ changes of the slow variable occur. 
As expected, the accuracy of the CMA algorithm improves as $L_c$ increases, 
at the cost of additional computational running time of the method. 
In comparison, our stochastic simulation free approach yields significantly 
more accurate results, with errors that are at least one order of magnitude 
lower than the CMA method with $L=20,000$. We plot in Figures \ref{figure5p5} 
and \ref{figure5p6} (top rows) the estimated stationary distribution 
of the CMA method for both chemical systems considered throughout this paper, 
for several values of the $L$ parameter. The bottom rows of the same 
Figures \ref{figure5p5} and \ref{figure5p6} show the estimated 
distribution, after smoothing out by the Kernel Density Estimation (KDE).

\begin{table}[t]
\centerline{
\begin{tabular}{| l ||   l|l|l|l|l|l|l|| l|}
\hline
	& $L_c=100$   & $L_c=500$  & $L_c=2,000$   
	& $L_c=5,000$ & $L_c=10,000$ & $L_c=20,000$ & ADM-CLE \\
\hline 
  CS-I   &  0.21768  &  0.089228 &  0.10723 
  & 0.045092 & 0.040542 &  0.066988 & 0.003389 \\
\hline
 CS-II & 0.66641    & 0.49063 &   0.14206 &  0.070711  
 &  0.12937 & 0.10995 &  0.028239 \\
\hline
\end{tabular}}
\caption{
{\it The distance (as measured by the error in $(\ref{errorDist})$)  between 
the estimated and the ground truth probability distributions of the slow 
variable, for the CMA algorithm which runs the CSSA algorithm for each 
value of the slow variable $S=s$, until $L_c$ changes of the slow variable 
occur. The rightmost column shows the recovery errors for our proposed 
Markov-based approach.}
}
\label{tabpi}
\end{table}

\section{Summary and discussion}   
\label{SummaryDiscussion}

In this paper we have introduced an ADM-CLE approach  for detecting intrinsic 
slow variables in high-dimensional dynamic data, generated by stochastic dynamical 
systems. In the original ADM framework, the local bursts of simulations initiated 
at each data point to estimate the local covariances are computationally expensive,
a shortcoming we avoid by using an approximation of the CLE. A second innovation
that further improved the computational performance relates to the underlying similarity
graph, a starting point for the diffusion map approach. By exploiting the spectrum
of each local covariance matrix, we built a sparse ellipsoid-like neighborhood graph 
at each point in the data set, with the end result of being able to build a sparse
similarity graph that requires the computation of a much smaller number of distances, 
which makes the ADM-CLE approach scalable to networks with thousands or even tens of 
thousands of nodes. For the two illustrative examples considered in this paper, the 
size of the resulting graphs is $N = 10,201$ for CS-I, and $N = 12,100$ for CS-II, 
respectively. Had these graphs been complete graphs, the number of resulting 
weighted edges would be over 50 million, while in our computations, the number 
of edges is approximately 2.9 million for CS-I, and 3.9 million for CS-II.

We have proposed a spectral-based method for inferring the slow variable 
present within the chemical system without any prior knowledge of its structure,
and a Markov-based approach for estimating its stationary distribution. We augment
the proposed algorithmic approach with numerical simulations that confirm that 
the ADM-CLE approach can compare favorably for some systems to the CMA for estimating 
the stationary distribution of such slow variables.     
The ADM-CLE approach can also be applied to systems with a low number
of states of slow variables. The CMA, as introduced in~\cite{cssa}, is more suitable
for systems where the slow variable(s) can take many different values, because
the CMA uses an underlying SDE approximation for the behaviour of the slow variables.
One option to overcome this problem is to estimate effective propensity functions
of the slow subsystem~\cite{cssa2014sim}.
An open question is to extend the ADM-CLE to systems where the range of 
the $X_i$ variables is very large. In the ADM-CLE approach applied to CS-I
or CS-II, we associate a state (i.e., node in the initial graph) to each possible 
combination of pairs of states $(x_1,x_2)$, an approach no longer feasible whenever 
the range of the variables is large. To bypass this problem, one could change 
the discretization of the state space and modify accordingly the 
Markov chain based approach (Figure \ref{figure5p2}) used in the ADM-CLE.

The ADM-CLE couples a method for finding slow variables (ADM) with an approach
to compute the stationary distribution of a multiscale chemical reaction network.
Chemical systems depend on a number of parameters (e.g. kinetic rate constants)
and an open question is to extend the ADM approach to situations where one (or more) 
parameters are varied, i.e. to perform bifurcation analysis of multiscale stochastic
chemical systems~\cite{liao}. 

Finally, we point out recent work of Dsilva et al.~\cite{DsilvaEtAl}, who 
also rely on the ADM framework to discover nonlinear intrinsic variables 
in high-dimensional data in the context of multiscale simulations, with the 
task of merging different simulations ensembles or partial observations of 
such ensembles in a globally consistent manner. Their work is motivated by 
the fact that often one is not merely interested in extracting the hidden 
(slow) variables from the underlying low-dimensional manifold,
given partial observations $\mathbf{x}^{(i)}, i=1,2,\ldots,N$ 
as in the ADM setting, but also in extending high-dimensional 
functions on a set of points lying in a low-dimensional manifold. 
Their proposed approach relies on the so called Laplacian 
Pyramids~\cite{LaplacianPyramids}, a multiscale algorithm for extending 
a high-dimensional function defined on a set of points in the space 
of intrinsic variables to a second set of points not in the data set, by using Laplacian 
kernels of decreasing bandwidths (the $\varepsilon$ parameter 
in (\ref{defWij})).

\end{document}